\documentstyle[floats,tighten,preprint,eqsecnum,prd,aps,graphicx]{revtex}

\def\diag{\mathop{\rm diag}}
\def\R#1{{\cal R}[#1]}
\def\tr{\mathop{\rm tr}}

\begin{document}  


\preprint{\parbox[t]{15em}{\raggedleft
FERMILAB-PUB-01/394-T \\
KEK-CP-119 \\
YITP-01-90 \\
hep-lat/0112045\\[2.0em]}}
\draft 

\title{Application of heavy-quark effective theory to lattice QCD:\\
III.~Radiative corrections to heavy-heavy currents}

\author{Junpei Harada,$^1$
Shoji~Hashimoto,$^2$ 
Andreas~S.~Kronfeld,$^3$ and
Tetsuya~Onogi,$^{1,4}$\cite{TO}}

\address{$^1$Department of Physics, Hiroshima University,
	Higashi-Hiroshima 739-8526, Japan \\
$^2$High Energy Accelerator Research Organization (KEK),
	Tsukuba 305-0801, Japan \\
$^3$Theoretical Physics Department,
	Fermi National Accelerator Laboratory, Batavia, Illinois 60510 \\
$^4$Yukawa Institute for Theoretical Physics, Kyoto University, 
	Sakyo-ku, Kyoto 606-8502, Japan}

\date{21 December 2001}
\maketitle

\begin{abstract}
We apply heavy-quark effective theory (HQET) to separate long- and
short-distance effects of heavy quarks in lattice gauge theory.
In this paper we focus on flavor-changing currents that mediate
transitions from one heavy flavor to another.
We stress differences in the formalism for heavy-light currents, which 
are discussed in a companion paper, showing how HQET provides a 
systematic matching procedure.
We obtain one-loop results for the matching factors of lattice 
currents, needed for heavy-quark phenomenology, such as the 
calculation of zero-recoil form factors for the semileptonic decays 
$B\to D^{(*)}l\nu$.
Results for the Brodsky-Lepage-Mackenzie scale~$q^*$ are also given.
\end{abstract} 



\section{Introduction}
\label{sec:intro}

This paper applies heavy-quark effective theory (HQET) to study the
renormalization in lattice gauge theory of currents containing heavy
quarks.
It is a sequel to our papers on power corrections~\cite{Kronfeld:2000ck}
and on radiative corrections to heavy-light
currents~\cite{Harada:2001hl}.
Here we treat the case where one heavy quark flavor decays to another.
To make contact with heavy-quark phenomenology, we denote the two
flavors by the labels~$b$ and~$c$.
In the description of lattice gauge theory with HQET, discretization
effects of the heavy quarks are absorbed into the short-distance
coefficients of the effective Lagrangian and effective currents.
The key difference between this work and Ref.~\cite{Harada:2001hl} is 
that now, with two heavy flavors, HQET is used to describe both heavy 
quarks.
Thus, the short-distance coefficients are functions of $m_ba$ and
$m_ca$, where $m_b$ and $m_c$ are the heavy quark masses, and $a$~is
the lattice spacing.

Heavy-heavy currents are needed to calculate the zero-recoil form
factors of the semi-leptonic decays $B\to Dl\nu$ and $B\to D^*l\nu$,
as well as the change in the form factors away from zero recoil.
These decays are of great phenomenological interest, because with
reliable lattice calculations of $B\to D^{(*)}$ transitions one can
make a model-independent determination of the element $V_{cb}$ of the
Cabibbo-Kobayashi-Maskawa (CKM) matrix.
For further background on the impact of lattice matrix elements on CKM
phenomenology see, for example, Refs.~\cite{Kronfeld:1993jf}.

The application of HQET to lattice gauge theory began with the
consideration of power corrections~\cite{Kronfeld:2000ck}, building on
the demonstration~\cite{El-Khadra:1997mp} that Wilson's formulation of
lattice fermions~\cite{Wilson:1975hf} have a well-defined heavy quark
limit.
In particular, the Isgur-Wise heavy quark symmetries~\cite{Isgur:1989vq}
emerge whenever the heavy-quark mass $m_h\gg\Lambda_{\text{QCD}}$,
even if $m_h\sim a^{-1}$.
(The label~$h$ denotes a generic heavy flavor.)
Because Wilson fermions have the the same symmetries and fields as 
continuum heavy quarks, on-shell lattice correlation functions can be 
described with HQET.
This point of view is sometimes called the ``non-relativistic
interpretation'' of lattice QCD~\cite{El-Khadra:1997mp}.

In describing either lattice gauge theory or continuum QCD with~HQET,
physics at short distances is lumped into short-distance coefficients.
Several short distances---two inverse heavy quark masses and (on the
lattice) the lattice spacing---arise in the decay of one heavy flavor
to another.
The coefficients depend not only on $m_ba$ and $m_ca$, but also on other
``irrelevant'' parameters of the lattice action and currents.
By adjusting these parameters of the action and currents, lattice
gauge theory can be tuned, term by term, to the heavy-quark expansion
of continuum QCD~\cite{El-Khadra:1997mp,Kronfeld:2000ck,Harada:2001hl}.
In this adjustment, details of HQET, such as its renormalization scheme,
drop out.
Matching with HQET is, thus, an intermediate conceptual step that
explains how to match lattice gauge theory to QCD when $m_ha\not\ll1$.
It provides an attractive extension of more familiar matching
procedures, such as those based on the Symanzik effective field theory~%
\cite{Symanzik:1979ph,Symanzik:1983dc,Luscher:1996sc,Jansen:1996ck},
which usually assume $m_ha\ll1$.

In addition to developing the formalism, which holds to all orders
in perturbation theory, we compute the one-loop terms of the matching
coefficients of the leading dimension vector and axial vector currents.
We present compact expressions of the integrands of the momentum
integration, for the Fermilab action and
currents~\cite{El-Khadra:1997mp}.
We also present numerical results for the
Sheikholeslami-Wohlert (SW)~\cite{Sheikholeslami:1985ij} and
Wilson~\cite{Wilson:1975hf} actions, which are the special cases most
often used in the (non-relativistic interpretation of the) Fermilab
heavy-quark method.
We include the so-called rotation
terms~\cite{Kronfeld:1995nu,El-Khadra:1997mp} in the currents, which
are needed for tree-level matching at dimension~four.
It is also possible to obtain most of the normalization
non-perturbatively~\cite{Hashimoto:2000yp,Hashimoto:2001ds,El-Khadra:2001rv},
and we give separately the residual short-distance correction.

Our perturbative results have been used in the calculation of the 
zero-recoil form factor for $B\to D^*l\nu$~\cite{Hashimoto:2001ds} 
and, indirectly, in the calculation for form factors for
$B\to\pi l\nu$~\cite{El-Khadra:2001rv}.
One-loop results without the rotation were given in a preliminary report
of this work~\cite{Kronfeld:1999tk}, and used in the calculation of
the zero-recoil form factors for $B\to Dl\nu$~\cite{Hashimoto:2000yp}.

This paper is organized along the lines Ref.~\cite{Harada:2001hl}.
Section~\ref{sec:hqet} discusses how to use HQET to separate long- and
short-distance physics with (continuum) effective field theories.
In particular, we obtain a definition of the matching factors
of the heavy-heavy vector and axial vector currents.
Section~\ref{sec:lattice} introduces a specific definition of lattice 
currents suited to improvement in the HQET matching procedure; these 
currents generalize those used recently for $B\to D^{(*)}$ matrix 
elements~\cite{Hashimoto:2000yp,Hashimoto:2001ds}.
In Sec.~\ref{sec:loop} we present one-loop results for the
matching factors, including the scale~$q^*$ in the 
Brodsky-Lepage-Mackenzie (BLM) scale-setting 
prescription~\cite{Brodsky:1983gc,Lepage:1993xa}.
Some concluding remarks are made in Sec.~\ref{sec:conclusions}.
Three appendices contain details of the one-loop calculation,
including an outline of a method to obtain compact expressions,
and explicit results for the one-loop Feynman integrands for the
vertex functions with the Fermilab action.

Instead of printing tables of the numerical results in
Sec.~\ref{sec:loop}, we are making a suite of programs freely
available~\cite{p:epaps}.
This suite includes programs for the heavy-light currents treated in
our companion paper~\cite{Harada:2001hl}.

\section{Matching with HQET}
\label{sec:hqet}

In our companion paper~\cite{Harada:2001hl} we reviewed how the standard
Symanzik description of cutoff effects breaks down when $m_ha\not\ll1$.
It is worth re-emphasizing that it is the \emph{description} that breaks
down---particularly the Taylor series of short-distance coefficients
in powers of $m_ha$.
Lattice gauge theory remains well-defined for all~$m_ha$, but one needs
other tools to understand how to relate lattice observables to
continuum~QCD.
The obvious alternatives are the effective field theories 
HQET~\cite{Eichten:1987xu,Eichten:1990zv,Grinstein:1990mj,%
Georgi:1990um,Eichten:1990vp} and 
NRQCD~\cite{Caswell:1986ui,Lepage:1987gg}, which exploit the simpler 
dynamics of systems with one or more heavy quarks.
The simpler dynamics also emerge in lattice gauge
theory with massive fermions, so the effective
theories also can be re-applied to understand lattice
observables~\cite{Kronfeld:2000ck,Harada:2001hl,El-Khadra:1997mp}.

In this section we show, in the case of heavy-heavy currents,
how to use HQET to match lattice gauge theory to continuum~QCD.
We first recall the HQET description of heavy-heavy currents in
continuum QCD, paralleling the discussion in Ref.~\cite{Harada:2001hl}.
We then explain what changes are needed to describe 
lattice gauge theory with heavy fermions.
We focus on HQET because of the phenomenological importance of 
$B\to D^{(*)}$ transitions; for quarkonium the logic could be repeated 
with NRQCD velocity-counting rules~\cite{Caswell:1986ui,Lepage:1987gg}.
Unlike the standard Symanzik 
program~\cite{Symanzik:1979ph,Symanzik:1983dc,Luscher:1996sc,Jansen:1996ck},
the HQET approach works even in the region where $m_ha$ is no longer small.
Like the usual HQET, however, it requires ($\bbox{p}$ is a typical 
momentum)
\begin{equation}
	m_h \gg \bbox{p},\, \Lambda_{\text{QCD}},
	\label{eq:when}
\end{equation}
but once this condition holds (and $\bbox{p}a\ll1$), our formalism 
provides a systematic description of lattice observables for 
all~$m_ha$, $h=b$,~$c$.

The conventions for HQET are the same as those given Sec.~III of
Ref.~\cite{Kronfeld:2000ck}.
Each HQET quark field carries a velocity label.
For the time being we use two different velocities, $v$ and~$v'$, for 
the two flavors.
The velocities can be chosen somewhat arbitrarily, but HQET is a good
description of QCD if each is close to the velocity of the hadron
containing the heavy quark.%
\footnote{In NRQCD, the relative velocity between the heavy
quark and heavy anti-quark of quarkonium should not be confused with
the velocities~$v$ and~$v'$ introduced here.
Note that it is also possible to formulate NRQCD with a total velocity
label like~$v$~\cite{Hashimoto:1996in,Sloan:1998fc}.}
The fourth Euclidean component $v_4=iv^0$, so in the rest frame
$v=(i,\bbox{0})$, and similarly for $v'$.
The metric is taken to be $\diag(\pm 1,1,1,1)$, with the upper (lower)
sign for Euclidean (Minkowski) spacetime.
In either case, $v^2={v'}^2=-1$.
We denote the two HQET fields as $\bar{c}_{v'}$ and~$b_v$.
When the flavor of quark is unimportant, we write formulas with the
more generic symbol~$h_v$.
The HQET field $h_v$ satisfies the constraint
$\case{1}{2}(1-i\kern+0.1em /\kern-0.55em v)h_v = h_v$, or
\begin{equation}
	            \kern+0.1em /\kern-0.55em v b_v = ib_v, \quad
	\bar{c}_{v'}\kern+0.1em /\kern-0.55em v'    = i\bar{c}_{v'}.
	\label{eq:Pvh}
\end{equation}
Physically this constraint means that only quarks, but not anti-quarks,
are described.
The tensor $\eta^\mu_\nu=\delta^\mu_\nu+v^\mu v_\nu$ projects onto
components orthogonal to~$v$.
For a vector~$p$, the component orthogonal to~$v$ is
$p_\perp^\mu=\eta^\mu_\nu p^\nu=p^\mu+v^\mu v\cdot p$.
In the rest frame, these are the spatial components.
Similarly, ${\eta'}^\mu_\nu=\delta^\mu_\nu+{v'}^\mu v'_\nu$,
and $p_{\perp'}^\mu={\eta'}^\mu_\nu p^\nu$.

HQET describes the dynamics of heavy-light bound states with an
effective Lagrangian built from~$c_{v'}$ and~$b_v$.
For each flavor one writes
\begin{equation}
	{\cal L}_{\text{QCD}} = - \bar{q}({\kern+0.1em /\kern-0.65em D}+m)q
		\doteq {\cal L}_{\text{HQET}},
	\label{eq:QCD=HQET}
\end{equation}
where the symbol $\doteq$ means ``has the same on-shell matrix 
elements as''.
The HQET Lagrangian consists of a series of sets of terms
\begin{equation}
	{\cal L}_{\text{HQET}} =
	{\cal L}^{(0)} +
	{\cal L}^{(1)} +
	{\cal L}^{(2)} + \cdots,
	\label{eq:L}
\end{equation}
where ${\cal L}^{(s)}$ consists of all operators of dimension~$s+4$
built out of $\bar{h}_v$ and $h_v$ (and gluons and light quarks).
The ultraviolet regulator and renormalization scheme of the
two sides of Eq.~(\ref{eq:QCD=HQET}) need not be the same.
For the aims of this paper it is enough to consider the first two
terms, ${\cal L}^{(0)}$ and ${\cal L}^{(1)}$, but the generalization 
to higher dimension should be clear.

The leading, dimension-four term is 
\begin{equation}
	{\cal L}^{(0)} = \bar{h}_v(iv\cdot D - m_h)h_v,
	\label{eq:L0}
\end{equation}
where $m_h$ is the mass of flavor~$h$.
${\cal L}^{(0)}$ is a good starting point for the heavy-quark expansion,
which treats the higher-dimension operators as small.
The mass term in ${\cal L}^{(0)}$ is usually omitted, because it 
obscures the heavy-quark flavor symmetry (though only 
slightly~\cite{Kronfeld:2000ck}).
By heavy-quark symmetry, it has an effect neither on bound-state wave
functions nor, consequently, on matrix elements.
It does affect the mass spectrum, but only additively.

When the mass term is included, higher-dimension operators are
constructed with ${\cal D}^\mu=D^\mu-iMv^\mu$ and
${\cal D}'=D^\mu-iMv'$.
Here $M$ selects the mass of flavor~$h$: $Mh_v=m_hh_v$.
To describe on-shell matrix elements one may omit operators that
vanish by the equations of motion, $-iv\cdot{\cal D}b_v=0$ and
$\bar{c}_{v'}iv'\cdot\loarrow{\cal D}'=0$,
In practice, higher dimension operators are constructed with
${\cal D}^\mu_\perp=D^\mu_\perp$ 
(or, for velocity~$v'$, $D^\mu_{\perp'}$)
and $[{\cal D}^\mu,{\cal D}^\nu]=[D^\mu,D^\nu]=F^{\mu\nu}$.
The dimension-five terms in the Lagrangian are
\begin{equation}
	{\cal L}^{(1)} = {\cal C}_2 {\cal O}_2 +
		{\cal C}_{\cal B} {\cal O}_{\cal B},
	\label{eq:L1}
\end{equation}
where ${\cal C}_2$ and ${\cal C}_{\cal B}$ are short-distance
coefficients, and
\begin{eqnarray}
	{\cal O}_2 & = &
		\bar{h}_vD_\perp^2 h_v, \label{eq:O2} \\
	{\cal O}_{\cal B} & = & 
		\bar{h}_v s_{\alpha\beta}B^{\alpha\beta}h_v,
		\label{eq:OB}
\end{eqnarray}
with $s_{\alpha\beta}=-i\sigma_{\alpha\beta}/2$
and  $B^{\alpha\beta}=\eta^\alpha_\mu\eta^\beta_\nu F^{\mu\nu}$.
In operators with two HQET fields (for flavor-changing currents, say)
$\loarrow{\cal D}'$ and ${\cal D}$ appear together.

In Eq.~(\ref{eq:L0}) one should think of the quark mass $m_h$ as 
a short-distance coefficient.
By reparametrization invariance~\cite{Luke:1992cs} and Lorentz 
invariance of (continuum) QCD, the same mass appears in the 
short-distance coefficient of the kinetic energy~${\cal C}_2{\cal O}_2$, 
namely,
\begin{equation}
	{\cal C}_2 = \frac{1}{2m_h}.
\end{equation}
If the operators of HQET are renormalized with a minimal subtraction
in dimensional regularization, then $m_h$ is the (perturbative) pole
mass.
The chromomagnetic operator~${\cal O}_{\cal B}$ depends on the
renormalization point~$\mu$ of HQET, and that dependence is
canceled by
\begin{equation}
	{\cal C}_{\cal B}(\mu) = \frac{z_{\cal B}(\mu)}{2m_h}.
\end{equation}

The description of flavor-changing currents proceeds along similar 
lines.
Let
\begin{equation}
	{\cal V}^\mu=\bar{c}i\gamma^\mu b
\end{equation}
be the flavor-changing vector current, where $\bar{c}$ and $b$ without
the velocity subscripts are QCD fields obeying the Dirac equation.
In HQET ${\cal V}^\mu$ is described by
\begin{equation}
	{\cal V}^\mu \doteq \bar{C}_{V_\parallel} v^\mu \bar{c}_{v'}b_v  +
		\bar{C}_{V_\perp} \bar{c}_{v'}i\gamma^\mu_\perp b_v +
		\bar{C}_{V_{v'}} v^{\prime\mu}_\perp \bar{c}_{v'}b_v -
		\sum_{i=1}^{14} \bar{B}_{Vi} \bar{\cal Q}^\mu_{Vi} + \cdots,
	\label{eq:VcontHQET}
\end{equation}
where $\bar{c}_{v'}$ and $b_v$ are HQET fields, which satisfy
Eq.~(\ref{eq:L}) and whose dynamics are given
by~${\cal L}_{\text{HQET}}$.
The fourteen dimension-four operators are
\begin{eqnarray}
	\bar{\cal Q}^\mu_{V1} & = & - v^\mu \bar{c}_{v'}
		{\kern+0.1em /\kern-0.65em D}_\perp b_v, \label{eq:Q1} \\
	\bar{\cal Q}^\mu_{V2} & = & \bar{c}_{v'}i\gamma^\mu_\perp
		{\kern+0.1em /\kern-0.65em D}_\perp b_v, \label{eq:Q2} \\
	\bar{\cal Q}^\mu_{V3} & = & \bar{c}_{v'}iD^\mu_\perp b_v,
		\label{eq:Q3} \\
	\bar{\cal Q}^\mu_{V4} & = & + v^\mu \bar{c}_{v'}
		{\kern+0.1em /\kern-0.65em \loarrow{D}}_{\perp'} b_v,
		\label{eq:Q4} \\
	\bar{\cal Q}^\mu_{V5} & = & \bar{c}_{v'}
		{\kern+0.1em /\kern-0.65em \loarrow{D}}_{\perp'}
		i\gamma^\mu_{\perp'} b_v, 
		\label{eq:Q5} \\
	\bar{\cal Q}^\mu_{V6} & = & \bar{c}_{v'}i\loarrow{D}^\mu_{\perp'} b_v,
		\label{eq:Q6} \\
	\bar{\cal Q}^\mu_{V7} & = & - v^{\prime\mu}_\perp \bar{c}_{v'}
	    {\kern+0.1em /\kern-0.65em D}_\perp b_v,
		\label{eq:Q7} \\
	\bar{\cal Q}^\mu_{V8} & = & - v^{\prime\mu}_\perp \bar{c}_{v'}
		{\kern+0.1em /\kern-0.65em \loarrow{D}}_{\perp'} b_v,
		\label{eq:Q8} \\
	\bar{\cal Q}^\mu_{V9} & = & - v^\mu \bar{c}_{v'}
		iv'\cdot D_\perp b_v,
		\label{eq:Q9} \\
	\bar{\cal Q}^\mu_{V10} & = & \bar{c}_{v'}i\gamma^\mu_\perp
		iv'\cdot D_\perp b_v,
		\label{eq:Q10} \\
	\bar{\cal Q}^\mu_{V11} & = & - v^\mu \bar{c}_{v'}
		iv\cdot \loarrow{D}_{\perp'} b_v, \label{eq:Q11} \\
	\bar{\cal Q}^\mu_{V12} & = & \bar{c}_{v'}
		iv\cdot \loarrow{D}_{\perp'}
		i\gamma^\mu_{\perp'} b_v,  \label{eq:Q12} \\
	\bar{\cal Q}^\mu_{V13} & = & - v^{\prime\mu}_\perp \bar{c}_{v'}
	    iv'\cdot D_\perp b_v,
		\label{eq:Q13} \\
	\bar{\cal Q}^\mu_{V14} & = & - v^{\prime\mu}_\perp \bar{c}_{v'}
		iv\cdot \loarrow{D}_{\perp'} b_v. \label{eq:Q14}
\end{eqnarray}
Further dimension-four operators are omitted, because they are linear
combinations of those listed and others that vanish by the equations
of motion.
In developing the heavy-quark expansion the appropriate equations of
motion for the heavy quarks are $(-iv\cdot{\cal D})b_v=0$
and $\bar{c}_{v'}(iv'\cdot\loarrow{\cal D}')=0$,
derived from the respective~${\cal L}^{(0)}$.

The QCD axial vector current is
\begin{equation}
	{\cal A}^\mu=\bar{c}i\gamma^\mu\gamma_5b,
\end{equation}
where $\bar{c}$ and $b$ are again Dirac fields.
${\cal A}$ has an HQET description analogous to Eq.~(\ref{eq:VcontHQET}),
\begin{equation}
	{\cal A}^\mu \doteq
		\bar{C}_{A_\perp} \bar{c}_{v'}i\gamma^\mu_\perp \gamma_5b_v -
		\bar{C}_{A_\parallel} v^\mu \bar{c}_{v'}\gamma_5b_v  -
		\bar{C}_{A_{v'}} v^{\prime\mu}_\perp \bar{c}_{v'}\gamma_5b_v -
		\sum_{i=1}^{14} \bar{B}_{Ai} \bar{\cal Q}_{Ai}^\mu + \cdots.
	\label{eq:AcontHQET}
\end{equation}
By convention, each operator $\bar{\cal Q}_{Ai}^{\mu}$ is obtained
from $\bar{\cal Q}_{Vi}^{\mu}$ by replacing $\bar{c}_{v'}$ with
$-\bar{c}_{v'}\gamma_5$.

The coefficients $\bar{C}_J$ and $\bar{B}_{Ji}$ and the 
operators~$\bar{\cal Q}_{Ji}$ play an analogous role to the 
coefficients $C_J$ and $B_{Ji}$ and the operators~${\cal Q}_{Ji}$ 
introduced in Ref.~\cite{Harada:2001hl}, but they are not the same.
The latter are defined in an effective theory using a Dirac 
field~$\bar{q}$ (without a velocity label) to describe the light(er) 
quark, whereas the barred symbols are defined in a theory using an 
HQET field~$\bar{c}_{v'}$ (with a velocity label) to describe the 
lighter (heavy) quark.
The bars are used here to emphasize the difference.

There are many fewer operators when $v'=v$.
One set of operators vanishes because $v'_\perp\to v_\perp=0$, namely,
$\bar{\cal Q}^\mu_{J7}$, $\bar{\cal Q}^\mu_{J8}$, $\bar{\cal Q}^\mu_{J13}$.
$\bar{\cal Q}^\mu_{J14}$; another set vanishes because
$v'\cdot D_\perp,\;v\cdot D_{\perp'}\to v\cdot D_\perp=0$, namely,
$\bar{\cal Q}^\mu_{Jj}$, $j\ge9$.
A~last set vanishes because $P_+(v)\Gamma P_+(v)=0$, where $\Gamma$
stands for the full Dirac structure sandwiched by $\bar{c}_v$ and $b_v$.
In the end, one is left with only $v^\mu\bar{c}_vb_v$,
$\bar{\cal Q}^\mu_{V2}$, $\bar{\cal Q}^\mu_{V3}$, $\bar{\cal Q}^\mu_{V5}$,
and $\bar{\cal Q}^\mu_{V6}$ for the vector current, and
$\bar{c}_vi\gamma^\mu_\perp\gamma_5b_v$, $\bar{\cal Q}^\mu_{A1}$,
and $\bar{\cal Q}^\mu_{A4}$ for the axial vector
current.

The coefficients of the HQET operators depend on the heavy-quark
masses~$m_b$ and~$m_c$ (to balance dimensions), as well as $m_c/m_b$ 
and $\mu/m_b$, where $\mu$ is the renormalization scale.
At the two-loop level and beyond, dependence on light quark masses
$m_q$ also appears.
With two velocities, the coefficients also depend on~$w=-v'\cdot v$.
Although they are not needed below, it is instructive to give the 
coefficients of the dimension-three terms.
At the tree level 
$\bar{C}_{J_\parallel}^{[0]}=\bar{C}_{J_\perp}^{[0]}=1$ 
and $\bar{C}_{J_{v'}}^{[0]}=0$.
Through one loop order, for $v'=v$,
\begin{eqnarray}
	\bar{C}_{V_\parallel} & = & 1 + C_F\frac{g^2(\mu)}{16\pi^2}
		3 f(m_c/m_b),
	\label{eq:Cparallel} \\
	\bar{C}_{A_\perp}     & = & 1 + C_F\frac{g^2(\mu)}{16\pi^2}\left[
		3 f(m_c/m_b) - 2 \right],
	\label{eq:Cperp}
\end{eqnarray}
where
\begin{equation}
	f(z) = \frac{z+1}{z-1}\ln z - 2.
	\label{eq:f}
\end{equation}
The important properties of $f(z)$ are $f(1)=0$, $f(1/z)=f(z)$.
At this order, $\mu$ dependence appears only in the renormalized
coupling~$g^2(\mu)$.
The ``non-$f$'' part of $\bar{C}_{A_\perp}$ given here assumes
that the renormalized axial current satisfies the axial Ward
identity~\cite{Trueman:1995ca}.
At the tree level, the coefficients of the dimension-four currents are
\begin{eqnarray}
	 \bar{B}_{V1}^{[0]} =  \bar{B}_{V2}^{[0]} & = & \frac{1}{2m_b} = 
	+\bar{B}_{A1}^{[0]} = +\bar{B}_{A2}^{[0]}
		 \label{eq:B1B2} \\
	 \bar{B}_{V4}^{[0]} =  \bar{B}_{V5}^{[0]} & = & \frac{1}{2m_c} = 
	-\bar{B}_{A4}^{[0]} = -\bar{B}_{A5}^{[0]}
		\label{eq:B4B5} \\
	 \bar{B}_{Ji}^{[0]} & = & 0, \quad \text{otherwise}, \label{eq:B3B6}
\end{eqnarray}
but all become non-trivial when radiative corrections are included.

HQET provides a systematic way to separate the short distances~$1/m_c$ 
and $1/m_b$ from the long distance~$\Lambda_{\text{QCD}}$,
as long as the condition~(\ref{eq:when}) holds for both flavors.
HQET can also be applied to lattice gauge theory with the same structure
and logic, treating $a$ as another short distance.
The strategy is viable as long as condition~(\ref{eq:when}) holds and,
of course, $\bbox{p}a\ll1$.
For lattice NRQCD applied to heavy-light systems, HQET is just the
corresponding Symanzik effective field theory.
HQET may also be applied to Wilson fermions, because they
have the correct particle content and heavy-quark symmetries as
QCD~\cite{Kronfeld:2000ck}.
Thus, for a lattice gauge theory with either NRQCD or
Wilson quarks
\begin{equation}
	{\cal L}_{\text{lat}} \doteq {\cal L}_{\text{HQET}}.
\end{equation}
The effective Lagrangian~${\cal L}_{\text{HQET}}$ has the
same long-distance operators as in Eq.~(\ref{eq:QCD=HQET}).
The lattice modifies the short-distance behavior, so lattice
artifacts of the heavy quarks should be lumped into the HQET
coefficients~\cite{Kronfeld:2000ck}.
For example, in the dimension-four HQET Lagrangian~${\cal L}^{(0)}$, 
one must replace $m_h$ with the lattice rest mass~$m_{1h}$.

Starting with ${\cal L}^{(3)}$ one must introduce operators to 
describe lattice violations of rotational invariance.
An example is the dimension-seven operator
${\cal O}_4=\sum_\mu\bar{h}_v(D_\perp^\mu)^4h_v$.
Such operators do not, of course, appear in the HQET description of 
continuum QCD, but they do not upset the general framework.
They can still be defined as insertions in a continuum-regulated
theory (analogously to operators like $\sum_\mu\bar{q}(D^\mu)^4q$
in a Symanzik effective field theory).
Because the symmetry-breaking arises at short-distances, it is more 
useful to focus on the operators' coefficients, noting that
${\cal C}^{\text{lat}}_4\neq 0$, instead of ${\cal C}_4=0$ as for 
continuum~QCD.

Returning to ${\cal L}^{(0)}$ and ${\cal L}^{(1)}$, the coefficient of 
the kinetic energy becomes
\begin{equation}
	{\cal C}_2^{\text{lat}} = \frac{1}{2m_{2h}},
\end{equation}
and $m_{2h}$ is called the kinetic mass.
If the HQET operators are defined by minimal subtraction in 
dimensional regularization, then both $m_{1h}$ and $m_{2h}$ can be 
computed from the pole in the perturbative quark 
propagator~\cite{Mertens:1998wx}, and they are infrared finite and 
gauge independent~\cite{Kronfeld:1998di}.
The lattice breaks Lorentz (or Euclidean) invariance, so
reparametrization invariance no longer requires $m_{2h}$ to 
equal~$m_{1h}$.

For the flavor-changing currents, one forms bilinears of lattice
fermions fields with the right quantum numbers.
Any such lattice currents for $b\to c$ can be described by
\begin{eqnarray}
	V^\mu_{\text{lat}} & \doteq &
		\bar{C}_{V_\parallel}^{\text{lat}} 
			v^\mu \bar{c}_{v'}b_v  +
		\bar{C}_{V_\perp}^{\text{lat}} 
			\bar{c}_{v'}i\gamma^\mu_\perp b_v +
		\bar{C}_{V_{v'}}^{\text{lat}} 
			v^{\prime\mu}_\perp \bar{c}_{v'}b_v -
		\sum_{i=1}^{14} \bar{B}_{Vi}^{\text{lat}} 
			\bar{\cal Q}^\mu_{Vi} + \cdots,
	\label{eq:VlatHQET} \\
	A^\mu_{\text{lat}} & \doteq &
		\bar{C}_{A_\perp}^{\text{lat}} 
			\bar{c}_{v'}i\gamma^\mu_\perp \gamma_5b_v -
		\bar{C}_{A_\parallel}^{\text{lat}} 
			v^\mu \bar{c}_{v'}\gamma_5b_v  -
		\bar{C}_{A_{v'}}^{\text{lat}} 
			v^{\prime\mu}_\perp \bar{c}_{v'}\gamma_5b_v -
		\sum_{i=1}^{14} \bar{B}_{Ai}^{\text{lat}} 
			\bar{\cal Q}_{Ai}^\mu + \cdots.
	\label{eq:AlatHQET}
\end{eqnarray}
An explicit construction of $V^\mu_{\text{lat}}$ and 
$A^\mu_{\text{lat}}$ through dimension~4 is given for Wilson fermions 
in Sec.~\ref{sec:lattice}, 
and for lattice NRQCD already in Ref.~\cite{Boyle:2000fi}.
On the right-hand side the HQET operators~$\bar{\cal Q}_{Ji}$ are the 
same as in Eqs.~(\ref{eq:VcontHQET}) and~(\ref{eq:AcontHQET}), but the 
short-distance coefficients 
$\bar{C}_{J_{\parallel,\perp,{v'}}}^{\text{lat}}$ and
$\bar{B}_{Ji}^{\text{lat}}$ are modified.
In particular, they depend on the lattice spacing~$a$ through~$m_ba$ 
or $m_ca$, and also on adjustable improvement parameters in 
$V^\mu_{\text{lat}}$ and $A^\mu_{\text{lat}}$.
At and beyond the dimension six, there are HQET operators to describe 
violations of rotational invariance in the lattice currents, but, as 
above, they do not spoil the overall framework.

Since Eqs.~(\ref{eq:Q1})--(\ref{eq:Q14}) give a complete set of 
dimension-four HQET currents, the coefficients 
$\bar{C}_{J_\parallel,J_\perp}^{\text{lat}}$ and 
$\bar{B}_{Ji}^{\text{lat}}$ contain short-distance effects from both 
heavy quarks.
For HQET operators of dimension~$s+3$, the corresponding coefficient
must contain $s$ powers of the short distances $m_b^{-1}$, $m_c^{-1}$
or~$a$.
Since they are functions of all ratios of short distances, it is a matter
of choice which dimensionful quantity is factored out.
The key point is that, even when applied to lattice gauge theory, the
heavy-quark expansion is still a successive approximation:
higher terms in the expansion are suppressed by powers of
$\Lambda_{\text{QCD}}$, with the dimensions balanced by the short
distances.

Lattice artifacts from gluons and light quarks can be described by
the Symanzik
formalism~\cite{Symanzik:1979ph,Symanzik:1983dc,Luscher:1996sc,Jansen:1996ck}.
At some level the light quarks will influence the HQET coefficients
of the heavy quarks, and the heavy quarks will influence the Symanzik
coefficients of the gluons and light quarks.
These are, however, details that do not obstruct the central idea of
using effective field theories to separate long- and short-distance
physics.

The abstract ideas can be made more concrete by comparing the HQET
descriptions of continuum and lattice matrix elements.
The continuum matrix element of $v\cdot{\cal V}$ is
\begin{eqnarray}
	\langle D|v\cdot {\cal V}|B\rangle & = & -
		\bar{C}_{V_\parallel} \langle D_{v'}^{(0)}|
			\bar{c}_{v'}b_v|B_v^{(0)}\rangle
			- 
		\sum_{i\in S_\parallel} \bar{B}_{Vi} \langle D_{v'}^{(0)}|
			v\cdot \bar{\cal Q}_{Vi}|B_v^{(0)}\rangle
			\nonumber \\ & - &
		{\cal C}_{2c}        \bar{C}_{V_\parallel} \int d^4x
			\langle D_{v'}^{(0)}|T\,{\cal O}_{2c}(x)
			\bar{c}_{v'}b_v|B_v^{(0)}\rangle^\star
			\nonumber \\ & - &
		{\cal C}_{{\cal B}c} \bar{C}_{V_\parallel} \int d^4x
			\langle D_{v'}^{(0)}|T\,{\cal O}_{{\cal B}c}(x)
			\bar{c}_{v'}b_v|B_v^{(0)}\rangle^\star 
			\nonumber \\ & - &
		{\cal C}_{2b}        \bar{C}_{V_\parallel} \int d^4x
			\langle D_{v'}^{(0)}|T\,\bar{c}_{v'}b_v
			{\cal O}_{2b}(x)|B_v^{(0)}\rangle^\star
			\nonumber \\ & - &
		{\cal C}_{{\cal B}b} \bar{C}_{V_\parallel} \int d^4x
			\langle D_{v'}^{(0)}|T\,\bar{c}_{v'}b_v
			{\cal O}_{{\cal B}b}(x) |B_v^{(0)}\rangle^\star 
			+ 
		O(\Lambda^2/m^2), \label{eq:VHQE}
\end{eqnarray}
where the set $S_\parallel=\{1,4,9,11\}$, and $B$ ($D$) is any 
$b$-flavored (charmed) hadronic state.
(The $\star$-ed $T$~product is defined in Ref.~\cite{Kronfeld:2000ck};
this detail is unimportant here.)
On the left-hand  side the states are QCD eigenstates; on the right-hand
side (with velocity subscripts) they are the corresponding eigenstates
of~${\cal L}^{(0)}$, the leading HQET Lagrangian.
The extra subscripts on the coefficients and operators 
from~${\cal L}^{(1)}$ denote flavor.
The corresponding lattice matrix element is similar, but slightly
different:
\begin{eqnarray}
	\langle D|v\cdot V_{\text{lat}}|B\rangle & = & -
		\bar{C}^{\text{lat}}_{V_\parallel} 
			\langle D_{v'}^{(0)}|\bar{c}_{v'}b_v|B_v^{(0)}\rangle
			- 
		\sum_{i\in S_\parallel} \bar{B}^{\text{lat}}_{Vi} 
			\langle D_{v'}^{(0)}|v\cdot \bar{\cal Q}_{Vi}|B_v^{(0)}\rangle
			\nonumber \\ & - &
		{\cal C}^{\text{lat}}_{2c}        \bar{C}^{\text{lat}}_{V_\parallel}
			\int d^4x
			\langle D_{v'}^{(0)}|T\,{\cal O}_{2c}(x) 
			\bar{c}_{v'}b_v|B_v^{(0)}\rangle^\star
			\nonumber \\ & - &
		{\cal C}^{\text{lat}}_{{\cal B}c} \bar{C}^{\text{lat}}_{V_\parallel}
			\int d^4x
			\langle D_{v'}^{(0)}|T\,{\cal O}_{{\cal B}c}(x)
			\bar{c}_{v'}b_v|B_v^{(0)}\rangle^\star 
			\nonumber \\ & - &
		{\cal C}^{\text{lat}}_{2b}        \bar{C}^{\text{lat}}_{V_\parallel}
			\int d^4x
			\langle D_{v'}^{(0)}|T\,\bar{c}_{v'}b_v
			{\cal O}_{2b}(x)|B_v^{(0)}\rangle^\star
			\nonumber \\ & - &
		{\cal C}^{\text{lat}}_{{\cal B}b} \bar{C}^{\text{lat}}_{V_\parallel}
			\int d^4x
			\langle D_{v'}^{(0)}|T\,\bar{c}_{v'}b_v
			{\cal O}_{{\cal B}b}(x) |B_v^{(0)}\rangle^\star 
			\nonumber \\ & + &
		K_{\sigma\cdot F} C^{\text{lat}}_{V_\parallel} \int\! d^4x
			\langle D_{v'}^{(0)}|T\,\bar{c}_{v'}b_v
			\bar{q}i\sigma Fq(x)|B_v^{(0)}\rangle^\star 
			+ 
		O(\Lambda^2a^2b(ma)).
\end{eqnarray}
Now the states on the left-hand side are lattice eigenstates.
But on the right-hand side the differences compared to
Eq.~(\ref{eq:VHQE}) are all in the coefficients:
except for the last matrix element, they are, term by term, the same.
The last $T$-product arise from the Symanzik local effective 
Lagrangian for light quarks, cf.\ Ref.~\cite{Harada:2001hl}.

Similar formulas hold for matrix elements of ${\cal V}^\mu_\perp$ and
$V^\mu_{\text{lat}\perp}$, and for the axial vector current.
If one multiplies the lattice matrix elements with
\begin{eqnarray}
	\bar{Z}_{J_\parallel} & = &
		\frac{\bar{C}_{J_\parallel}}{\bar{C}_{J_\parallel}^{\text{lat}}},
	\label{eq:defZVpara} \\
	\bar{Z}_{J_\perp}     & = &
		\frac{\bar{C}_{J_\perp}}{\bar{C}_{J_\perp}^{\text{lat}}},
	\label{eq:defZVperp}
\end{eqnarray}
and subtracts the lattice from the continuum equations, a simple picture
of cutoff effects emerges: lattice artifacts of the heavy quark are 
isolated into the mismatch of the short-distance coefficients, namely,
\begin{eqnarray}
	\delta{\cal C}_i = {\cal C}_i^{\text{lat}} & - & {\cal C}_i,
	\label{eq:deltaC} \\
	\delta \bar{B}_{Ji} =
		\bar{Z}_{Ji}  \bar{B}_{Ji}^{\text{lat}} & - & \bar{B}_{Ji}.
	\label{eq:deltaB}
\end{eqnarray}
In the lattice term of $\delta \bar{B}_{Ji}$, matching factors 
$\bar{Z}_{Ji}$ appear to restore a canonical normalization to the 
lattice currents.
One has $\bar{Z}_{Ji}=\bar{Z}_{J_\parallel}$ for $i\in S_\parallel$,
and     $\bar{Z}_{Ji}=\bar{Z}_{J_\perp}$     for $i\in
S_\perp=\{2,3,5,6,7,8,10,12,13,14\}$.

\begin{figure}
    \centering
	\begin{picture}(200,120)(50,0)
		\thicklines
		\put(102,100){lattice}
		\put(115,95){\vector(0,-1){73}}
		\put(206,68){\vector(-2,1){68}}
		\put(200,54){HQET}
		\put(206,49){\vector(-2,-1){68}}
		\put(105,10){QCD}
		\put(75,54){$\bar{C}/\bar{C}^{\text{lat}}$}
		\put(177,88){$\bar{C}^{\text{lat}}$}
		\put(177,23){$\bar{C}$}
	\end{picture}
    \caption[fig:matching]{Diagram illustrating how the matching
    factors $\bar{C}^{\text{lat}}$, $\bar{C}$, and 
    $\bar{Z}=\bar{C}/\bar{C}^{\text{lat}}$ match
    lattice gauge theory and QCD to HQET, and to each other.}
    \label{fig:matching}
\end{figure}
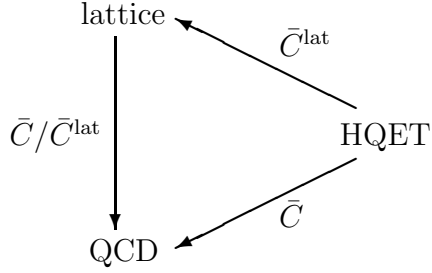

The matching factors~$\bar{Z}_{J_\parallel}$ and~$\bar{Z}_{J_\perp}$ 
play the following role, sketched in Fig.~\ref{fig:matching}.
The denominator converts a lattice-regulated scheme to a renormalized
HQET scheme, and the numerator converts the latter to a renormalized
(continuum) QCD scheme.
As long as the same HQET scheme is used, HQET drops out of the
calculation of $\bar{Z}_{J_\parallel}$ and~$\bar{Z}_{J_\perp}$.
Moreover, changes in continuum renormalization conventions modify only
the numerator, and changes in the lattice action or currents modify only
the denominator.
Thus, the matching factors~$\bar{Z}_{J_\parallel}$ and~$\bar{Z}_{J_\perp}$ 
play the same role as the factors~$Z_{J_\parallel}$ and~$Z_{J_\perp}$ in 
HQET matching for heavy-light currents~\cite{Harada:2001hl}.

When HQET matching is applied to Wilson quarks, it is possible that 
the lighter heavy quark satisfies $m_ca\ll 1$.
Then one could equally well apply the heavy-``light'' formalism of
Ref.~\cite{Harada:2001hl}, describing the charmed quark \`a la Symanzik
with a Dirac field.
This regime is interesting because in practice one often has 
$m_ca<\case{1}{3}$ and $\Lambda/2m_ca<\case{1}{3}$,
so both the Symanzik and the HQET descriptions are reasonably accurate.

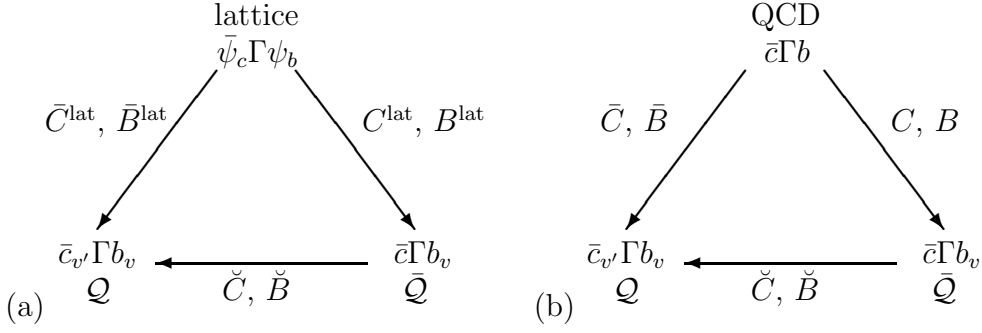
\begin{figure}
    \centering
	\begin{picture}(400,140)(-200,-20)
		\thicklines
		\put(-195,0){(a)}
		\put(-116,110){lattice}
		\put(-115,96){$\bar{\psi}_c\Gamma\psi_b$}
		\put(-85,93){\vector(3,-4){45}}
		\put(-48,20){$\bar{c}\Gamma b_v$}
		\put(-45,6){$\bar{\cal Q}$}
		\put(-115,93){\vector(-3,-4){45}}
		\put(-175,20){$\bar{c}_{v'}\Gamma b_v$}
		\put(-165,6){${\cal Q}$}
		\put(-58,20){\vector(-1,0){80}}
		\put(-60,70){$C^{\text{lat}}$, $B^{\text{lat}}$}
		\put(-180,70){$\bar{C}^{\text{lat}}$, $\bar{B}^{\text{lat}}$}
		\put(-113,6){$\breve{C}$, $\breve{B}$}
		\put(5,0){(b)}
		\put(87,110){QCD}
		\put(92,96){$\bar{c}\Gamma b$}
		\put(115,93){\vector(3,-4){45}}
		\put(152,20){$\bar{c}\Gamma b_v$}
		\put(155,6){$\bar{\cal Q}$}
		\put(85,93){\vector(-3,-4){45}}
		\put(25,20){$\bar{c}_{v'}\Gamma b_v$}
		\put(35,6){${\cal Q}$}
		\put(142,20){\vector(-1,0){80}}
		\put(140,70){$C$, $B$}
		\put(30,70){$\bar{C}$, $\bar{B}$}
		\put(87,6){$\breve{C}$, $\breve{B}$}
	\end{picture}
    \caption[fig:2step]{Diagrams illustrating how the matching
    factors $\bar{C}$, $C$, and $\breve{C}$ match the underlying
	theory to effective theories where the lighter quark is described
	either with a Dirac field~$\bar{c}$ or an HQET field~$\bar{c}_{v'}$
	(note subscript in HQET case); for
    (a) lattice gauge theory, (b) continuum~QCD.}
    \label{fig:2step}
\end{figure}

By comparing the two matching procedures, one can derive relations 
between the two sets of short-distance coefficients.%
\footnote{One must take the same lattice currents.
The improved currents of Ref.~\cite{Harada:2001hl} and 
Sec.~\ref{sec:lattice} are nearly the same.}
To proceed, one must introduce coefficients to relate the 
heavy-``light'' theory with the Dirac field $\bar{c}$ to the 
heavy-``heavy'' theory with the HQET  field~$\bar{c}_{v'}$:
\begin{eqnarray}
	\bar{c}b_v  & \doteq & \breve{C}_{V_\parallel} \bar{c}_{v'}b_v +
		\sum_{i\in S_\parallel} \breve{B}_{Vi} \; 
		v\cdot \bar{\cal Q}_{Vi} + \cdots,
		\label{eq:breveparallel} \\
	\bar{c}i\gamma^\mu_\perp b_v  & \doteq &
		\breve{C}_{V_\perp} \bar{c}_{v'}i\gamma^\mu_\perp  b_v +
		\breve{C}_{V_{v'}} v^{\prime\mu}_\perp \bar{c}_{v'}b_v -
		\sum_{i\in S_\perp}     \breve{B}_{Vi} \;
		\eta^\mu_\nu \bar{\cal Q}_{Vi}^\nu + \cdots,
	\label{eq:breveperp} \\
	{\cal Q}_{Vi} & \doteq &
		\sum_{j=1}^{14} \breve{C}_{Vij} \bar{\cal Q}_{Vj} + \cdots,
	\quad i=1,6,
	\label{eq:breveQ}
\end{eqnarray}
and similarly for the axial vector current.
The relation between the three sets of short-distance coefficients---%
$\bar{C}$ and $\bar{B}$, $C$ and $B$, and $\breve{C}$ and $\breve{B}$---%
is sketched in Fig.~\ref{fig:2step}.
By substituting Eqs.~(\ref{eq:breveparallel})--(\ref{eq:breveQ})
into Eq.~(2.40) of Ref.~\cite{Harada:2001hl} and comparing to the above
Eq.~(\ref{eq:VlatHQET}), one finds (when $m_ca\ll 1$)
\begin{eqnarray}
	\bar{C}_{J_\parallel}^{\text{lat}}     & = & 
		C_{J_\parallel}^{\text{lat}} \breve{C}_{J_\parallel}
		\label{eq:CJpara} \\
	\bar{C}_{J_\perp}^{\text{lat}}         & = & 
		C_{J_\perp}^{\text{lat}}     \breve{C}_{J_\perp}
		\label{eq:CJperp} \\
	\bar{C}_{J_{v'}}^{\text{lat}}         & = & 
		C_{J_\perp}^{\text{lat}}     \breve{C}_{J_{v'}}
		\label{eq:CJv'} \\
	\bar{B}_{Jj}^{\text{lat}} & = & \sum_{i=1}^6
		B_{Ji}^{\text{lat}} \breve{C}_{Jij} +
		C_{J_\parallel}^{\text{lat}} \breve{B}_{Jj},
		\quad j\in S_\parallel , \label{eq:B14911} \\
	\bar{B}_{Jj}^{\text{lat}} & = & \sum_{i=1}^6
		B_{Ji}^{\text{lat}} \breve{C}_{Jij} +
		C_{J_\perp}^{\text{lat}} \breve{B}_{Jj},
		\quad j\in S_\perp .     \label{eq:Brest} 
\end{eqnarray}
The same relations hold for the coefficients describing continuum QCD, 
i.e., dropping the superscript ``lat''.
The same coefficients $\breve{C}$ and~$\breve{B}$ appear, 
because no lattice fields appear in
Eqs.~(\ref{eq:breveparallel})--(\ref{eq:breveQ}).
In Fig.~\ref{fig:2step}, the bottom sides of the triangles are 
oblivious to the underlying theory, be it lattice gauge theory or 
continuum QCD.
Thus, one can eliminate the $\breve{C}_J$ from
Eqs.~(\ref{eq:CJpara})--(\ref{eq:CJv'}) in favor of $\bar{C}/C$, 
yielding relations between the matching factors, when $m_ca\ll1$,
\begin{eqnarray}
	\bar{Z}_{J_\parallel} & = & Z_{J_\parallel}, \label{eq:ZJpara} \\
	\bar{Z}_{J_\perp}     & = & Z_{J_\perp},     \label{eq:ZJperp} \\
	\bar{Z}_{J_\perp} \bar{C}_{J_{v'}}^{\text{lat}} & = & \bar{C}_{J_{v'}}.
\end{eqnarray}
In particular, Eqs.~(\ref{eq:ZJpara}) and~(\ref{eq:ZJperp}) show that 
the heavy-heavy matching factors~$\bar{Z}_{J_\parallel}$ 
and~$\bar{Z}_{J_\perp}$ are a natural extension, into the 
regime~$m_ca\not\ll1$, of the heavy-light matching 
factors~$Z_{J_\parallel}$ and~$Z_{J_\perp}$.
Consequently, in the following sections we shall drop the distinction 
and omit the bar for the heavy-heavy matching factors.

One also finds relations among the coefficients of dimension-four
currents:
\begin{eqnarray}
	\bar{Z}_{J_\parallel} \bar{B}_{Jj}^{\text{lat}} - \bar{B}_{Jj} & = &  
		\sum_{i=1}^6
		(Z_{J_\parallel} B_{Ji}^{\text{lat}} - B_{Ji}) \breve{C}_{Jij},
		\quad j\in S_\parallel , \label{eq:ZB14911} \\
	\bar{Z}_{J_\perp}     \bar{B}_{Jj}^{\text{lat}} - \bar{B}_{Jj} & = &  
		\sum_{i=1}^6
		(Z_{J_\perp}     B_{Ji}^{\text{lat}} - B_{Ji}) \breve{C}_{Jij},
		\quad j\in S_\perp     . \label{eq:ZBrest} 
\end{eqnarray}
The terms on the right-hand sides are proportional to mismatches in the
heavy-light HQET.
Each $Z_JB_{Ji}^{\text{lat}}-B_{Ji}$ is suppressed by a 
power of $a$, multiplied by a function of $m_ba$.
Similar constraints can be derived for higher-dimension coefficients.
Following this reasoning for the whole tower of higher dimension
operators leads to the conclusion that, when $m_ca\ll1$, the mismatch
of the heavy-``heavy'' currents reduces to that of the heavy-``light''
case.
In this way, lattice gauge theory reproduces the entire $1/m_c$ 
expansion, apart from deviations that are suppressed by powers 
of the other short distances, $a$ and~$1/m_b$.

Deep in the $a\to 0$ limit, where $m_ca\le m_ba\ll 1$ one can relate 
the short-distance coefficients of the HQET formalism to those in the 
Symanzik formalism.%
\footnote{One must, of course, take the same lattice currents.
The improved currents of Ref.~\cite{Luscher:1996sc}
and of Sec.~\ref{sec:lattice} are {\em not} the same.}
For simplicity let us take $v=v'$.
For the [axial] vector current, one inserts Eq.~(\ref{eq:VcontHQET}) 
[Eq.~(\ref{eq:AcontHQET})] into Eq.~(2.5) [Eq.~(2.6)] of 
Ref.~\cite{Harada:2001hl}, neglects terms of order~$m^2a^2$,
and compares with Eq.~(\ref{eq:VlatHQET}) [Eq.~(\ref{eq:AlatHQET})].
One also must match the tensor bilinear to HQET at the dimension-three 
level,
\begin{equation}
	\bar{c}i\sigma^{\mu\nu}b \doteq
		\bar{C}_T \eta^\mu_\alpha \eta^\nu_\beta
			\bar{c}_v i\sigma^{\alpha\beta} b_v + \cdots,
\end{equation}
where one introduces another short-distance coefficient~$\bar{C}_T$.
At the tree level, $\bar{C}_T^{[0]}=1$.
The pseudoscalar bilinear is described by HQET operators of dimension 
four and higher, because $P_+(v)\gamma_5P_+(v)=0$.
After carrying out these steps, one finds that 
\begin{eqnarray}
	\bar{Z}_{V_\parallel} & = & Z_V, \label{eq:ZVpara} \\
	\bar{Z}_{V_\parallel} \bar{B}_{Vi}^{\text{lat}} & = & 
		\bar{B}_{Vi} + a Z_V K_V\bar{C}_T , \quad i=2,6, \label{eq:ZVB26} \\
	\bar{Z}_{V_\parallel} \bar{B}_{Vi}^{\text{lat}} & = & 
		\bar{B}_{V3} - a Z_V K_V\bar{C}_T , \quad i=3,5, \label{eq:ZVB35}
\end{eqnarray}
from matching the vector current (with $Z_V$ and $K_V$ as defined in 
Ref.~\cite{Harada:2001hl}), and
\begin{eqnarray}
	\bar{Z}_{A_\perp}     & = & Z_A, \label{eq:ZAperp} \\
	\bar{Z}_{A_\perp}     \bar{B}_{Ai}^{\text{lat}} & = & 
		\bar{B}_{Ai} + O(a^2),              \quad i=1,4, \label{eq:ZAB14}
\end{eqnarray}
from matching the axial vector current (with $Z_A$ as defined in 
Ref.~\cite{Harada:2001hl}).
Consequently, one sees
\begin{equation}
	\lim_{a\to 0} \bar{Z}_{Ji} \bar{B}_{Ji}^{\text{lat}} = \bar{B}_{Ji},
\end{equation}
and the limit is accelerated for standard $O(a)$ improvement.
Similar relations apply for the short-distance coefficients of HQET
currents of dimension five and higher.

Let us summarize the main results of this section.
We have given a description of heavy-heavy lattice currents in HQET, 
which parallels the description of continuum~QCD.
The parallel structure shows that cutoff effects are isolated into the 
mismatch of short-distance coefficients for lattice gauge theory and 
continuum QCD, Eqs.~(\ref{eq:deltaB}) and~(\ref{eq:deltaC}).
If the lattice currents (and Lagrangian) have enough free parameters,
one can reduce the mismatch and, in this way, bring lattice calculations
closer to continuum QCD, term by term in the heavy-quark expansion.
In the practical region for the charmed quark, $m_ca\sim\case{1}{3}$,
it is also worth remarking that (for Wilson fermions) one could also 
apply the heavy-light formalism of our companion 
paper~\cite{Harada:2001hl}.
Then one sees, extending Eqs.~(\ref{eq:ZB14911}) and~(\ref{eq:ZBrest})
to higher dimension, how the whole heavy-quark expansion for a quark
with mass $m_ca<1$ is recovered.

The remainder of this paper pursues the program of HQET matching in 
perturbation theory.
One-loop corrections to the rest mass~$m_1$ and the kinetic mass~$m_2$
have been considered already in Ref.~\cite{Mertens:1998wx}.
We construct lattice currents suitable for matching through 
dimension-five operators in the Lagrangian and dimension-four in the 
currents.
Our heavy-heavy lattice currents turn out to be nearly the same as our 
heavy-light lattice currents.
We then calculate the matching factors $Z_{V_\parallel}$ and
$Z_{A_\perp}$ at the one-loop level, which are needed to fix the overall
normalization of the heavy-light currents.

\section{Lattice Action and Currents}
\label{sec:lattice}

In this section our aim is to define heavy-heavy currents with Wilson
fermions that are suited to the HQET matching formalism.
Our construction is similar to that for heavy-light
currents~\cite{Harada:2001hl}.
The descriptive part of the HQET formalism applies as long as the 
lattice theory has the right particle content and obeys the 
heavy-quark symmetries, as it does with Wilson fermions.
To use HQET systematically to improve lattice gauge theory, however,
one should make choices to ensure that $\delta{\cal C}_i$ and
$\delta\bar{B}_{Ji}$ [cf.\ Eqs.~(\ref{eq:deltaC}) and (\ref{eq:deltaB})]
remain bounded in the infinite-mass limit.
For convenience we focus on the case when $v'=v$.
Then good behavior is attained by mimicking the structure of
Eqs.~(\ref{eq:Q1})--(\ref{eq:Q6}), and free parameters in the currents
can be adjusted so that $\delta{\cal C}_i$ and $\delta\bar{B}_{Ji}$
(approximately) vanish.
We show how to do so in perturbation theory, obtaining
$\bar{B}^{\text{lat}}_{Ji}$ at the tree level and, in
Sec.~\ref{sec:loop}, the matching factors $Z_{V_\parallel}$
and~$Z_{A_\perp}$ at the one-loop level.

A suitable lattice Lagrangian was introduced in
Ref.~\cite{El-Khadra:1997mp} and reviewed in the context
of HQET matching in our companion paper on heavy-light
currents~\cite{Harada:2001hl}.
We will not repeat the formulas here, but refer the reader to
Ref.~\cite{Harada:2001hl}.
It is enough to recall that the Lagrangian has several 
couplings---$r_s$, $\zeta$, $c_B$ and~$c_E$.
The coupling~$r_s$ is redundant, in the technical sense, but the
others can be tuned to match lattice gauge theory to continuum QCD;
in this way they become non-trivial functions of the heavy quark's 
bare mass $m_0a$~\cite{El-Khadra:1997mp}.
In particular, one may define the renormalized heavy quark mass by
identifying ${\cal C}_2^{\text{lat}}={\cal C}_2$, and one can adjust
$c_B$ so that ${\cal C}_{\cal B}^{\text{lat}}={\cal C}_{\cal B}$.
(If one also requires $m_1=m_2$, one can adjust both $\zeta$ and the
bare mass~\cite{El-Khadra:1997mp,Sroczynski:2000hg}.)

The corresponding currents for transitions from one heavy fermion to
another are defined as follows.
First define a ``rotated'' field~\cite{Kronfeld:1995nu,El-Khadra:1997mp}
\begin{equation}
	\Psi_q = \left[1 + ad_{1q}
		\bbox{\gamma}\cdot\bbox{D}_{\text{lat}}
		\right] \psi_q,
	\label{eq:rotate}
\end{equation}
where $\psi_q$ is the field in ${\cal L}_0$ of flavor~$q$ ($q=c$, $b$),
and $D_{\text{lat}}$ is the symmetric covariant difference operator.
(The lattice Lagrangian ${\cal L}_0$ is given in Eq.~(3.1) of 
Ref.~\cite{Harada:2001hl}.)
The bilinears
\begin{eqnarray}
	V^\mu_0 & = & \bar{\Psi}_ci\gamma^\mu \Psi_b,
	\label{eq:Vlat} \\
	A^\mu_0 & = & \bar{\Psi}_ci\gamma^\mu\gamma_5 \Psi_b.
	\label{eq:Alat}
\end{eqnarray}
have the correct quantum numbers, but not enough freedom to match at
the dimension-four level of HQET.
Thus, these currents must be improved.
We focus on matching with $v'=v$, so we take
\begin{eqnarray}
	V^\mu_{\text{lat}} & = & V^\mu_0 -
		\sum_{i\in\{2,3,5,6\}} b_{Vi} Q^\mu_{Vi},
	\label{eq:Vimp} \\
	A^\mu_{\text{lat}} & = & A^\mu_0 -
		\sum_{i\in\{1,4\}} b_{Ai} Q^\mu_{Ai},
	\label{eq:Aimp}
\end{eqnarray}
where the $b_{Ji}$ are adjustable,
and the higher-dimension lattice operators are
\begin{eqnarray}
	Q^\mu_{V2} & = & \bar{\psi}_ci\gamma^\mu_\perp
		{\kern+0.1em /\kern-0.65em D_\perp}_{\text{lat}} \psi_b,
		\label{eq:Q2lat} \\
	Q^\mu_{V3} & = & \bar{\psi}_c i {D^\mu_\perp}_{\text{lat}} \psi_b,
		\label{eq:Q3lat} \\
	Q^\mu_{V5} & = & \bar{\psi}_c
		{\kern+0.1em /\kern-0.65em \loarrow{D}_\perp}_{\text{lat}}
		i\gamma^\mu_\perp \psi_b,  \label{eq:Q5lat} \\
	Q^\mu_{V6} & = & \bar{\psi}_ci{\loarrow{D}^\mu_\perp}_{\text{lat}}
		\psi_b, \label{eq:Q6lat}
\end{eqnarray}
and
\begin{eqnarray}
	Q^\mu_{A1} & = & - v^\mu \bar{\psi}_c i{\kern+0.1em /\kern-0.55em v}
		\gamma_5 {\kern+0.1em /\kern-0.65em D_\perp}_{\text{lat}} \psi_b,
		\label{eq:Q1lat} \\
	Q^\mu_{A4} & = & +v^\mu \bar{\psi}_c
		{\kern+0.1em /\kern-0.65em \loarrow{D}_\perp}_{\text{lat}}
		i{\kern+0.1em /\kern-0.55em v} \gamma_5 \psi_b. \label{eq:Q4lat}
\end{eqnarray}
Lattice quark fields do not satisfy Eq.~(\ref{eq:Pvh}), so
${\kern+0.1em /\kern-0.55em v}$ appears explicitly.
The dimension-four lattice operators~$Q^\mu_{Ji}$ are the same as for 
heavy-light currents, but now only corrections orthogonal (parallel)
to~$v$ for the vector (axial vector) current are needed. 
An analogous construction for lattice NRQCD has been given by
Boyle and Davies~\cite{Boyle:2000fi}, who found the same pattern of
dimension-four terms.

It is worth emphasizing the difference between Eq.~(\ref{eq:VlatHQET}),
the corresponding equation for heavy-light currents
(Eq.~(2.40) of Ref.~\cite{Harada:2001hl}), and Eq.~(\ref{eq:Vimp}).
Equation~(\ref{eq:VlatHQET}) is a general HQET description of any
heavy-heavy lattice current.
Equation~(\ref{eq:Vimp}) is a definition of a specific lattice
current, namely the one used in this paper (and in calculations of
$B\to D^{(*)}$ transition matrix elements).
In the same vein, the $\bar{\cal Q}_{Ji}$ in
Eqs.~(\ref{eq:Q1})--(\ref{eq:Q14})
are HQET operators, whereas the $Q_{Ji}$
in Eqs.~(\ref{eq:Q2lat})--(\ref{eq:Q4lat}) are lattice operators.
Correspondingly, the coefficients $\bar{B}^{\text{lat}}_{Ji}$ are
the output of a matching calculation: they depend on the~$b_{Ji}$,
which must be adjusted to make~$\delta\bar{B}_{Ji}$ vanish.
The difference between $B_{Ji}$ and ${\cal Q}_{Ji}$ ($i=1$,\ldots,~6) 
from Ref.~\cite{Harada:2001hl} on the one hand, and $\bar{B}_{Ji}$ and 
$\bar{\cal Q}_{Ji}$ ($i=1$,\ldots,~14) on the other, was discussed 
above.
In the former, the light(er) quark is described by a Dirac field;
in the latter, the lighter (heavy) quark is described by an HQET field.

The difference between the definition of the lattice currents and the
description with HQET can be illustrated by giving the matching
coefficients at the tree level.
A~simple calculation of the on-shell matrix elements
$\langle c|J^\mu_{\text{lat}}|b\rangle$ yields
\begin{eqnarray}
	\bar{C}^{\text{lat}[0]}_{V_\parallel} & = &
		e^{-(m_{1q}^{[0]}+m_{1b}^{[0]})a/2} ,
	\label{eq:matchCV[0]} \\
	\bar{B}^{\text{lat}[0]}_{V2} & = &
		e^{-(m_{1q}^{[0]}+m_{1b}^{[0]})a/2} \left(
		\frac{1}{2m^{[0]}_{3b}} + b^{[0]}_{V2}\right),
	\label{eq:matchB2[0]}  \\
	\bar{B}^{\text{lat}[0]}_{V5} & = &
		e^{-(m_{1q}^{[0]}+m_{1b}^{[0]})a/2} \left(
		\frac{1}{2m^{[0]}_{3c}} + b^{[0]}_{V5}\right),
	\label{eq:matchB5[0]}  \\
	\bar{B}^{\text{lat}[0]}_{Vi} & = &
		e^{-(m_{1q}^{[0]}+m_{1b}^{[0]})a/2} b^{[0]}_{Vi},
		\quad i=3,6, \label{eq:matchB3[0]} 
\end{eqnarray}
for the vector current, and
\begin{eqnarray}
	\bar{C}^{\text{lat}[0]}_{A_\perp} & = &
		e^{-(m_{1q}^{[0]}+m_{1b}^{[0]})a/2} ,
	\label{eq:matchCA[0]} \\
	\bar{B}^{\text{lat}[0]}_{A1} & = & +
		e^{-(m_{1q}^{[0]}+m_{1b}^{[0]})a/2} \left(
		\frac{1}{2m^{[0]}_{3b}} + b^{[0]}_{A1}\right),
	\label{eq:matchB1[0]} \\
	\bar{B}^{\text{lat}[0]}_{A4} & = & -
		e^{-(m_{1q}^{[0]}+m_{1b}^{[0]})a/2} \left(
		\frac{1}{2m^{[0]}_{3c}} + b^{[0]}_{A4}\right),
	\label{eq:matchB4[0]}
\end{eqnarray}
for the axial vector current.
The exponentials here contain the tree-level rest mass
\begin{equation}
	m_{1h}^{[0]}a = \log(1 + m_{0h}a),
\end{equation}
which enters through the wave-function normalization.
The easiest way to derive these results is to combine the Feynman rule 
for the current in Appendix~\ref{app:feynman} with the Feynman rules 
for external lines in Appendix~C of Ref.~\cite{El-Khadra:1997mp}, and 
expand the matrix element 
$\langle c(\bbox{p}')|J^\mu_{\text{lat}}|b(\bbox{p})\rangle$ to 
first order in~$\bbox{p}$ and~$\bbox{p}'$.
The zeroth order yields the $\bar{C}$ coefficients, and the first 
order the $\bar{B}$ coefficients with, for our lattice Lagrangian and 
currents,
\begin{equation}
	\frac{1}{2m^{[0]}_{3h}a} = 
		\frac{\zeta(1+m_{0h}a)}{m_{0h}a(2+m_{0h}a)} - d_{1h}
 	\label{eq:m3}
\end{equation}
for each flavor~$h$.

One should compare Eqs.~(\ref{eq:matchCV[0]})--(\ref{eq:matchB4[0]}) 
with Eqs.~(\ref{eq:B1B2})--(\ref{eq:B3B6}).
The results for $\bar{C}^{\text{lat}[0]}_{V_\parallel}$ and 
$\bar{C}^{\text{lat}[0]}_{A_\perp}$, together with 
$\bar{C}_{V_\parallel}^{[0]}=\bar{C}_{A_\perp}^{[0]}=1$, show that 
$Z_{V_\parallel}^{[0]}=Z_{A_\perp}^{[0]}=%
e^{(m_{1c}^{[0]}+m_{1b}^{[0]})a/2}$.
[Recall, in view of Eqs.~(\ref{eq:ZJpara}) and~(\ref{eq:ZJperp}),
that we drop the bars from~$Z_J$.]
To obtain $\delta\bar{B}_{Ji}^{[0]}=0$, one adjusts $d_1$ and 
the~$b_{Ji}$.
The way to adjust $d_1$, at the tree level, is to set $m^{[0]}_3$ equal
to the (tree-level) heavy-quark mass.
Since the rest mass plays only a trivial role in heavy-quark physics,
the appropriate (tree-level) matching condition is
$m^{[0]}_3=m^{[0]}_2$, which implies
\begin{equation}
	d_1 = \frac{\zeta(1+m_0a-\zeta)}{m_0a(2+m_0a)} - 
		\frac{r_s\zeta}{2(1+m_0a)}.
	\label{eq:d1(m)}
\end{equation}
If $d_1$ is adjusted in this way, one can then take all~$b_{Ji}^{[0]}=0$.

Beyond the tree level, it is convenient to define $d_1$ so that spatial
component of the degenerate-mass vector current is correctly normalized.
Then $Q_{V2}$ and $Q_{V5}$ would be superfluous for equal mass,
although for unequal masses both are still needed.
For the axial current $Q_{A1}$ and $Q_{A4}$ are required even for
equal masses.

For equal-mass currents it is possible to compute $Z_{V_\parallel^{hh}}$
nonperturbatively for all~$m_ha$.
One may therefore prefer to write
\begin{equation}
	Z_{J^{cb}} = \sqrt{ Z_{V_\parallel^{cc}} Z_{V_\parallel^{bb}} }
		\rho_{J^{cb}}
	\label{eq:rho}
\end{equation}
and compute only the factor $\rho_{J^{cb}}$ in perturbation theory.
This split is very useful in numerical calculations of matrix
elements~\cite{Hashimoto:2000yp,Hashimoto:2001ds,El-Khadra:2001rv}.
The strong mass dependence of the $Z_J$s cancels, as do the 
contributions of tadpole diagrams: both enter through the self energy, 
which is common to all $Z_J$ factors.
Hence, the expansion coefficients for $\rho_J$ are expected to be small,
and, as a consequence, the uncertainty in $\rho_J$ from truncating
perturbation theory at fixed order is under better control than the
uncertainty in the corresponding~$Z_J$.
The one-loop calculation of $Z_{V_\parallel^{hh}}^{[1]}$ is given in 
Sec.~\ref{subsec:degenerate}.
It has been used in Ref.~\cite{Harada:2001hl} to obtain 
analogous~$\rho_J$ factors for heavy-light currents.
For both heavy-heavy and heavy-light currents our calculations verify 
that, generically, the one-loop terms~$\rho_J^{[1]}$ are smaller than 
the corresponding~$Z_J^{[1]}$.

\section{One-Loop Results}
\label{sec:loop}

In this section we present results for the matching factors at the
one-loop level in perturbation theory.
For the Wilson and Sheikholeslami-Wohlert (SW) actions, the one-loop 
contributions were given in Ref.~\cite{Kronfeld:1999tk}.
At that stage we omitted the rotation term in the current, 
which is needed to match dimension-four currents at the tree 
level~\cite{Kronfeld:1995nu,El-Khadra:1997mp}.
Now we complete the work started in Ref.~\cite{Kronfeld:1999tk} and 
report results with the rotation.
For comparison we also present our results without the rotation,
both with and without the clover term.
After some general remarks in Sec.~\ref{subsec:general}, we present 
numerical results for the one-loop terms.
To illustrate the general features, we present three special cases: 
equal initial- and final-state masses in Sec.~\ref{subsec:degenerate},
unequal masses with a fixed mass ratio in 
Secs.~\ref{subsec:unequal:fixed-ratio}, 
and unequal masses with a fixed final-state mass in 
Sec.~\ref{subsec:unequal:fixed-mc}.

In the appendices, we present expressions for the Feynman rules and 
one-loop integrands, allowing both flavors of heavy quarks to have
independent and arbitrary values of all couplings.
A~computer code for generating these results is freely
available~\cite{p:epaps}.

\subsection{General remarks}
\label{subsec:general}

We shall apply the formalism of Sec.~\ref{sec:hqet} with 
$v'=v=(i,\bbox{0})$, which applies when the initial and final states 
are at rest, or nearly at rest.
The matching factors $Z_{V_\parallel}$ and $Z_{A_\perp}$ are simply
ratios of the lattice and continuum radiative corrections:
\begin{equation}
	Z_J = \frac{%
	\left[Z_{2b}^{1/2}\Lambda_J Z_{2c}^{1/2}\right]^{\rm cont}}{%
	\left[Z_{2b}^{1/2}\Lambda_J Z_{2c}^{1/2}\right]^{\rm lat}},
	\label{eq:ZJ}
\end{equation}
where $Z_{2b}$ and $Z_{2c}$ are wave-function renormalization factors 
of the heavy bottom and charmed quarks.
[Recall, in view of Eqs.~(\ref{eq:ZJpara}) and~(\ref{eq:ZJperp}),
that we drop the bars from~$Z_J$.]
The vertex function~$\Lambda_J$ is the sum of one-particle irreducible 
three-point diagrams, in which 
one point comes   from the current~$J$ ($J=V_\parallel$, $A_\perp$), 
and the other two from the external quark states.
The expression relating $Z_2$ to the lattice self energy, for all 
masses and gauge couplings, can be found in Ref.~\cite{Mertens:1998wx}.
In view of the mass dependence of the self-energy function,
$Z_2\propto e^{-m_1a}$, we write
\begin{equation}
	e^{-(m_{1c}^{[0]}a+m_{1b}^{[0]}a)/2} Z_J =
		1 + \sum_{l=1}^\infty g^{2l} Z_J^{[l]},
	\label{eq:ZJ[1]}
\end{equation}
so that the $Z_J^{[l]}$ are only mildly mass dependent.
(This convention follows that of Ref.~\cite{Harada:2001hl}, but is 
slightly different from the one in Ref.~\cite{Mertens:1998wx}.)
By construction, the exponential mass dependence in $\rho_J$ cancels out,
so we write
\begin{equation}
	\rho_J =
		1 + \sum_{l=1}^\infty g_0^{2l} \rho_J^{[l]}.
\end{equation}

In each of the following subsections we also present the information 
needed to obtain the BLM scale~\cite{Brodsky:1983gc,Lepage:1993xa}.
Because the ideas behind the BLM prescription are covered in our 
paper on heavy-light currents~\cite{Harada:2001hl}, we give only the 
essential formulas here.
Let~$\zeta^{[1]}$ be one of our one-loop terms.
From the Feynman diagram, it is expressed as
\begin{equation}
	\zeta^{[1]} = \int d^4k\,f(k),
\end{equation}
where $k$ is the gluon's momentum.
The BLM scale $q^*$ is given through
\begin{equation}
	\ln (q^*a)^2 = {}^*\zeta^{[1]}/\zeta^{[1]}.
	\label{eq:BLMq*}
\end{equation}
where
\begin{equation}
	^*\zeta^{[1]} = \int d^4k\,\ln(ka)^2\,f(k).
\end{equation}
Studies of perturbation theory indicate that using $g^2_V(q^*)$
an expansion parameter usually leads to well-behaved series, because 
the BLM prescription pulls in some of the higher orders.
Here $V$ denotes the scheme for which the heavy-quark potential is 
$V(q)=-C_Fg^2_V(q)/q^2$.
In other schemes the optimal scale is slightly different, so that 
$g^2_S(q^*_S)\approx g^2_V(q^*)$.

It is also interesting to see how $q^*$ changes under tadpole
improvement.
If one introduces the tadpole-improved matching factors
\begin{equation}
	\tilde{Z}_J = Z_J/u_0,
\end{equation}
where the mean link $u_0$ is any tadpole-dominated short-distance
quantity, the arguments of Ref.~\cite{Lepage:1993xa} suggest that
the perturbative series for $\tilde{Z}_J$ has smaller coefficients.
In analogy with Eq.~(\ref{eq:ZJ[1]}) we write
\begin{equation}
	e^{-(\tilde{m}^{[0]}_{1c}+\tilde{m}^{[0]}_{1b})a/2} \tilde{Z}_J =
		1 + \sum_{l=1}^\infty g_0^{2l} \tilde{Z}_J^{[l]},
	\label{eq:tildeZJ[1]}
\end{equation}
where
\begin{equation}
	\tilde{m}_1^{[0]}a = \ln[1 + m_0a/u_0]
\end{equation}
is the tadpole-improved rest mass.
Then
\begin{equation}
	\tilde{Z}_J^{[1]} = Z_J^{[1]} - \frac{1}{2}
		\left(\frac{1}{1+m_{0c}a} + \frac{1}{1+m_{0b}a} \right)
		u_0^{[1]},
	\label{eq:tildeZ[1]}
\end{equation}
and because $Z_J^{[1]}<0$ and $u_0^{[1]}<0$ one sees that the one-loop
coefficients are reduced.
Similarly, for computing the BLM scale
\begin{equation}
	{}^*\tilde{Z}_J^{[1]} = {}^*Z_J^{[1]} - \frac{1}{2}
		\left(\frac{1}{1+m_{0c}a} + \frac{1}{1+m_{0b}a} \right)
		{}^*u_0^{[1]}.
	\label{eq:*tildeZ[1]}
\end{equation}
To illustrate, we take $u_0$ from the average plaquette,
so $u_0^{[1]}=-C_F/16$ and $^*u_0^{[1]}=-0.204049(1)$.

The combinations of wave-function and vertex renormalization in~$Z_J$ 
are gauge invariant and ultraviolet and infrared finite.
Infrared cancellation between lattice and continuum integrands occurs
point-by-point if the masses in continuum propagators are set equal
to the corresponding kinetic masses.
Because there are no divergences, it is straightforward to weight the 
integrands with~$\ln(ka)^2$.

Although the main results of this section, given below, are the 
calculation of the full mass dependence of the one-loop terms, 
let us anticipate what to expect in limiting cases.
When both masses vanish, we must (and do) obtain the textbook result 
for matching in the Symanzik formalism.
If one mass vanishes, we must recover our results for heavy-light 
matching factors.
In fact, our codes are based on the same integrand functions, so this 
``check'' is automatic.

When both masses satisfy $m_h\gg a^{-1}$, both heavy-quark propagators
become static, and, consequently, the one-loop lattice radiative
corrections vanish.
Thus, the matching factors reduce to the continuum QCD radiative 
corrections.
Using Eqs.~(\ref{eq:Cparallel}) and~(\ref{eq:Cperp})
\begin{eqnarray}
	\lim_{m_ba,\,m_ca\to\infty} Z_{V_\parallel}^{[1]} \to
	\bar{C}_{V_\parallel}^{[1]} & = & C_F \,
		3 f(m_{2c}/m_{2b})/16\pi^2 ,
		\label{eq:etaV[1]} \\
	\lim_{m_ba,\,m_ca\to\infty} Z_{A_\perp}^{[1]}     \to
	\bar{C}_{A_\perp}^{[1]}     & = & C_F \left[
		3 f(m_{2c}/m_{2b}) - 2 \right]/16\pi^2 .
		\label{eq:etaA[1]}
\end{eqnarray}
The lattice contribution to the BLM numerator~$^*\zeta^{[1]}$ also 
vanishes in this limit, leaving
\begin{eqnarray}
	\lim_{m_ba,\,m_ca\to\infty} {}^*Z_{V_\parallel}^{[1]} \to
	{}^*\bar{C}_{V_\parallel}^{[1]} & = & C_F \,
		9 f(m_{2c}/m_{2b})/32\pi^2 +
		\bar{C}_{V_\parallel}^{[1]} \ln(m_{2b}am_{2c}a), 
		\label{eq:*etaV[1]} \\
	\lim_{m_ba,\,m_ca\to\infty} {}^*Z_{A_\perp}^{[1]}     \to
	{}^*\bar{C}_{A_\perp}^{[1]}     & = & C_F \left[
		\case{5}{2} f(m_{2c}/m_{2b}) - 1 \right]/16\pi^2 +
		\bar{C}_{A_\perp}^{[1]}     \ln(m_{2b}am_{2c}a),
		\label{eq:*etaA[1]}
\end{eqnarray}
where the right-hand sides are due to Ref.~\cite{Neubert:1995qt}.

When one mass becomes very large, but the other is held fixed, one
can still deduce the limiting behavior from
Eqs.~(\ref{eq:etaV[1]})--(\ref{eq:*etaA[1]}).
As the heavier mass increases, $m_{2b}a\to\infty$, the lattice Feynman
diagrams become independent of~$m_{2b}a$.
The logarithmic part then takes the form $3\ln(m_{2c}a)$.
This lattice logarithm cancels only part of the corresponding 
logarithm in~$3f(m_{2c}/m_{2b})$, leaving the matching factor 
with~$3\ln(m_{2b}a)$.
Then non-logarithmic constants depend, of course, on the fixed mass of 
the lighter quark.
In the BLM numerators, for $m_{2c}a$ held fixed and $m_{2b}a\to\infty$,
one obtains a quadratic in $\ln(m_{2b}a)$, in which the coefficient of
$\ln^2(m_{2b}a)$ is identical to those of the continuum radiative
corrections, and the coefficient of the single logarithm and the
non-logarithmic term itself are mildly dependent on~$m_{2c}a$.

Below we plot the mass dependence as a function of~$m_{1b}^{[0]}a$, 
because it brings out the asymptotic behavior for both $m_ha=0$ 
and~$m_ha\to\infty$:
when $m_ha\ll 1$, $m_{1h}^{[0]} \approx     m_{2h}^{[0]}$, but
when $m_ha\gg 1$, $m_{1h}^{[0]}a\approx \ln(m_{2h}^{[0]}a)$.
Thus, a plot against~$m_{1b}^{[0]}a$ makes it easy to look both at
slope and curvature in the small-mass region, and at the expected
logarithms in the large-mass region.

In presenting results below, we show together $Z_J^{[1]}$,
$^*Z_J^{[1]}$, $\rho_J^{[1]}$, and~$^*\rho_J^{[1]}$, for each current,
followed by the BLM scale for $Z_{V_\parallel}$ and $Z_{A_\perp}$.

\subsection{Degenerate masses}
\label{subsec:degenerate}

\begin{figure*}[p]
	\centering\small
	(a) \includegraphics[width=0.44\textwidth]{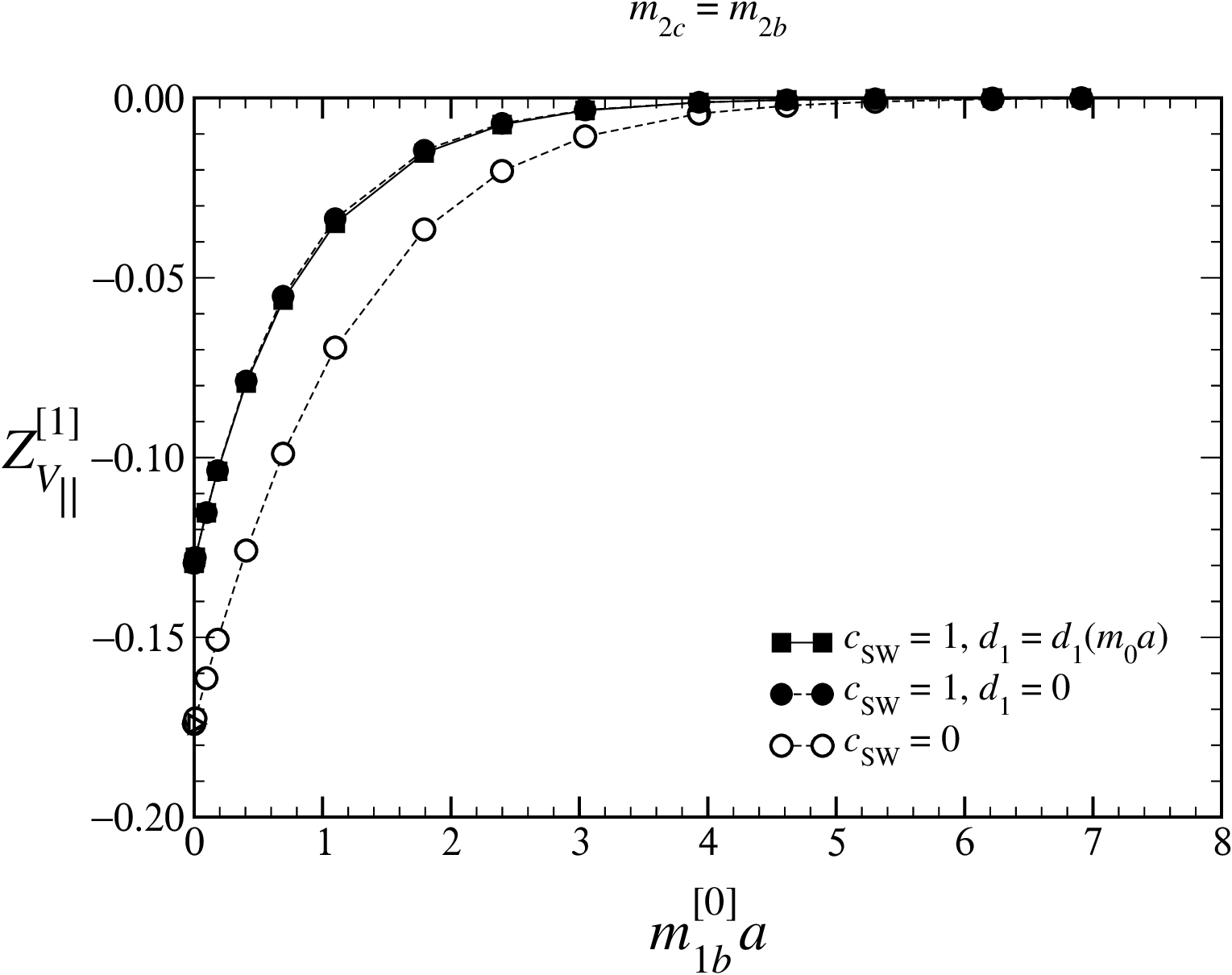} \hfill
	(b) \includegraphics[width=0.44\textwidth]{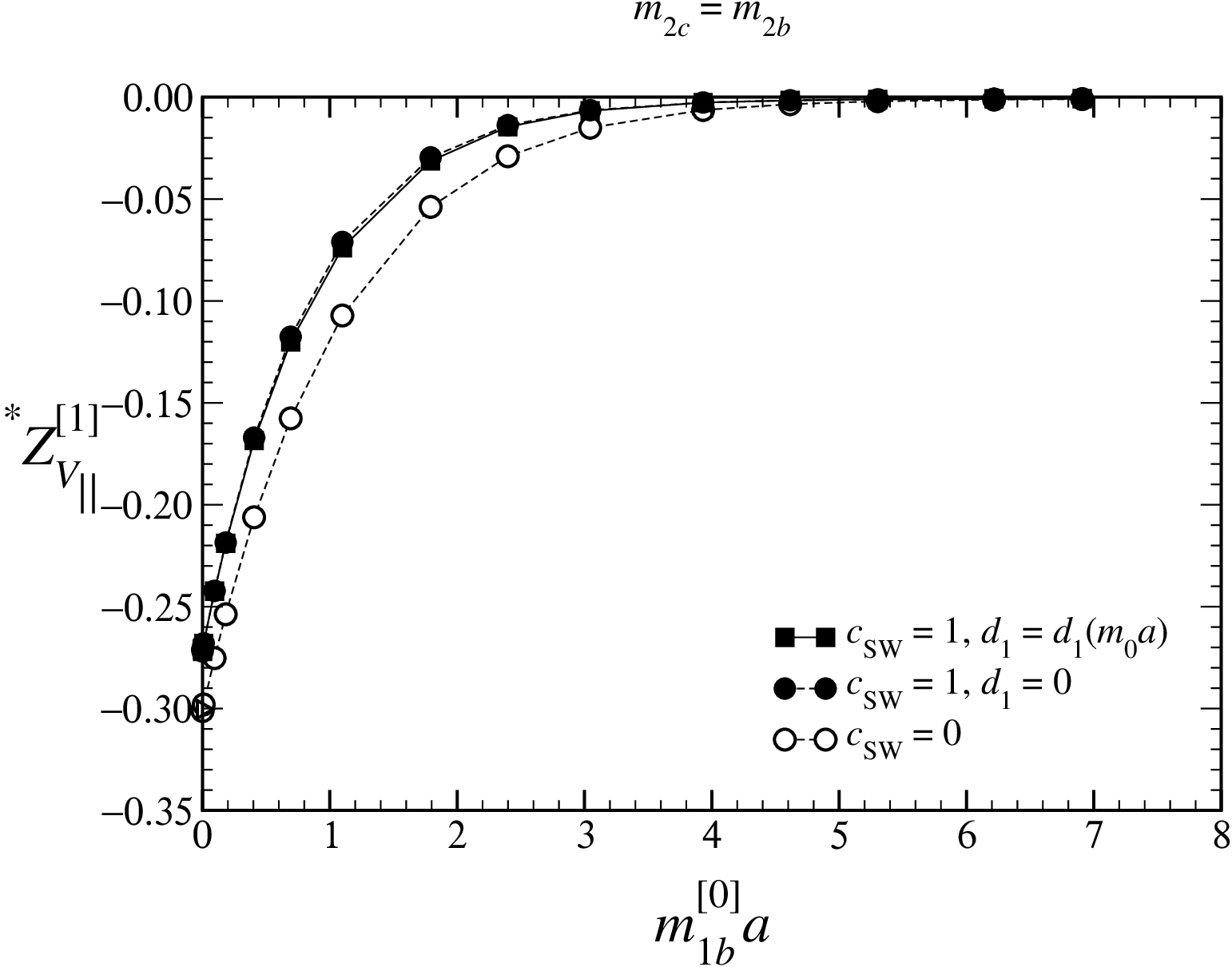}
	\caption[fig:ZVz]{Full mass dependence of the one-loop coefficients
	    of the matching factors of the vector current with equal masses:
	(a)~$Z_{V_\parallel}^{[1]}$ and (b)~$^*Z_{V_\parallel}^{[1]}$.
	Filled (open) symbols denote the SW (Wilson) action.
	Solid (dotted) lines connecting squares (circles) 
	indicate the rotation is included (omitted).}
	\label{fig:ZV1}
\end{figure*}

\begin{figure*}[p]
	\centering\small
	(a) \includegraphics[width=0.44\textwidth]{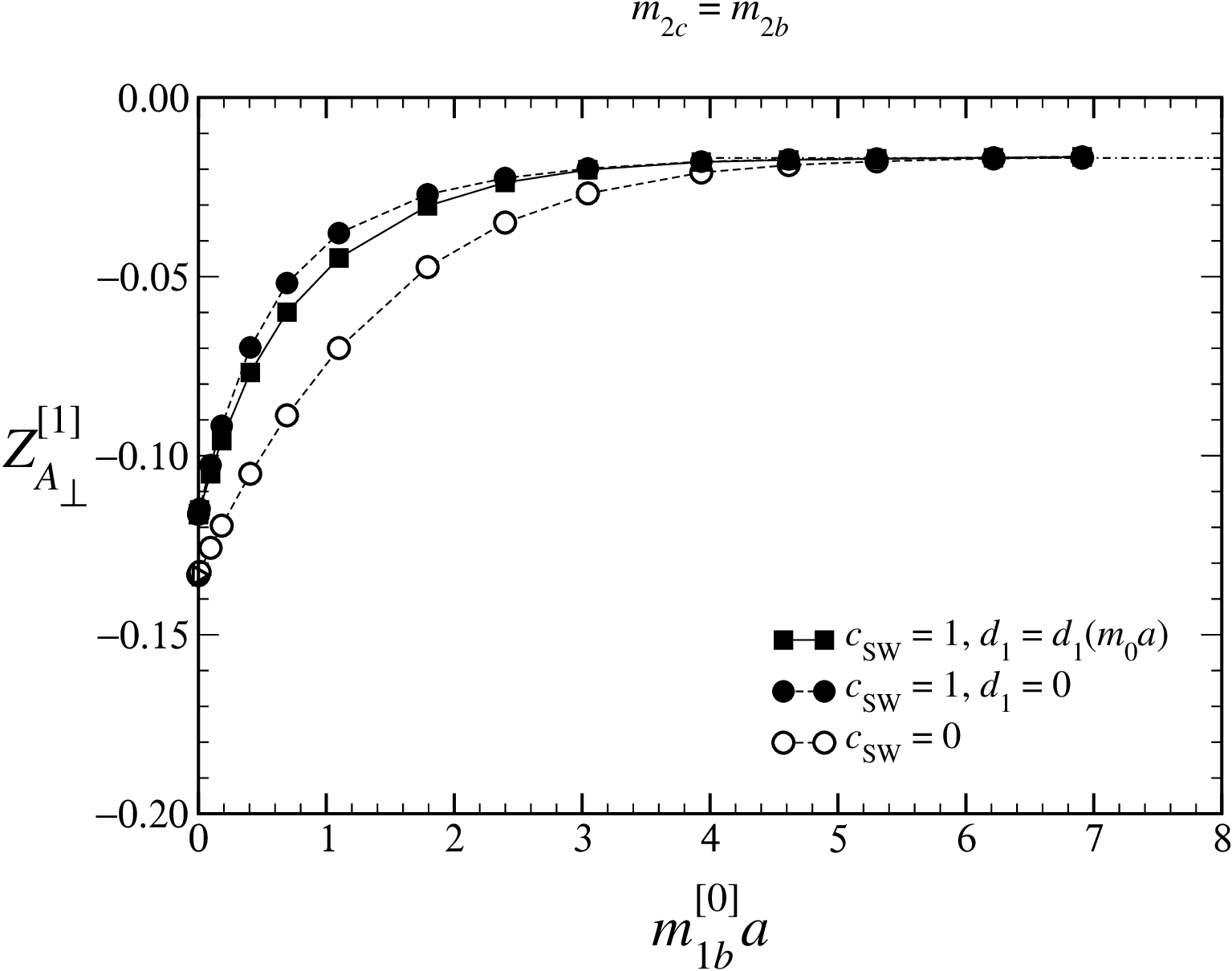} \hfill
	(b) \includegraphics[width=0.44\textwidth]{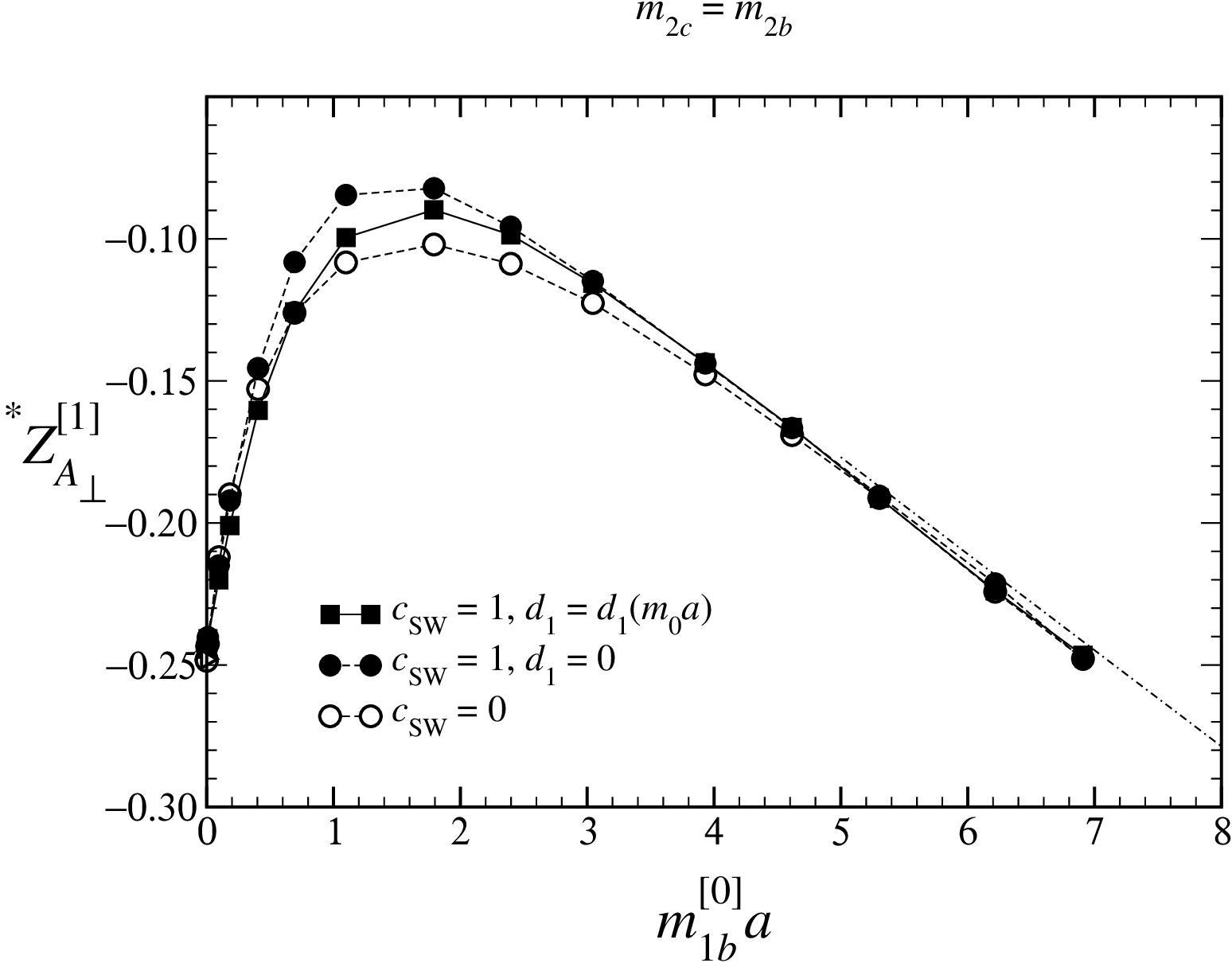} \\[0.5em]
	(c) \includegraphics[width=0.44\textwidth]{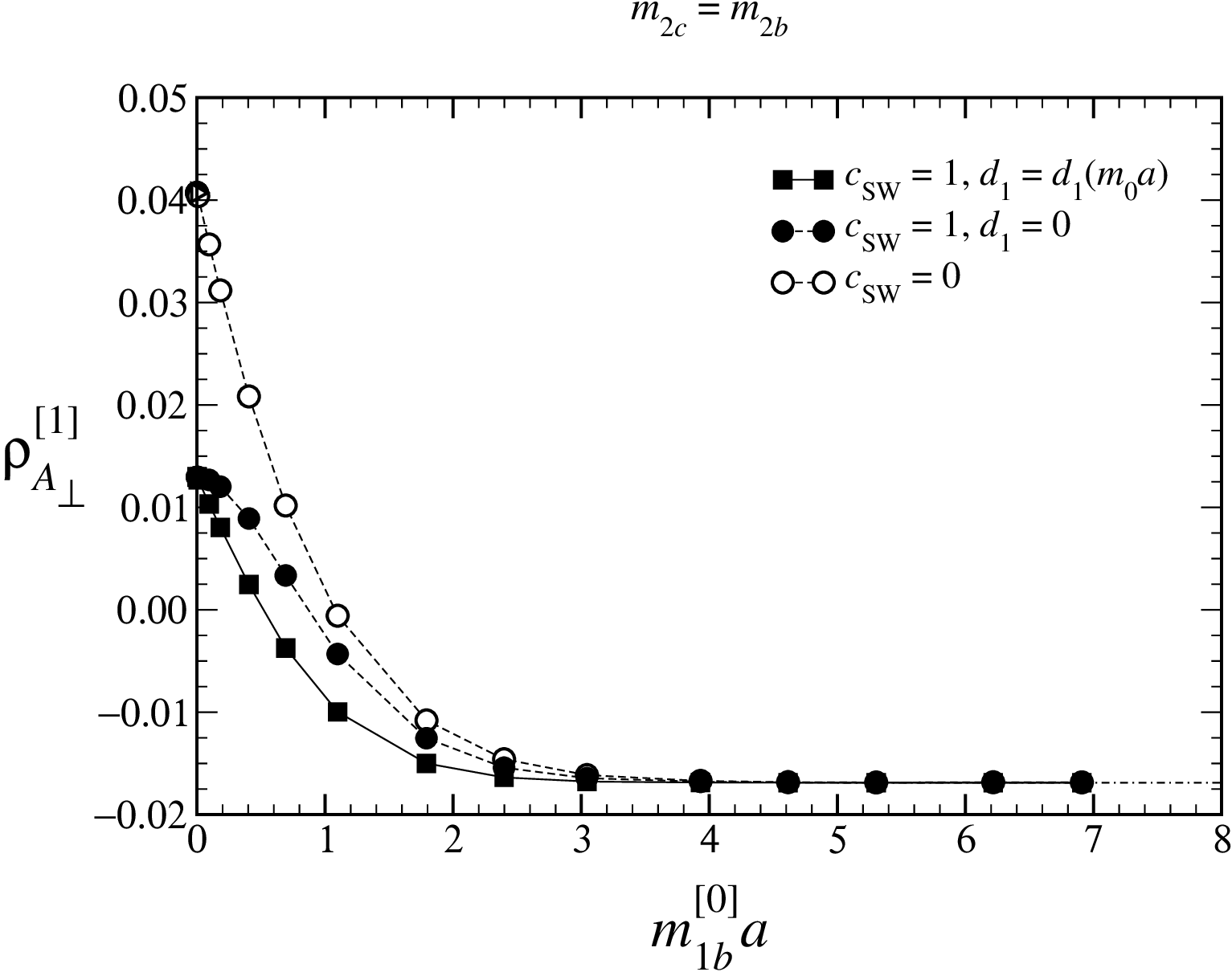} \hfill
	(d) \includegraphics[width=0.44\textwidth]{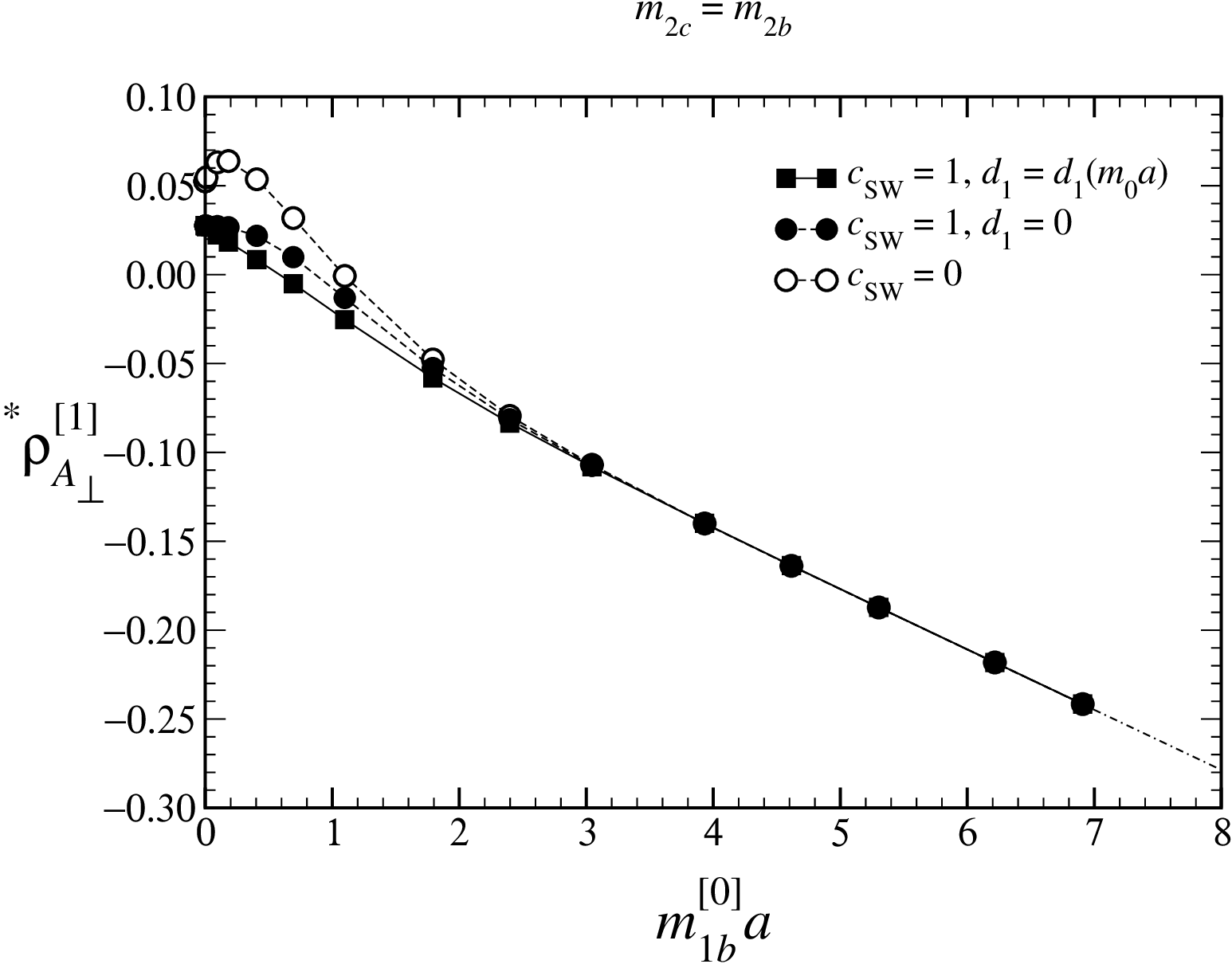}
	\caption[fig:ZA1]{Full mass dependence of the one-loop coefficients
	of the matching factors of the axial vector current with equal masses:
	   (a)~$Z_{A_\perp}^{[1]}$,        (b)~$^*Z_{A_\perp}^{[1]}$,
	(c)~$\rho_{A_\perp}^{[1]}$, and (d)~$^*\rho_{A_\perp}^{[1]}$.
	The symbols have the same meaning as in Fig.~\ref{fig:ZV1};
	dashed-dotted lines show the infinite-mass limit.}
	\label{fig:ZA1}
\end{figure*}

Figure~\ref{fig:ZV1} shows the full mass dependence of the matching 
factor~$Z_{V_\parallel}^{[1]}$ for the vector current, with $m_{2c}=m_{2b}$, 
and its BLM numerator~$^*Z_{V_\parallel}^{[1]}$.
Figure~\ref{fig:ZA1} shows the same for the axial vector current, 
including the one-loop term for~$\rho_{A_\perp}$ and its BLM 
numerator~$^*\rho_{A_\perp}$.
In this case, with equal masses, one has 
simply~$\rho_{A_\perp}=Z_{A_\perp}/Z_{V_\parallel}$.
As expected, all interpolate smoothly from small to large quark mass.
Except for $^*\rho_{A_\perp}^{[1]}$, the curves exhibit a ``knee'' 
around $m_1^{[0]}a\sim 1$--2 ($m_0a\approx m_2a\sim2$--6).
For $^*\rho_{A_\perp}^{[1]}$ the knee effect seems to cancel, and the 
function crosses over quickly to logarithmic behavior.

For $m_ca=m_ba=0$, our matching factors must (and do) reduce to the 
well-known massless limit, obtained long ago for
$c_{\text{SW}}=1$~\cite{Gabrielli:1991us} and even longer ago for
$c_{\text{SW}}=0$~\cite{Martinelli:1983mw}.
As discussed above, in the infinite mass limit (with $a$ fixed), 
the heavy-quark spin symmetry of Wilson fermions ensures that the 
lattice radiative corrections vanish.
The infinite mass limits are $Z_{V_\parallel}^{[1]}=0$, 
$^*Z_{V_\parallel}^{[1]}=0$, 
$Z_{A_\perp}^{[1]}=\rho_{A_\perp}^{[1]}=-2C_F/16\pi^2$, 
and $^*Z_{A_\perp}^{[1]}={}^*\rho_{A_\perp}^{[1]}=%
-C_F[4\ln(m_2a)+1]/16\pi^2$.
They are shown in each panel of Figs.~\ref{fig:ZV1} and~\ref{fig:ZA1} 
with a dashed-dotted line.
As a rule, the tendency to the infinite mass limit becomes obvious
for $m_1^{[0]}a>5$ ($m_0a\approx m_2a>150$).

For smaller masses ($m_ca,~m_ba\lesssim1$), one expects the one-loop 
term $\rho_{A_\perp}^{[1]}$ to be smaller than~$Z_{A_\perp}^{[1]}$, 
because it does not contain any tadpole diagrams.
This expectation is borne out in comparing Fig.~\ref{fig:ZA1}(a) and~(c).
For larger masses, the tadpole diagram is suppressed,
cf.\ Eqs.~(\ref{eq:tildeZ[1]}) and~(\ref{eq:*tildeZ[1]}),
and $\rho_{A_\perp}^{[1]}$ and~$Z_{A_\perp}^{[1]}$ have the same 
asymptotic value.
By comparing Fig.~\ref{fig:ZA1}(b) and~(d), one sees that 
$^*\rho_{A_\perp}^{[1]}$ is smaller than~$^*Z_{A_\perp}^{[1]}$ also.

\begin{figure}
	\centering\small
	(a) \includegraphics[width=0.44\textwidth]{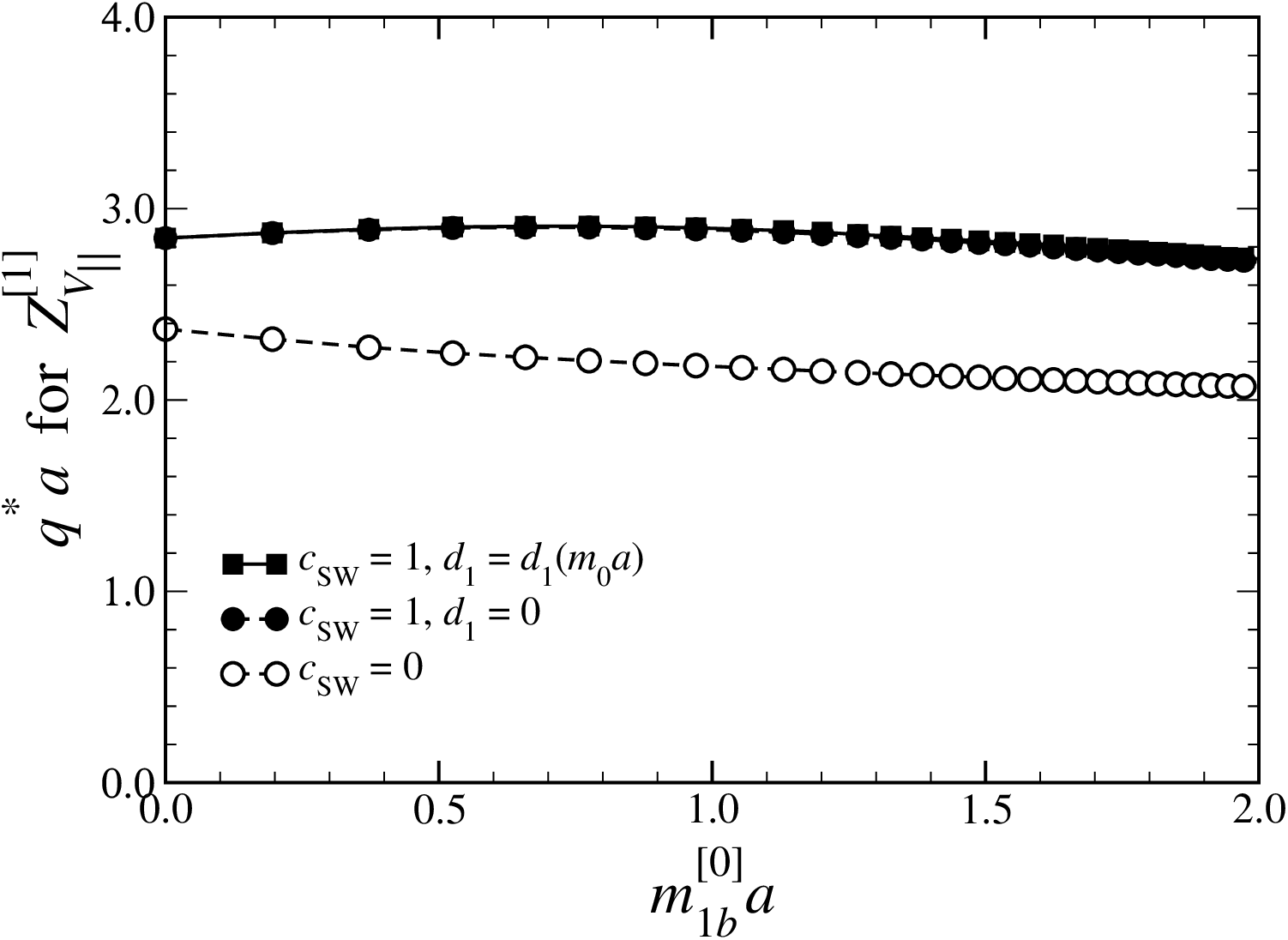} \hfill
	(b) \includegraphics[width=0.44\textwidth]{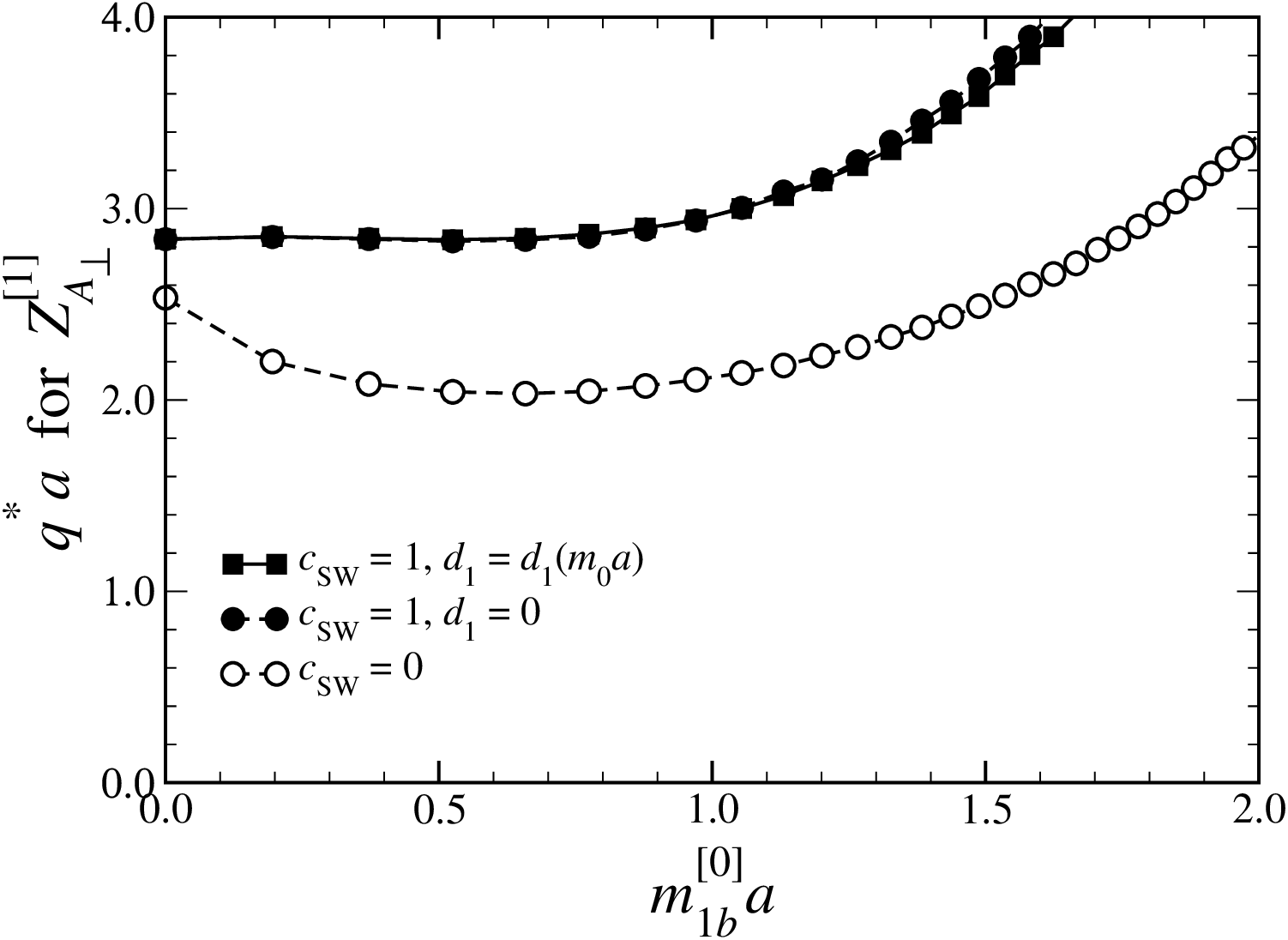}
	\caption[fig:qstar1]{Full mass dependence of the BLM scale~$q^*a$
	for the matching factors
	(a)~$Z_{V_\parallel}$ and (b)~$Z_{A_\perp}$,
	with $m_c=m_b$.
	The symbols have the same meaning as in Figs.~\ref{fig:ZV1}
	and~\ref{fig:ZA1}.}
	\label{fig:qstar1}
\end{figure}

\begin{figure}
	\centering\small
	(a) \includegraphics[width=0.44\textwidth]{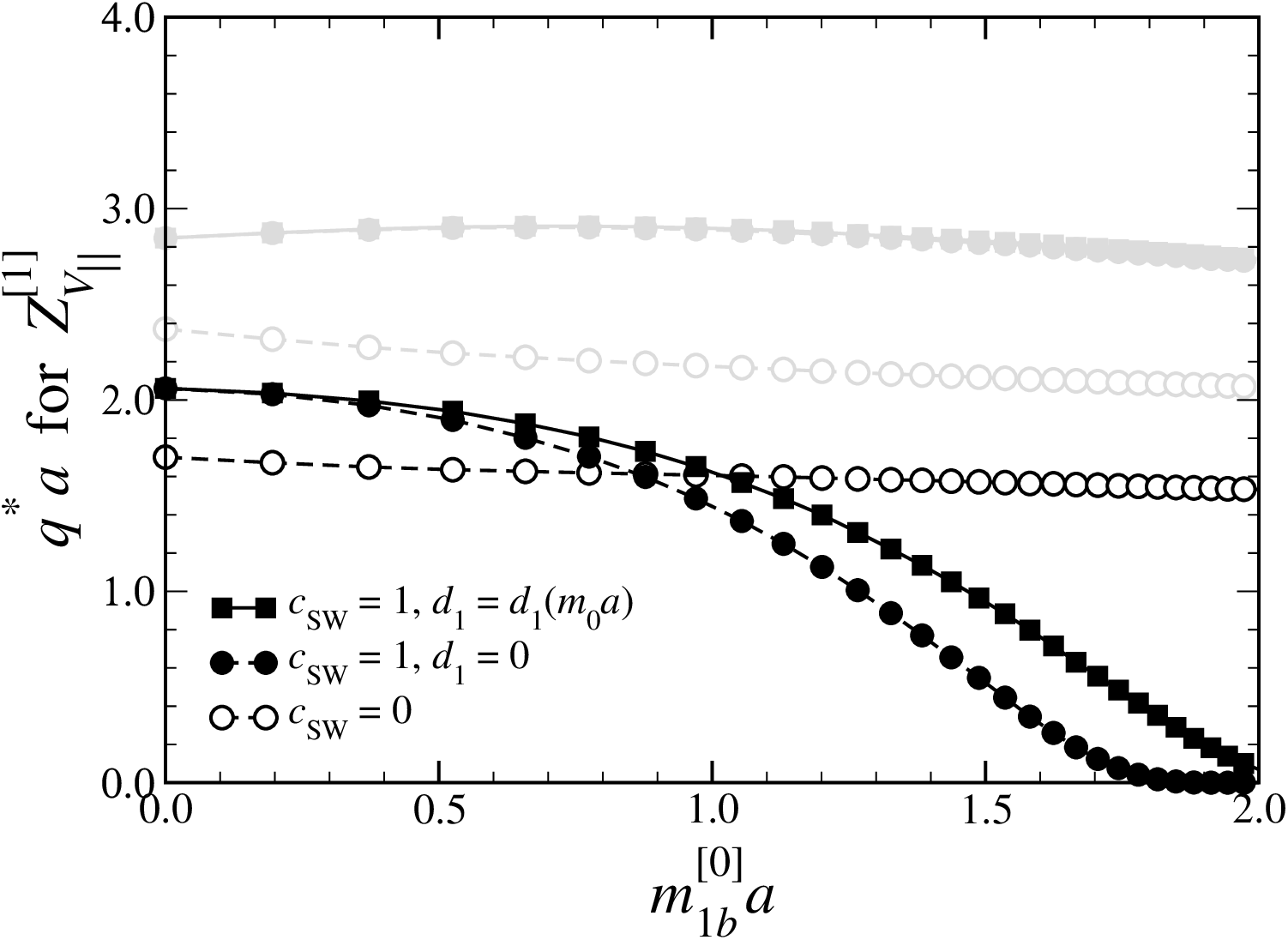} \hfill
	(b) \includegraphics[width=0.44\textwidth]{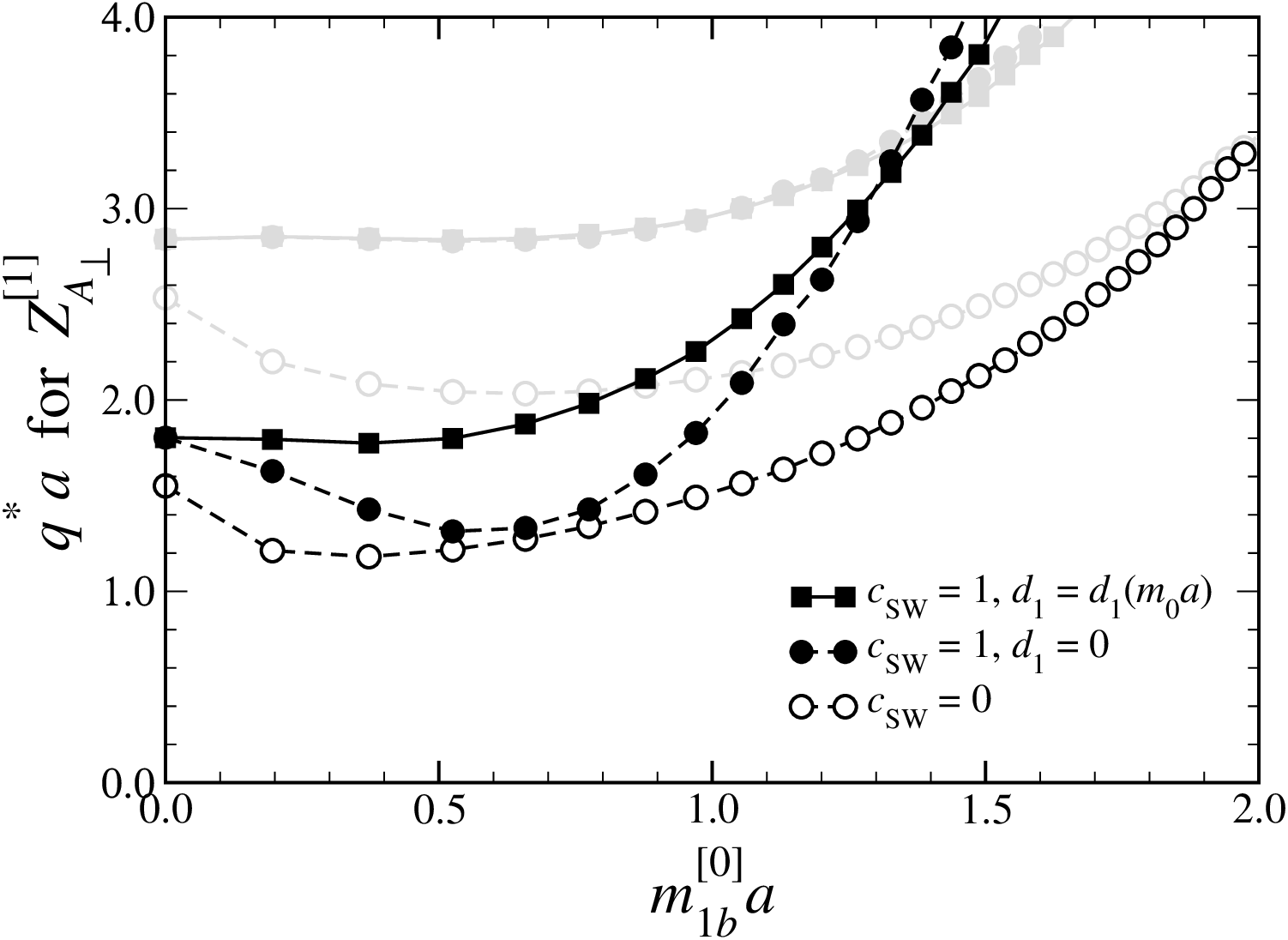}
	\caption[fig:tadqstar1]{Full mass dependence of the BLM scale~$q^*a$
	for the matching factors
	(a)~$Z_{V_\parallel}$ and (b)~$Z_{A_\perp}$,
	with $m_c=m_b$, after tadpole improvement.
	Curves from Fig.~\ref{fig:qstar1} are shown in grey.}
	\label{fig:tadqstar1}
\end{figure}

In Fig.~\ref{fig:qstar1} we combine the information from
Figs.~\ref{fig:ZV1} and~\ref{fig:ZA1} and show the BLM scale~$q^*a$ for
the vector and axial vector currents.
In the region of greatest interest, $m_{1b}\le1$ ($m_{2b}<1.5$), we find
$q^*a\approx2.8$ for the clover action.
In Fig.~\ref{fig:tadqstar1} we apply tadpole improvement.
In the region of greatest interest, $m_{1b}\le1$, we find a smaller
$q^*a\approx1.8$--2.0 for the clover action.
The singular behavior for $Z_{V_\parallel}$ for $m_{1b}\gtrsim1.5$
arises because the denominator vanishes.
If numerator and denominator in Eq.~(\ref{eq:BLMq*}) have the same
(opposite) sign, then $q^*a\to\infty$ ($q^*a\to0$).
In such a case the BLM prescription does not make sense and should
be modified~\cite{Hornbostel:2001ey}.

\begin{figure}
	\centering
	\includegraphics[width=0.54\textwidth]{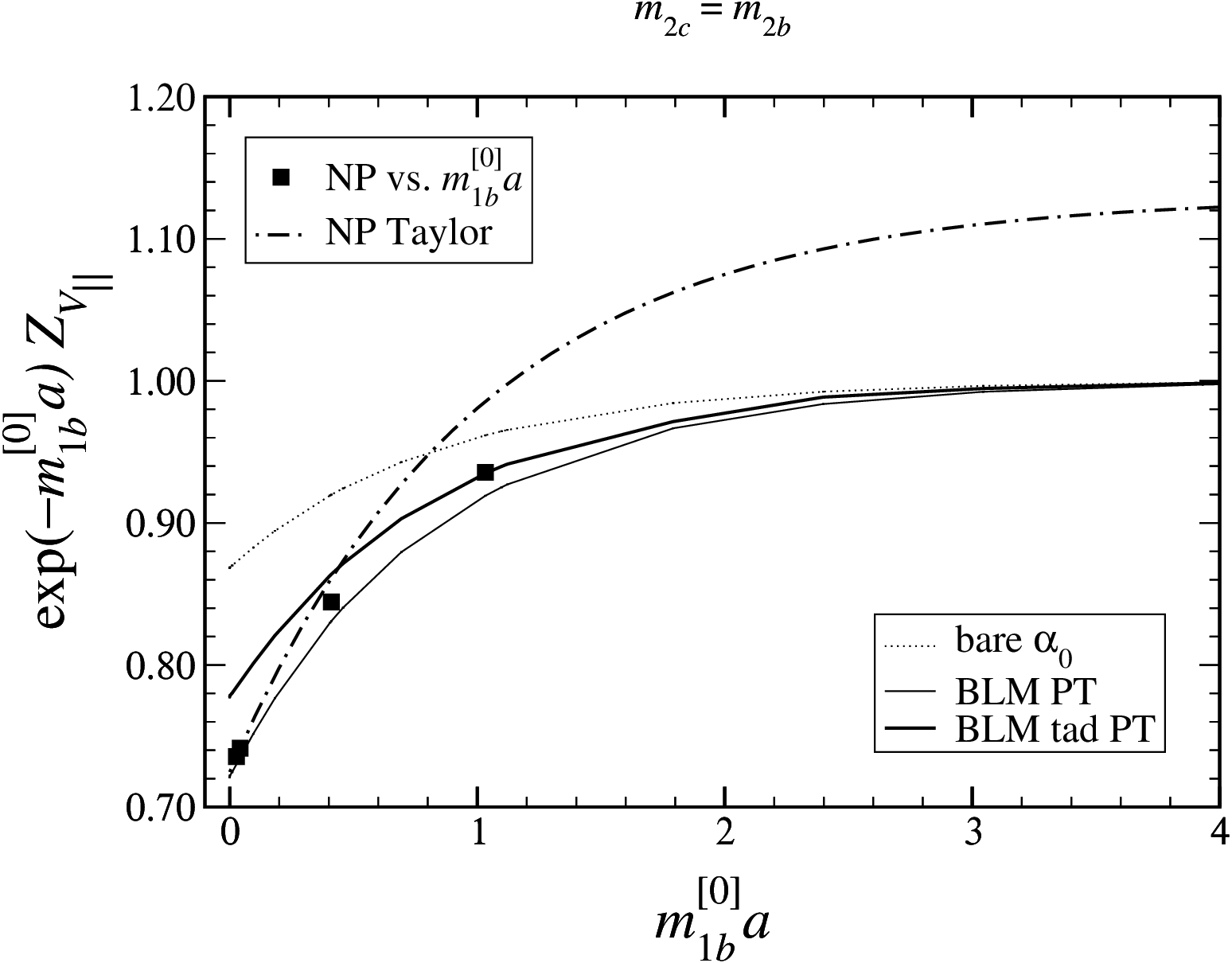}
	\caption[fig:NPvsPT]{Comparison of full mass dependence of the matching
	factor~$Z_{V_\parallel}$.
	The symbols are a non-perturbative, hadronic
	calculation~\cite{Simone:2001zv}.
	The dotted curve is bare perturbation theory.
	The solid curves are BLM perturbation theory;
	the heavy (light) line includes (omits) tadpole improvement.
	The dashed-double-dotted curve shows the first-order Taylor series,
	with intercept and slope determined non-perturbatively.}
	\label{fig:NPvsPT}
\end{figure}

With equal masses, the matching factor $Z_{V_\parallel}$ can be computed
non-perturbatively, allowing us to test how well BLM perturbation theory
works.
Figure~\ref{fig:NPvsPT} plots $\exp(-m_{1b}^{[0]}a)Z_{V_\parallel}$
vs.\ $m_{1b}^{[0]}a$ for several methods of calculating
$Z_{V_\parallel}$.
In perturbation theory, we truncate Eq.~(\ref{eq:ZJ[1]}) at the first
non-trivial term, and use either the bare coupling~$g_0^2$ or the BLM
prescription~$g_V^2(q^*)$.
We also truncate Eq.~(\ref{eq:tildeZJ[1]}) at the first non-trivial
term, which is tadpole improvement.
As a non-perturbative check, we define
\begin{equation}
	\frac{1}{Z_{V_\parallel}^{\text{NP}}} =
		\frac{\langle H| \bar{\Psi}_b\gamma_4\Psi_b |H\rangle%
		}{\langle H|H\rangle}
	 \label{eq:ZVNP}
\end{equation}
where $H$ is a $b$-flavored hadron.
Results for several $b$ masses at bare gauge coupling
$g_0^2=6/5.9$~\cite{Simone:2001zv} are shown in Fig.~\ref{fig:NPvsPT}.
We also plot the Taylor expansion for small mass
\begin{equation}
	Z_{V_\parallel}(m_0a) = Z_V\left[1 + b_V m_0a\right],
	\label{eq:AlphaZV}
\end{equation}
where the notation on the right-hand side follows
Ref.~\cite{Luscher:1996sc}.
From the two small-mass points, we determine
$Z_V=0.7247$ and $b_V=1.5588$
(for $g_0^2=6/5.9$ and $c_{\text{SW}}=1.50$).
Figure~\ref{fig:NPvsPT} reveals several lessons.
First, bare perturbation theory deviates from the non-perturbative
points by several percent.
Second, BLM perturbation theory is quite accurate for all masses.
The deviations are a few percent:
they are neither completely absent nor several percent.
Third, it is less accurate to neglect the full mass dependence (using
only $Z_V$ from $ma\to0$) than to use BLM perturbation theory,
already around the charmed quark mass ($m_{1}\sim0.5$).
Finally, although Eq.~(\ref{eq:AlphaZV}) remains within a few percent
of the full mass dependence for $m_{1}\le0.5$ (cf.\ dashed-double-dotted
curve), it does not have the correct large mass limit.
It gives an explicit example of the pitfalls that arise in applying
the improvement program of Refs.~\cite{Luscher:1996sc,Jansen:1996ck}
to heavy quarks.

\subsection{Unequal masses: fixed mass ratio}
\label{subsec:unequal:fixed-ratio}

In a typical usage of the matching factors, one wants to calculate
a physical process, such as $B\to D^{(*)}l\nu$.
To monitor lattice spacing effects, one would like to repeat the 
calculation at several lattice spacings with the ratio of the initial 
and final heavy quark masses held fixed.
It is therefore useful to see how the matching factors vary with 
$m_{1b}a$ at fixed~$z=m_{2c}^{[0]}/m_{2b}^{[0]}$.
For illustration we choose $z=0.256$, which was used in a recent
calculation~\cite{Hashimoto:2000yp} of the form factors for
$B\to Dl\nu$.

\begin{figure*}
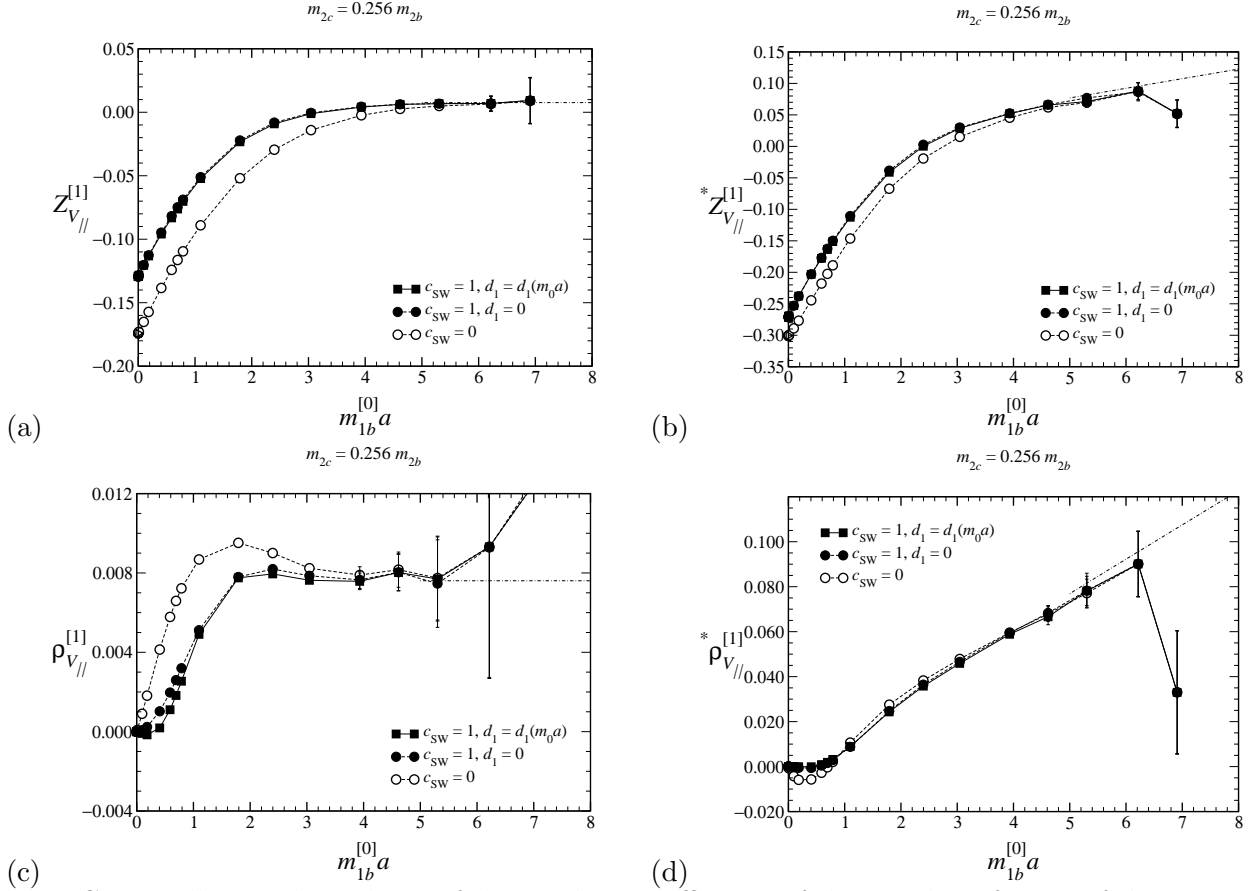

	\centering\small
	(a) \includegraphics[width=0.44\textwidth]{ZV4m1_z.eps} \hfill
	(b) \includegraphics[width=0.44\textwidth]{starZV4m1_z.eps} \\
	(c) \includegraphics[width=0.44\textwidth]{rhoV4m1_z.eps} \hfill
	(d) \includegraphics[width=0.44\textwidth]{starrhoV4m1_z.eps}
	\caption[fig:ZVz]{Full mass dependence of the one-loop coefficients
	of the matching factors of the vector current with $z=0.256$:
	(a)    $Z_{V_\parallel}^{[1]}$,     (b)    $^*Z_{V_\parallel}^{[1]}$,
	(c) $\rho_{V_\parallel}^{[1]}$, and (d) $^*\rho_{V_\parallel}^{[1]}$.
	The symbols have the same meaning as in Figs.~\ref{fig:ZV1} 
	and~\ref{fig:ZA1}.}
	\label{fig:ZVz}
\end{figure*}

\begin{figure*}
	\centering\small
	(a) \includegraphics[width=0.44\textwidth]{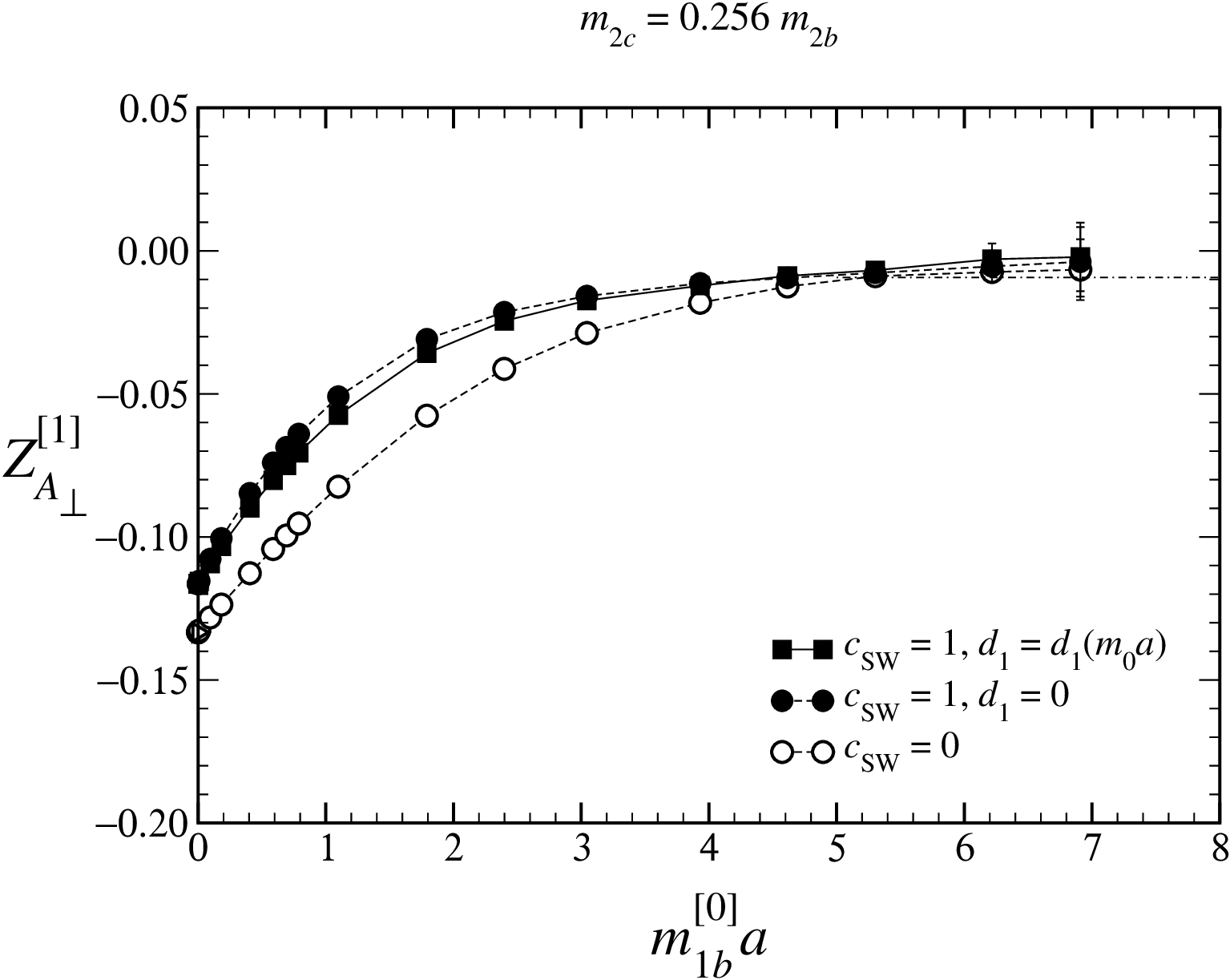} \hfill
	(b) \includegraphics[width=0.44\textwidth]{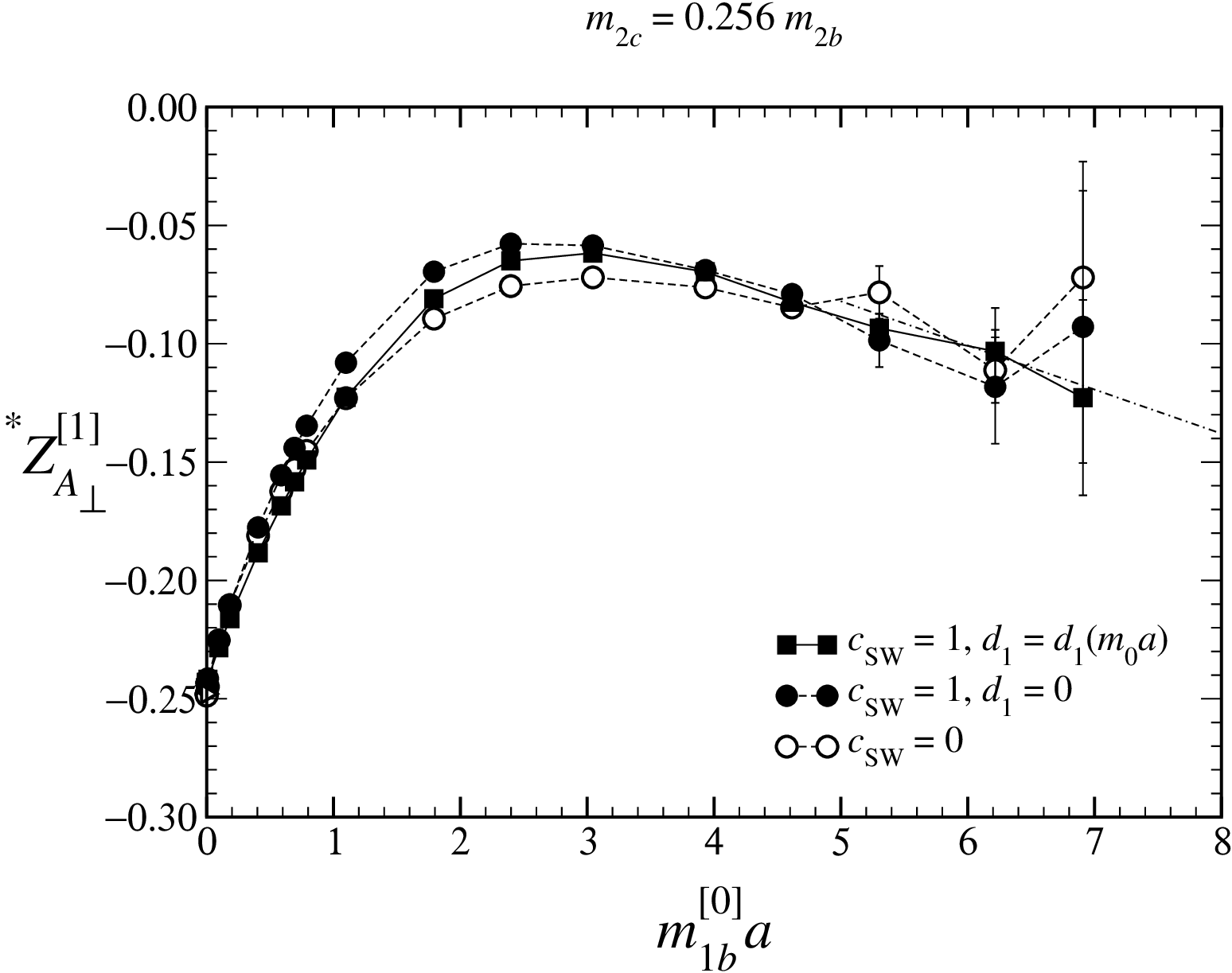} \\
	(c) \includegraphics[width=0.44\textwidth]{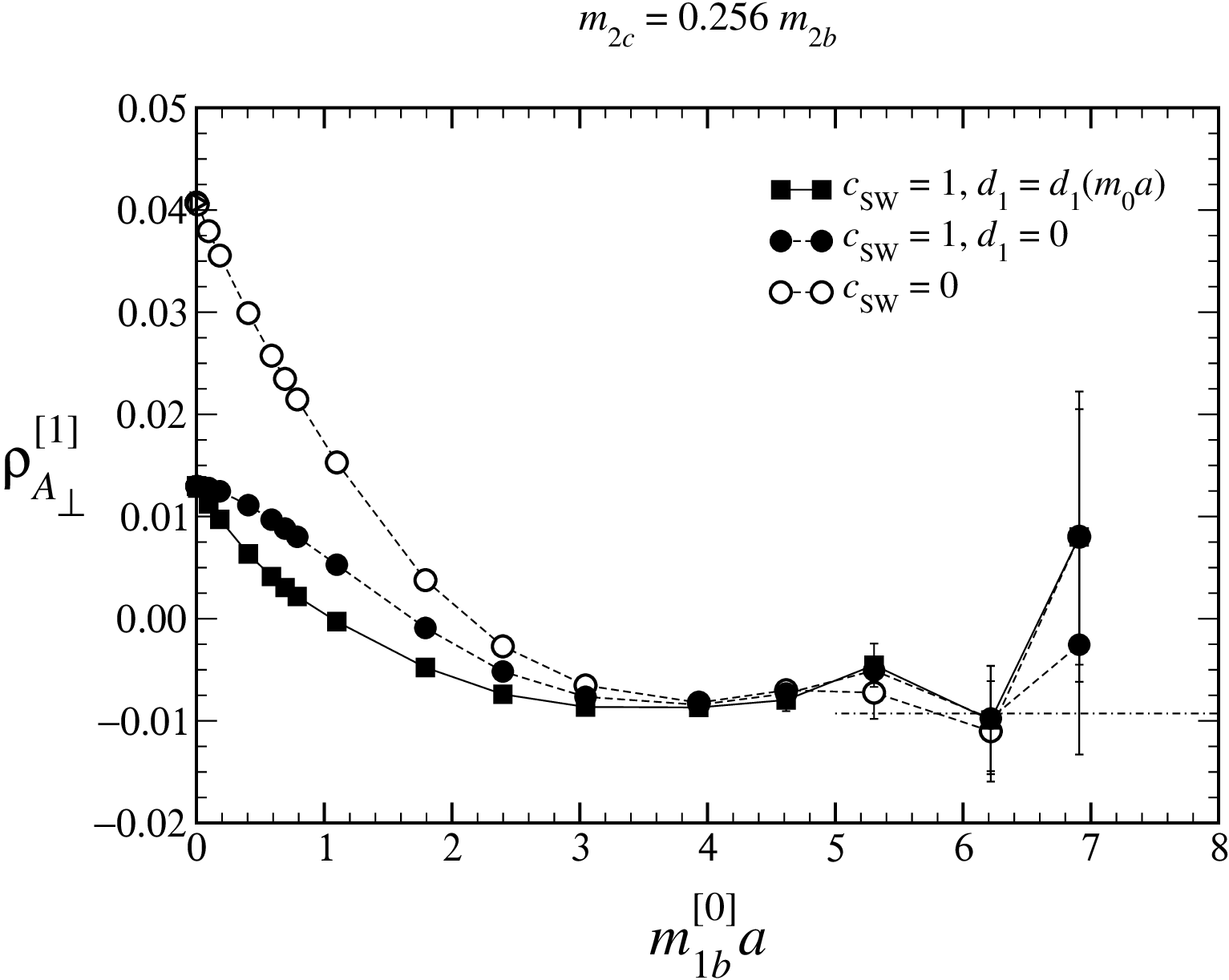} \hfill
	(d) \includegraphics[width=0.44\textwidth]{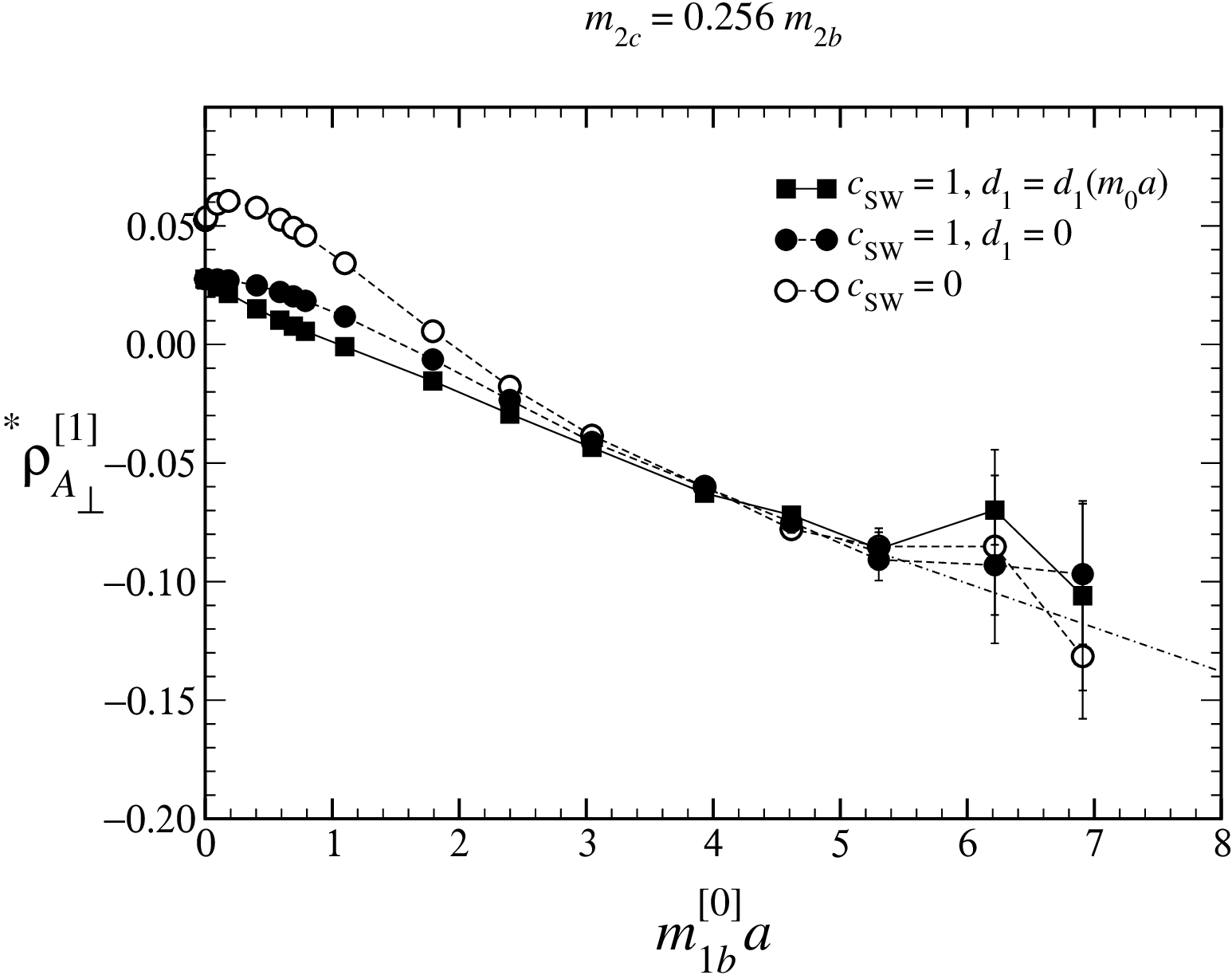}
	\caption[fig:ZAz]{Full mass dependence of the one-loop coefficients
	of the matching factors of the axial vector current with $z=0.256$:
	(a)    $Z_{A_\perp}^{[1]}$,     (b)    $^*Z_{A_\perp}^{[1]}$,
	(c) $\rho_{A_\perp}^{[1]}$, and (d) $^*\rho_{A_\perp}^{[1]}$.
	The symbols have the same meaning as in Figs.~\ref{fig:ZV1} 
	and~\ref{fig:ZA1}.}
	\label{fig:ZAz}
\end{figure*}

Figure~\ref{fig:ZVz} shows the full mass dependence 
of~$Z_{V_\parallel}^{[1]}$, $^*Z_{V_\parallel}^{[1]}$, 
$\rho_{V_\parallel}^{[1]}$, and~$^*\rho_{V_\parallel}^{[1]}$,
with $z=m_{2c}/m_{2b}$ held fixed at~0.256.
Figure~\ref{fig:ZAz} shows the same 
for~$Z_{A_\perp}^{[1]}$, $^*Z_{A_\perp}^{[1]}$, 
$\rho_{A_\perp}^{[1]}$, and~$^*\rho_{A_\perp}^{[1]}$.
The behavior is qualitatively similar to that with $z=1$.
After a knee, the expected infinite mass limit is again reached for 
$m_1^{[0]}a>5$.
The asymptotic behavior is shown with the dashed-dotted lines:
\begin{eqnarray}
	\rho_{V_\parallel}^{[1]} =
		Z_{V_\parallel}^{[1]} & = & +0.901C_F/16\pi^2, \\
	\rho_{A_\perp}^{[1]} = 
		Z_{A_\perp}^{[1]}     & = & -1.099C_F/16\pi^2,
\end{eqnarray}
and
\begin{eqnarray}
	{}^*\rho_{V_\parallel}^{[1]} =
		{}^*Z_{V_\parallel}^{[1]} & = &
			C_F[0.124 + 1.802\ln(m_2a)]/16\pi^2, \\
	{}^*\rho_{A_\perp}^{[1]} =
		{}^*Z_{A_\perp}^{[1]} & = &
			C_F[1.248 - 2.198\ln(m_2a)]/16\pi^2,
\end{eqnarray}
using $f(0.256)=0.3003$.
For both currents, the crossover in~$^*\rho_J^{[1]}$ is smooth enough 
that no knee is manifest.

At the largest masses our numerical integration
(with VEGAS~\cite{Lepage:1977sw} or BASES~\cite{Kawabata:1995th})
deteriorates, because ultraviolet divergences of the continuum vertex
and self-energy diagrams do not cancel point by point (as they do
when $z=1$).
If accurate values of the one-loop terms were essential, this numerical 
difficulty could be avoided. 
But it arises for values of $m_{1b}a$ where the quarks are essentially 
static, and one may just as well make them completely static.

\begin{figure}
	\centering\small
	(a) \includegraphics[width=0.44\textwidth]{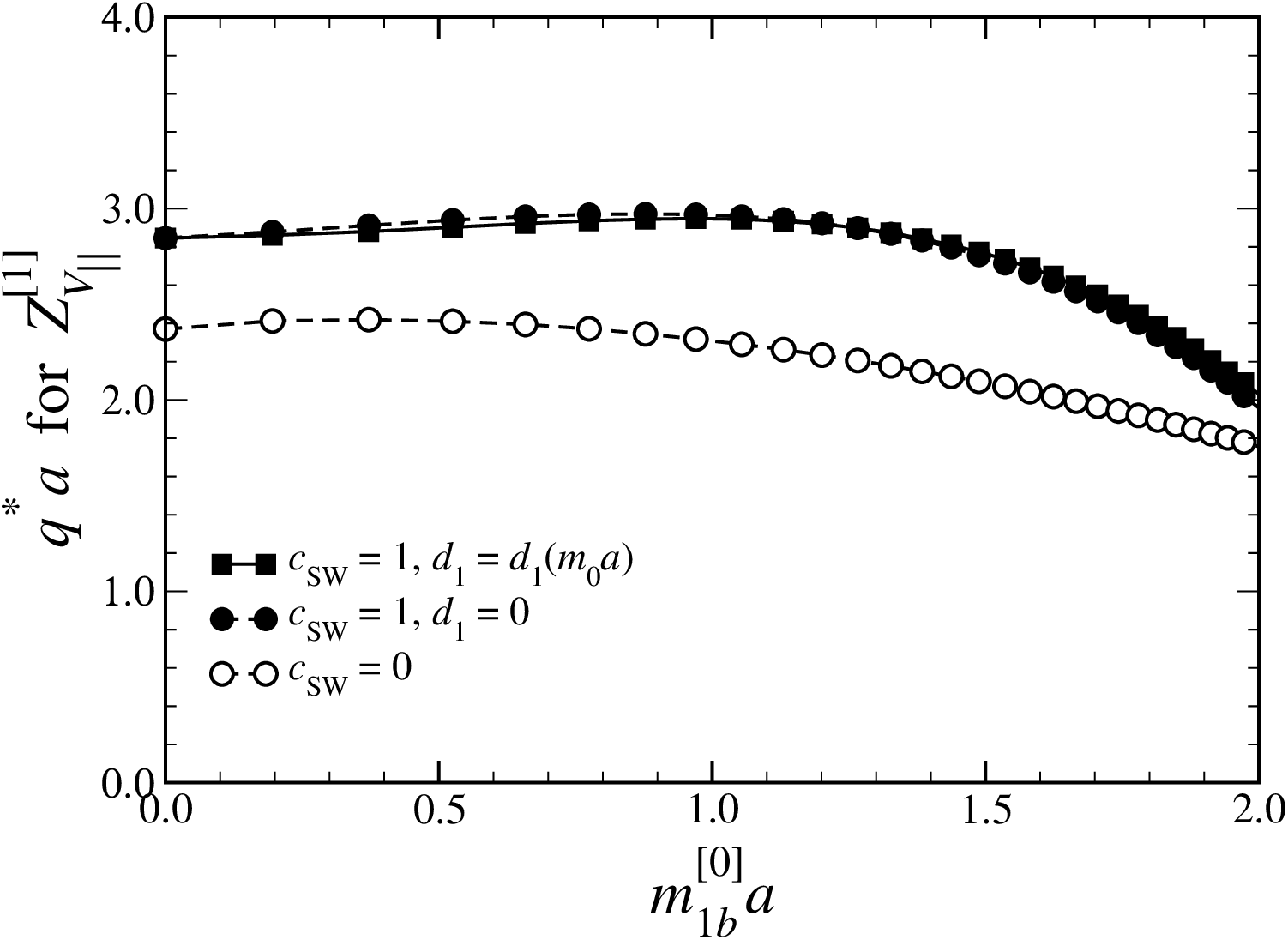} \hfill
	(b) \includegraphics[width=0.44\textwidth]{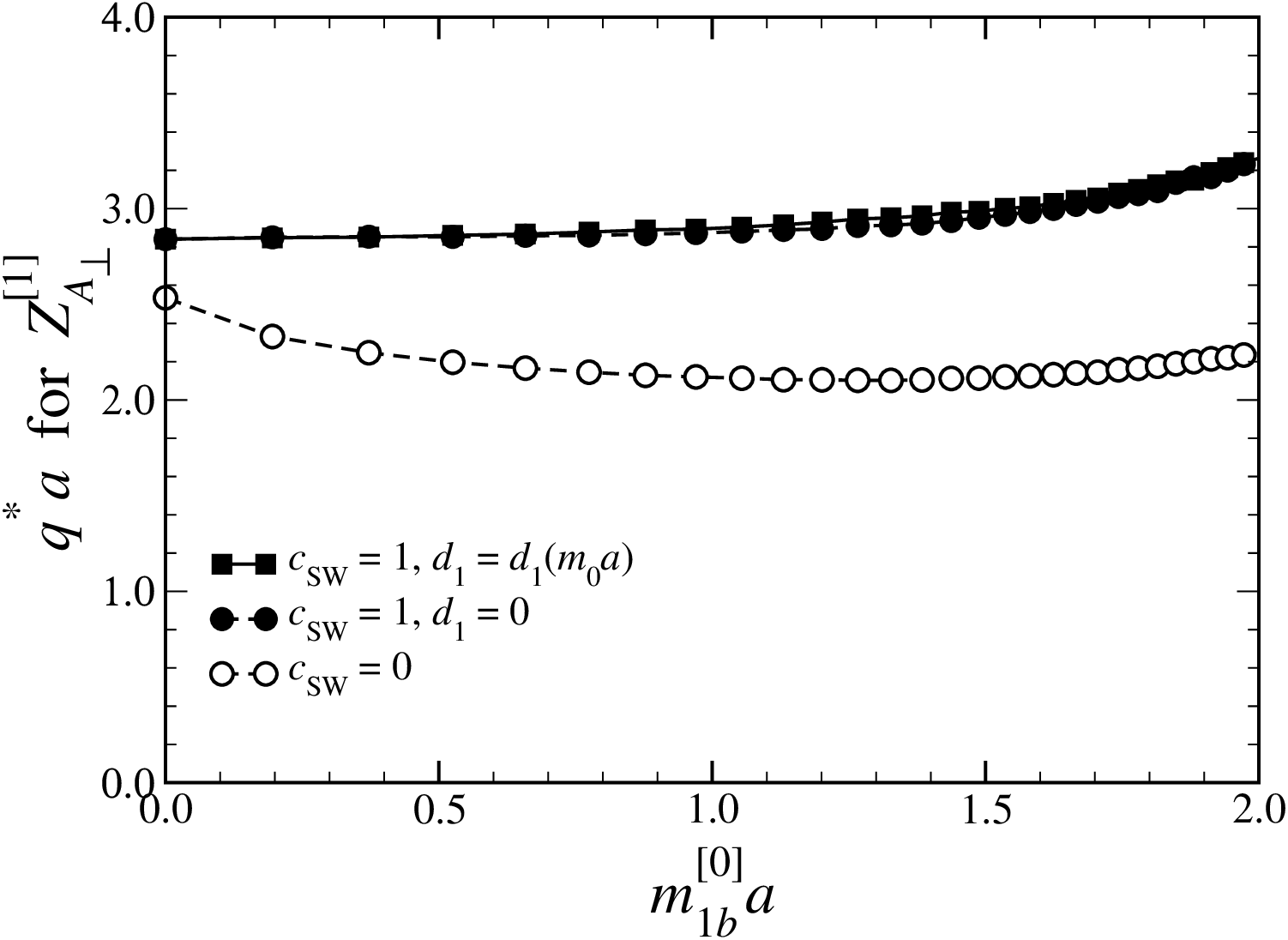}
	\caption[fig:qstarz]{Full mass dependence of the BLM scale~$q^*a$
	for the matching factors
	(a)~$Z_{V_\parallel}$ and (b)~$Z_{A_\perp}$,
	with $m_{2c}=0.256m_{2b}$.
	The symbols have the same meaning as in Figs.~\ref{fig:ZVz}
	and~\ref{fig:ZAz}.}
	\label{fig:qstarz}
\end{figure}

\begin{figure}
	\centering\small
	(a) \includegraphics[width=0.44\textwidth]{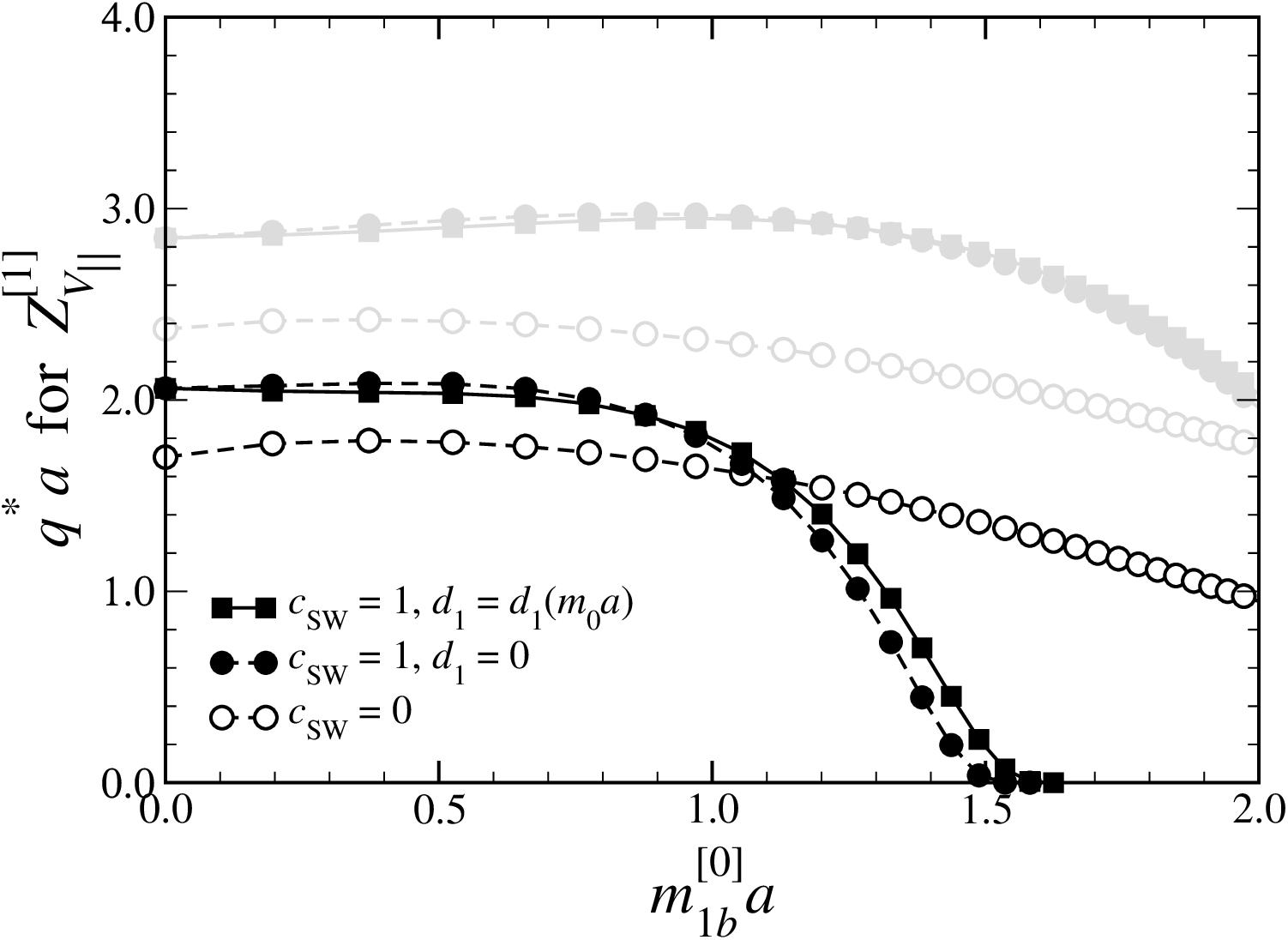} \hfill
	(b) \includegraphics[width=0.44\textwidth]{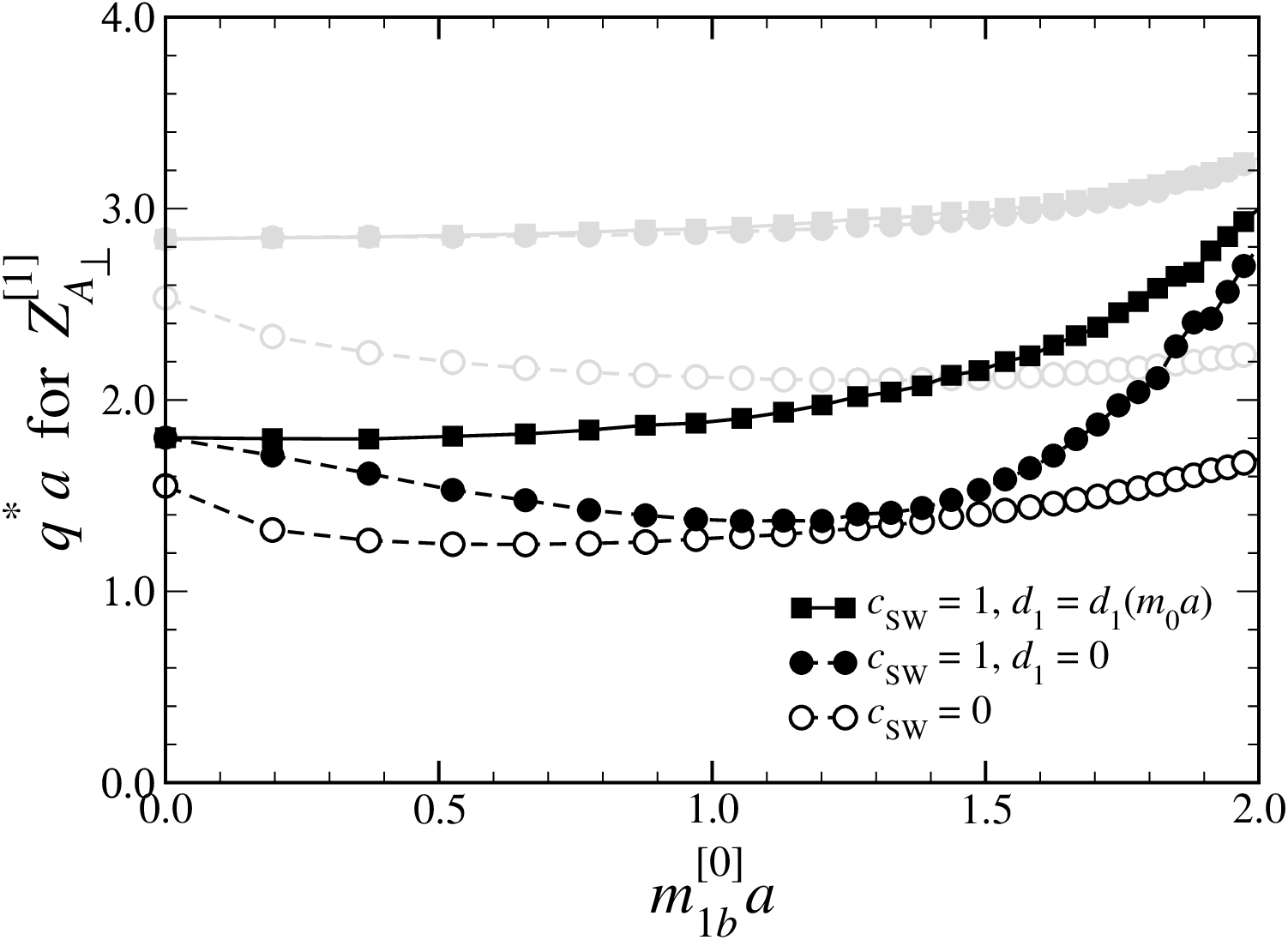}
	\caption[fig:tadqstarz]{Full mass dependence of the BLM scale~$q^*a$
	for the matching factors
	(a)~$Z_{V_\parallel}$ and (b)~$Z_{A_\perp}$,
	with $m_{2c}=0.256m_{2b}$, after tadpole improvement.
	Curves from Fig.~\ref{fig:qstarz} are shown in grey.}
	\label{fig:tadqstarz}
\end{figure}

In Fig.~\ref{fig:qstarz} we combine the information from
Figs.~\ref{fig:ZVz} and~\ref{fig:ZAz} and show the BLM scale~$q^*a$ for
the vector and axial vector currents.
In the region of greatest interest, $m_{1b}\le1$, we find
$q^*a\approx2.8$ for the clover action.
In Fig.~\ref{fig:tadqstarz} we apply tadpole improvement.
In the region of greatest interest, $m_{1b}\le1$, we find a smaller
$q^*a\approx1.8$--2.0 for the clover action.
The singular behavior for $Z_{V_\parallel}$ near $m_{1b}\sim1.5$
(and for $Z_{A_\perp}$ near $m_{1b}\sim2.5$) arises, again, because the
denominator vanishes, so one should switch to the modified prescription
of Ref.~\cite{Hornbostel:2001ey}.

\begin{figure}
	\centering
	\includegraphics[height=2.5in]{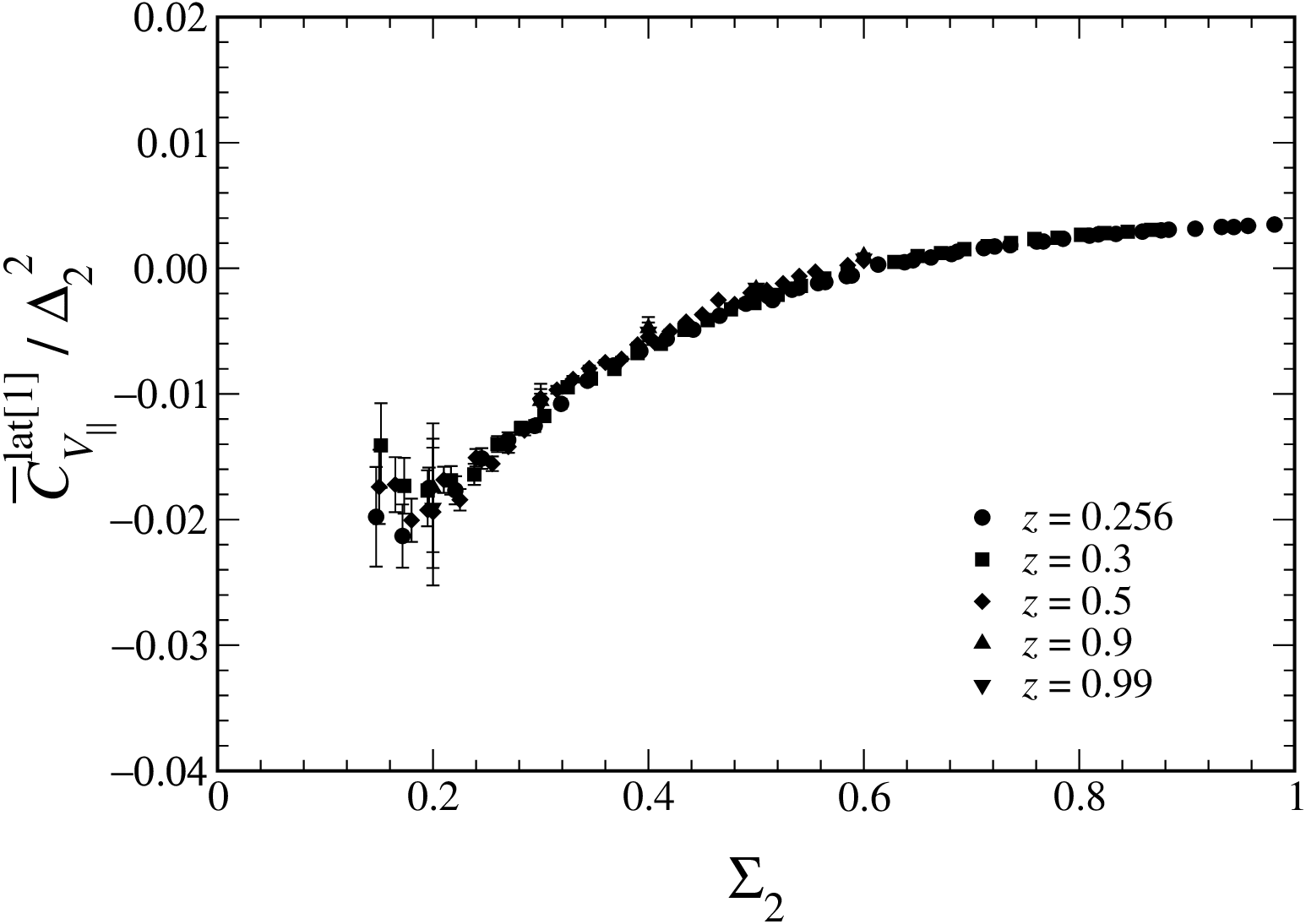}
	\caption[fig:luke]{Test of Luke's theorem, namely
	Eq.~(\ref{eq:luke}) at one loop, for several mass ratios
	$z=0.256$, $0.3$, $0.5$, $0.9$, $0.99$.}
	\label{fig:luke}
\end{figure}

As we have mentioned above, when $m_h\gg a^{-1}$ for both flavors,
heavy quark symmetries emerge~\cite{Kronfeld:2000ck}.
One consequence is that the one-loop lattice radiative corrections
vanish, as Figs.~\ref{fig:ZV1}, \ref{fig:ZA1}, \ref{fig:ZVz}, and
\ref{fig:ZAz} verify numerically.
There is an interesting further consequence for $\rho_{V_\parallel}$,
in which the unphysical $e^{m_1a}$ normalization drops out.
Its approach to the infinite-mass limit must satisfy%
\footnote{In Sec.~\ref{sec:hqet}, $\bar{C}_{V_\parallel}^{\text{lat}}$
is a generic notation for a matching coefficient in HQET.
Here it denotes the same coefficient for the specific current
$\sqrt{Z_{V_\parallel^{cc}}Z_{V_\parallel^{bb}}}%
\bar{\Psi}_c\gamma^4\Psi_b$.}
\begin{equation}
	\frac{\bar{C}_{V_\parallel}}{\rho_{V_\parallel}} \equiv
	\bar{C}_{V_\parallel}^{\text{lat}} = \Delta_2^2
		\left[ \bar{c}^{(2)} + \bar{c}^{(3)} \Sigma_2 \right] ,
	\label{eq:luke}
\end{equation}
where the $\bar{c}^{(n)}$ are independent of the masses, and
\begin{eqnarray}
	\Delta_2 = \frac{1}{2m_{2c}a} - \frac{1}{2m_{2b}a} ,
	\label{eq:Delta} \\
	\Sigma_2 = \frac{1}{2m_{2c}a} + \frac{1}{2m_{2b}a} .
	\label{eq:Sigma}
\end{eqnarray}
Equation~(\ref{eq:luke}) is a version of Luke's
theorem\cite{Luke:1990eg}.
It follows because the mass dependence must be symmetric under the
interchange $m_c\leftrightarrow m_b$, and because $\rho_{V_\parallel}$
must vanish when $m_c=m_b$.
Following Ref.~\cite{Boyle:2000fi} we test whether our one-loop
calculation satisfies the theorem for a variety of different mass
ratios.
Figure~\ref{fig:luke} plots
$\bar{C}_{V_\parallel}^{\text{lat}[1]}/\Delta_2^2$ vs.\ $\Sigma_2$
for $z=0.256,0.3,0.5,0.9,0.99$.
All mass combinations lie essentially on a single curve, verifying
Eq.~(\ref{eq:luke}).
Small deviations are possible (and noticeable in Fig.~\ref{fig:luke}),
arising from higher orders in the small $\bbox{p}$ exapnsion of the
quark propagator, where coefficients other than $1/(2m_2a)$ appear.

\subsection{Unequal masses: fixed daughter mass}
\label{subsec:unequal:fixed-mc}

In this subsection we study the matching factors, varying only the initial 
mass while holding the final mass fixed.
For illustration we set $m_{0c}a=1$.
As discussed above, one expects the large-mass limit to exhibit some
qualitative behavior seen in the heavy-light matching factors in
our heavy-light paper~\cite{Harada:2001hl}.
Thus, looking at this slice through the function of two variables can
illustrate how the heavy-heavy matching factors are an extension of the
heavy-light matching factors.

\begin{figure*}
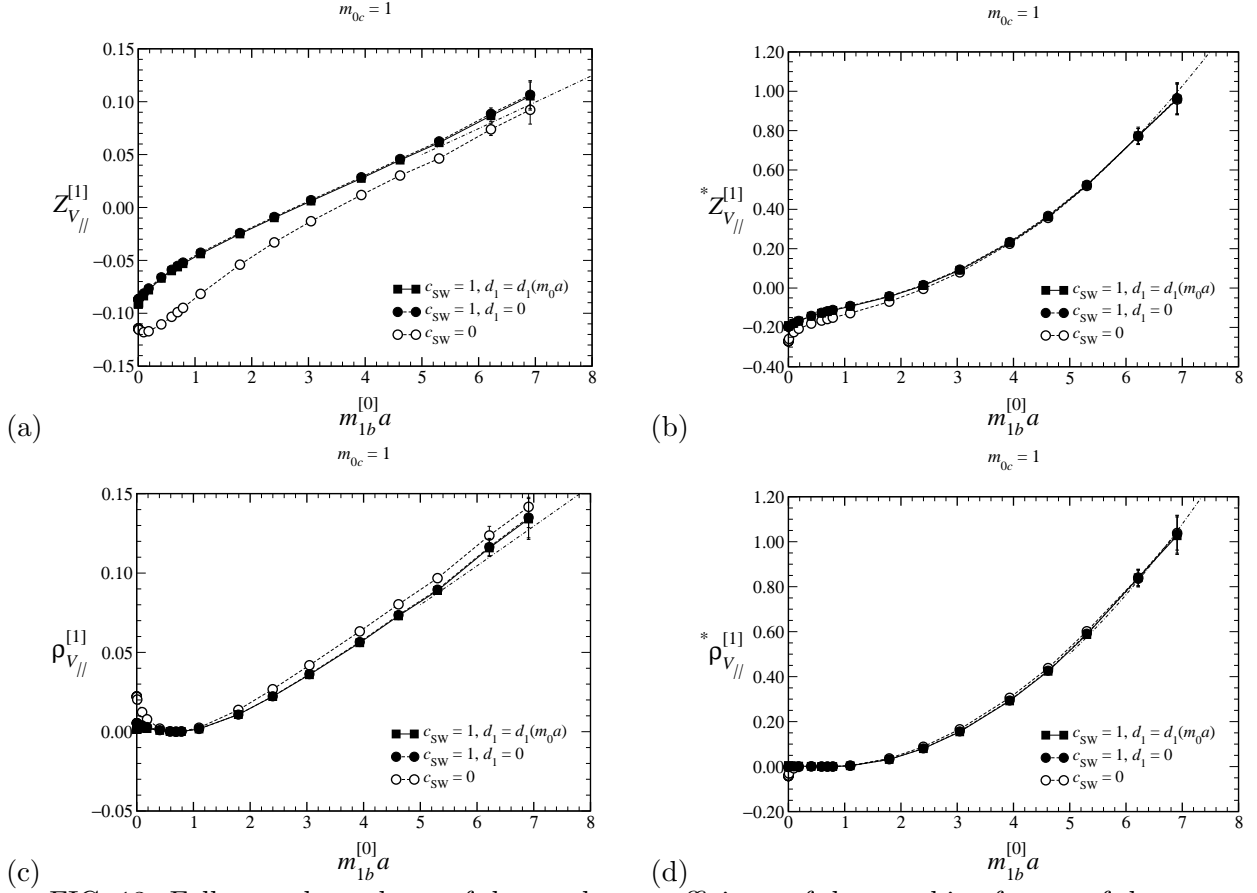

	\centering\small
	(a) \includegraphics[width=0.44\textwidth]{ZV4m1_mc.eps} \hfill
	(b) \includegraphics[width=0.44\textwidth]{starZV4m1_mc.eps} \\
	(c) \includegraphics[width=0.44\textwidth]{rhoV4m1_mc.eps} \hfill
	(d) \includegraphics[width=0.44\textwidth]{starrhoV4m1_mc.eps}
	\caption[fig:ZVmc]{Full mass dependence of the one-loop coefficients
	of the matching factors of the vector current with $m_{0c}=1$:
	(a)    $Z_{V_\parallel}^{[1]}$,     (b)    $^*Z_{V_\parallel}^{[1]}$,
	(c) $\rho_{V_\parallel}^{[1]}$, and (d) $^*\rho_{V_\parallel}^{[1]}$.
	The symbols have the same meaning as in Figs.~\ref{fig:ZV1} 
	and~\ref{fig:ZA1}.}
	\label{fig:ZVmc}
\end{figure*}

\begin{figure*}
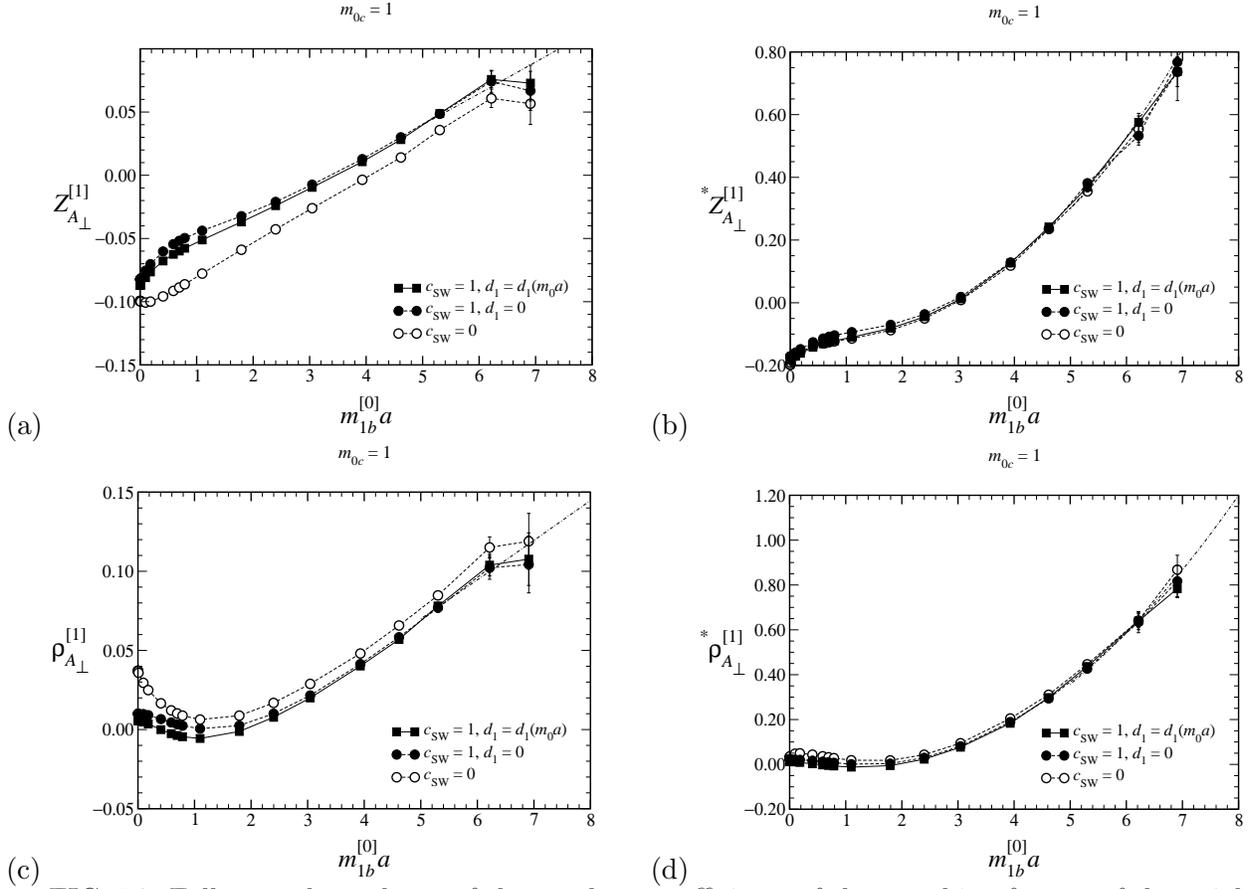

	\centering\small
	(a) \includegraphics[width=0.44\textwidth]{ZAjm1_mc.eps} \hfill
	(b) \includegraphics[width=0.44\textwidth]{starZAjm1_mc.eps} \\
	(c) \includegraphics[width=0.44\textwidth]{rhoAjm1_mc.eps} \hfill
	(d) \includegraphics[width=0.44\textwidth]{starrhoAjm1_mc.eps}
	\caption[fig:ZAmc]{Full mass dependence of the one-loop coefficients
	of the matching factors of the axial vector current with $m_{0c}=1$:
	(a)    $Z_{A_\perp}^{[1]}$,     (b)    $^*Z_{A_\perp}^{[1]}$,
	(c) $\rho_{A_\perp}^{[1]}$, and (d) $^*\rho_{A_\perp}^{[1]}$.
	The symbols have the same meaning as in Figs.~\ref{fig:ZV1} 
	and~\ref{fig:ZA1}.}
	\label{fig:ZAmc}
\end{figure*}

Figure~\ref{fig:ZVmc} shows full $m_ba$ dependence 
of~$Z_{V_\parallel}^{[1]}$, $^*Z_{V_\parallel}^{[1]}$, 
$\rho_{V_\parallel}^{[1]}$, and~$^*\rho_{V_\parallel}^{[1]}$,
with fixed $m_{0c}a=1$ ($m_{2c}a=0.857$).
Figure~\ref{fig:ZAmc} shows the same 
for~$Z_{A_\perp}^{[1]}$, $^*Z_{A_\perp}^{[1]}$, 
$\rho_{A_\perp}^{[1]}$, and~$^*\rho_{A_\perp}^{[1]}$.
As with the heavy-light matching factors~\cite{Harada:2001hl}, a knee is
not really visible, because soon after reaching $m_{1b}^{[0]}a\gtrsim2$
$(m_{2b}a\gtrsim6$), logarithmic behavior starts to dominate.
For small $m_ba$, one finds that
$\rho_{V_\parallel}(m_ba=0,m_{0c}a=1)\neq0$,
but with the SW action it is very small: 0.0015017(9) with the rotation
on the $c$~leg and~0.0054380(10) without.

\begin{figure}
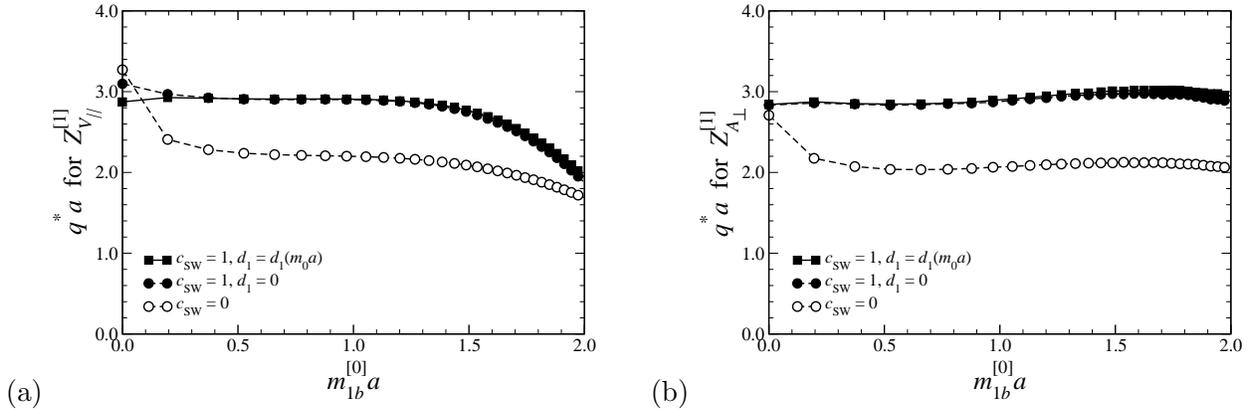

	\centering\small
	(a) \includegraphics[width=0.44\textwidth]{qstarZV4m1_mc.eps} \hfill
	(b) \includegraphics[width=0.44\textwidth]{qstarZAjm1_mc.eps}
	\caption[fig:qstarmc]{Full mass dependence of the BLM scale~$q^*a$
	for the matching factors
	(a)~$Z_{V_\parallel}$ and (b)~$Z_{A_\perp}$,
	with $m_{0c}a=1$.
	The symbols have the same meaning as in Figs.~\ref{fig:ZVmc}
	and~\ref{fig:ZAmc}.}
	\label{fig:qstarmc}
\end{figure}

\begin{figure}
	\centering\small
	(a) \includegraphics[width=0.44\textwidth]{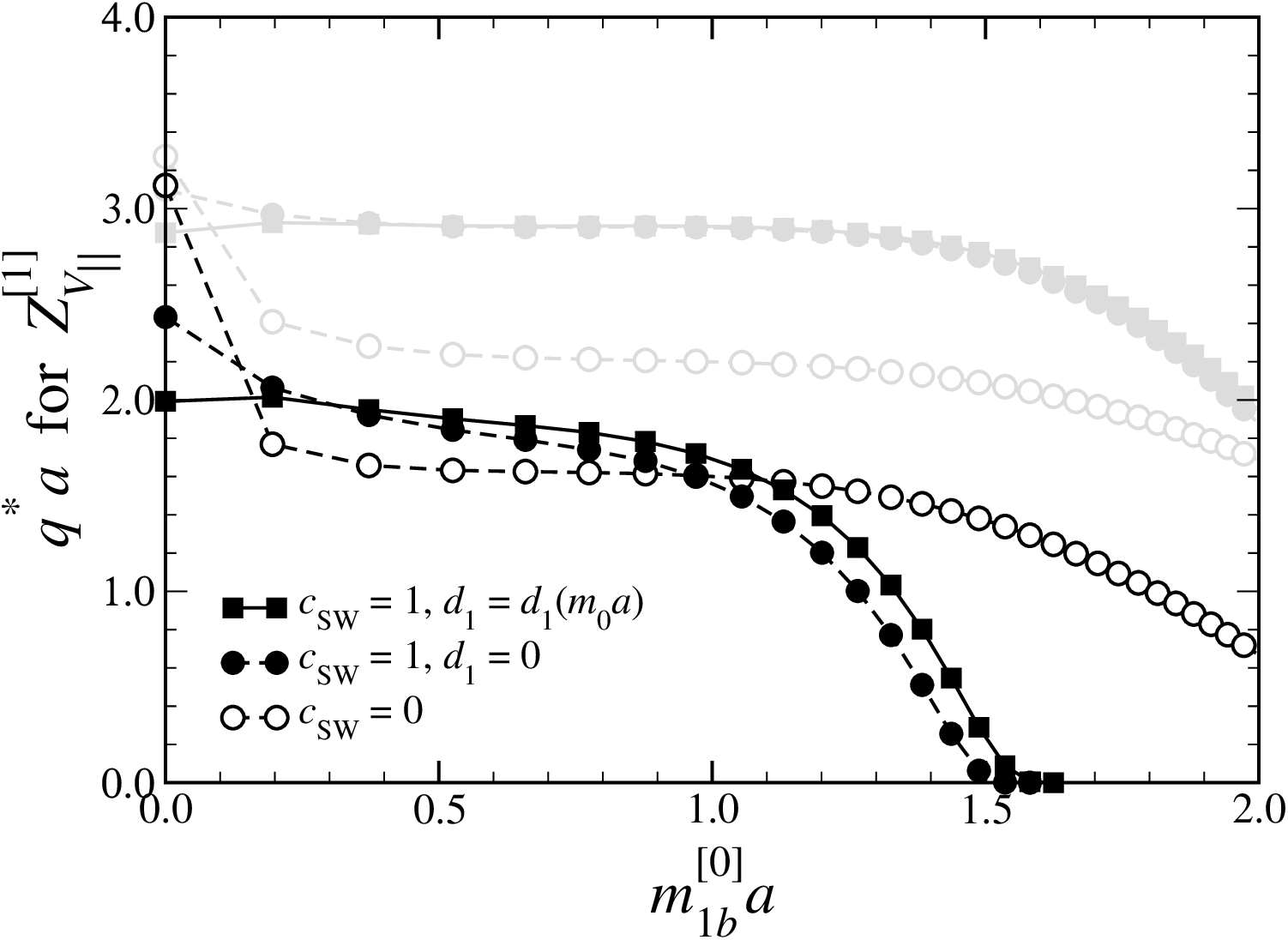} \hfill
	(b) \includegraphics[width=0.44\textwidth]{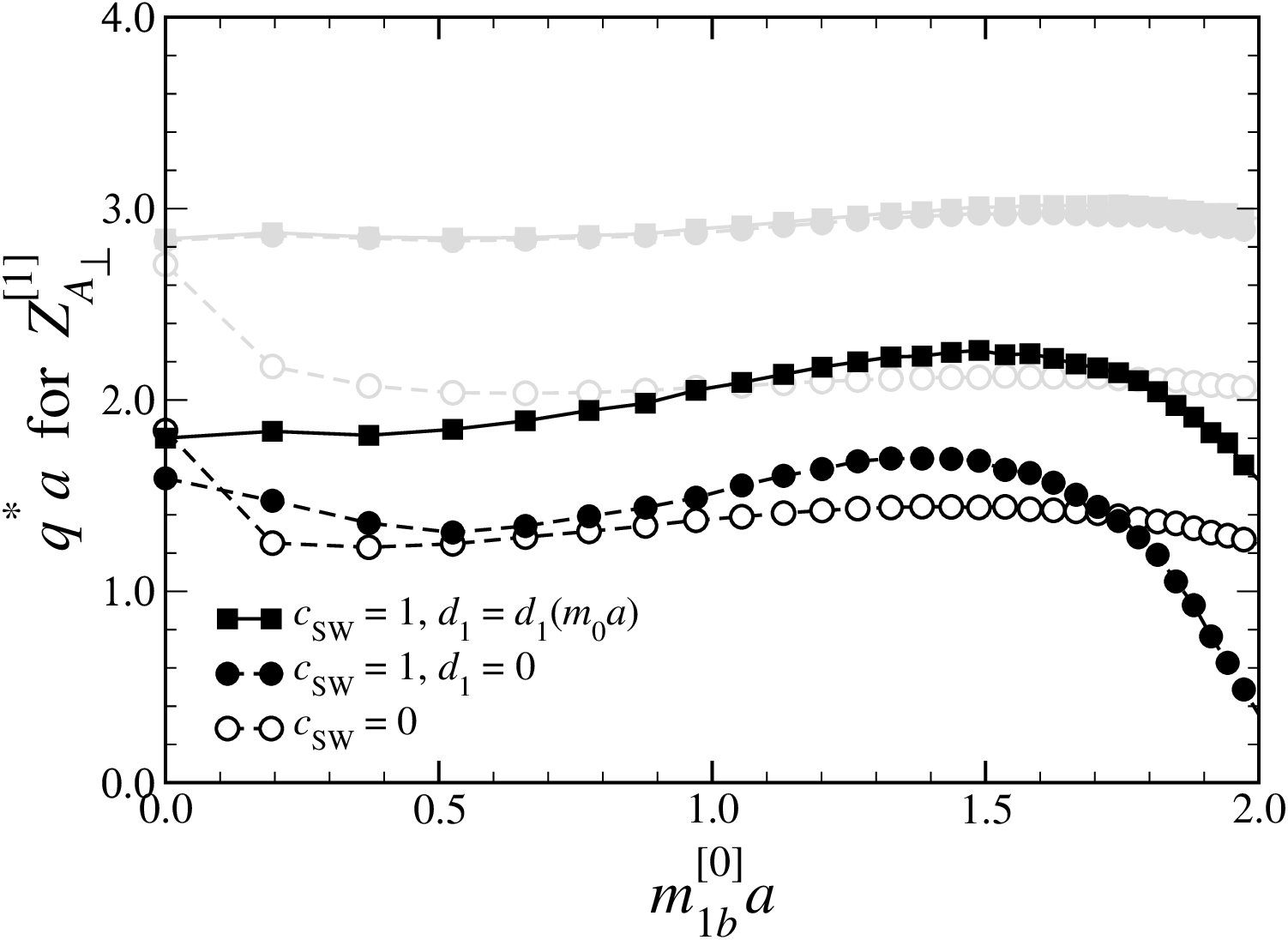}
	\caption[fig:tadqstarmc]{Full mass dependence of the BLM scale~$q^*a$
	for the matching factors
	(a)~$Z_{V_\parallel}$ and (b)~$Z_{A_\perp}$,
	with $m_{0c}a=1$, after tadpole improvement.
	Curves from Fig.~\ref{fig:qstarmc} are shown in grey.}
	\label{fig:tadqstarmc}
\end{figure}

In Fig.~\ref{fig:qstarmc} we combine the information from
Figs.~\ref{fig:ZVmc} and~\ref{fig:ZAmc} and show the BLM scale~$q^*a$ for
the vector and axial vector currents.
In the region of greatest interest, $m_{1b}\le1$, we find
$q^*a\approx2.8$--3.0 for the clover action
(and 2.0--2.2 for the Wilson action).
In Fig.~\ref{fig:tadqstarmc} we apply tadpole improvement.
In the region of greatest interest, $m_{1b}\le1$, we find a smaller
$q^*a\approx1.8$--1.0 for the clover action
(and 1.2--1.4 for the Wilson action).
The singular behavior for $Z_{V_\parallel}$ ($Z_{A_\perp}$) near
$m_{1b}\sim1.5$ (2.2) arises, once again, because the denominator in
Eq.~(\ref{eq:BLMq*}) vanishes, so one should switch to the modified
prescription of Ref.~\cite{Hornbostel:2001ey}.

\subsection{Summary}

Let us summarize the main points of this section.
The heavy-heavy matching factors are a function of two variables,
and we have plotted several one-dimensional slices to illustrate their
size and behavior.
The dependence on the masses smoothly connects the limit where
${m_ha\ll1}$ with the infinite-mass limit(s).
This is typical of radiative corrections in massive lattice gauge
theory~\cite{El-Khadra:1997mp}, and the phenomenon should, by now, be
familiar from several explicit
examples~\cite{Kronfeld:1993dd,Kuramashi:1998tt,Ishikawa:1997xh,%
Mertens:1998wx,Kronfeld:1999tk,Harada:2001ei,Harada:2001hl}.

There are several checks on our calculations.
Numerical results have been tested with two or more completely
independent programs.
The curves behave as expected for large mass.
When one mass is very small in lattice units, our results smoothly connect
to the heavy-light matching factors.
When both are small we reproduce results in the literature for 
massless fermions.
We are confident, therefore, that our programs compute the one-loop
coefficients correctly for intermediate values of the masses.

It does not seem especially useful to print tables of our results.
In practice~\cite{Hashimoto:2000yp,Hashimoto:2001ds}, one will want
to analyze many different mass combinations.
To facilitate this analysis, we are making one suite of programs freely
available~\cite{p:epaps}.
That will allow the user to obtain the coefficients to suit his or her
needs.
As shown in Appendix~\ref{app:dirac} we have found a way to incorporate
the rotation into the Dirac algebra in a simple way.
Since the public program is designed around this technique, it may be
relatively easy to extend it to other lattice actions, such as highly
improved actions.

\section{Conclusions}
\label{sec:conclusions}

In this paper we have set up a matching procedure, based on~HQET,
for heavy-heavy currents.
It is valid for all $m_ba$ and $m_ca$, where $m_b$ and $m_c$ are the
heavy quarks' masses and $a$ is the lattice spacing.
The procedure holds to all orders in perturbation theory and, 
therefore, potentially on a non-perturbative level also.
It could be applied to lattice NRQCD, although here it is applied to
Wilson fermions.
With Wilson fermions it is possible to consider the limit of small
$m_ha$.
Then heavy-heavy HQET matching agrees with heavy-light HQET matching
when the lighter quark satisfies $m_ca\ll1$, and with Symanzik matching
when both $m_ha\ll1$.
In this way, HQET matching is a natural and attractive extension into
the regime $ma\not\ll1$, which is needed for heavy-quark phenomenology
on currently available computers.

The connection to heavy-light matching has an important implication
for the charmed quark, whose mass is not necessarily large enough for
the leading~$1/m_c$ corrections to be accurate enough on their own.
At the same time, however, the charmed mass is small enough that
$m_ca\lesssim\case{1}{3}$, on currently available lattices.
In this regime the mismatches of short-distance coefficients of
higher-dimension operators in HQET are influenced by the standard
continuum limit (for the charmed quark), so they are easily kept down
to the level of $\alpha_sm_ca$ and~$(m_ca)^2$.
Thus, more and more of the heavy-quark expansion is recovered, for
masses where the higher-dimension terms may be significant.

Our one-loop results for the SW action are of immediate value for
lattice calculations of form factors for the semi-leptonic decays
$B\to D^{(*)} l\nu$.
Indeed, our earlier one-loop results~\cite{Kronfeld:1999tk} (which
omitted the ``rotation'' terms in the current) were used for $B\to D$
in Ref.~\cite{Hashimoto:2000yp}, and our new results with rotation
were used for $B\to D^*$ in Ref.~\cite{Hashimoto:2001ds}.
In particular, we have obtained the BLM scale~$q^*$ for the
renormalization factors, which reduces the uncertainty of one-loop
calculations.
Similarly, computing part of the normalization factor,
namely $\sqrt{Z_{V_\parallel^{cc}}Z_{V_\parallel^{bb}}}$,
non-perturbatively reduces the normalization uncertainty even
further~\cite{Hashimoto:2000yp,Hashimoto:2001ds,El-Khadra:2001rv}.

An outstanding problem at this time is the one-loop calculation of the
coefficients $\bar{B}_{Ji}^{\text{lat}}$ of the dimension-four terms
in the HQET description.
A~calculation of these coefficients, and the subsequent adjustment
of the parameters~$b_{Ji}$ in the lattice currents, would reduce
the uncertainties in (future) calculations of $B\to D^{(*)}$ matrix
elements.
Because heavy-quark symmetry forbids several power corrections (see
Refs.~\cite{Kronfeld:2000ck,Hashimoto:2001ds} for details), one must,
however, also start to consider some of the dimension-five corrections
to the currents.

\acknowledgments
Dedicated to the memory of William E. Caswell, 
co-inventor of NRQED and NRQCD, 
who died September 11, 2001.

A.S.K. would like to thank Akira Ukawa and the Center for Computational
Physics for hospitality while part of this work was being carried out,
and the Aspen Center for Physics for a stimulating atmosphere while
part of the paper was being written.
S.H., A.S.K., and T.O. would also like to thank the Insititute for 
Nuclear Theory at the University of Washington for hospitality while 
this paper was being finished.
S.H. and T.O. are supported by Grants-in-Aid of the Japanese Ministry of
Education (Nos.\ 11740162 and 12640279, respectively).
Fermilab is operated by Universities Research Association Inc.,
under contract with the U.S.\ Department of Energy.

\appendix

\section{Feynman Rules}
\label{app:feynman}

The needed propagators and vertices for quark-gluon interactions are
given already in Ref.~\cite{Mertens:1998wx}.
Here we give the additional Feynman rules induced by the rotation term
of the heavy quark.
The additional rules are easy to derive by expressing the translation
operator
\begin{equation}
	T_{\pm\mu} = t_{\pm\mu/2} e^{\pm g_0aA_\mu} t_{\pm\mu/2}
\end{equation}
and following the methods in Ref.~\cite{Kronfeld:1985zv}.

There are six rules to give, with 0, 1, and 2 gluons emerging from the
vertex; it is convenient to keep rules for gluons emitted from the 
incoming and outgoing rotations separate.
Let the Dirac matrix of the current be~$\Gamma$, and let
\begin{eqnarray}
	R(p)   & = & 1 + i d_1\sum_r\gamma_r\sin(p_r) ,
	\label{eq:Rrot}  \\
	R'(p') & = & 1 + id'_1\sum_r\gamma_r\sin(p'_r).
	\label{eq:Rrot'}
\end{eqnarray}
Then,
\begin{eqnarray}
	\mbox{0-gluon}   & = & R'(p') \Gamma R(p) ,
	\label{eq:0} \\
	\mbox{1-gluon}   & = & g_0 t^a d_1 \,
		R'(p') \Gamma \gamma_i \cos(p+\case{1}{2} k)_i ,
	\label{eq:1} \\
	\mbox{1-gluon}'  & = & g_0 t^a d'_1 \,
		\cos(p'-\case{1}{2} k)_i \gamma_i \Gamma R(p)  ,
	\label{eq:1'} \\
	\mbox{2-gluon}   & = & g_0^2 \case{1}{2}\{t^a, t^b\} \delta_{ij}
		d_1\, R'(p') \Gamma \gamma_i 
		i\sin(p+\case{1}{2} k+\case{1}{2}\ell)_i
	\label{eq:2} \\
	\mbox{2-gluon}'' & = & g_0^2 \case{1}{2}\{t^a, t^b\} \delta_{ij}
		d'_1\, i\sin(p'-\case{1}{2} k-\case{1}{2}\ell)_i 
		\gamma_i \Gamma R(p)
	\label{eq:2''} \\
	\mbox{2-gluon}'  & = & g_0^2 t^b t^a \; d'_1 d_1 
		\cos(p'-\case{1}{2}\ell)_j \gamma_j \Gamma
		\gamma_i \cos(p+\case{1}{2} k)_i
	\label{eq:2'}
\end{eqnarray}
where momentum $p$ ($p'$) is the quark momentum flowing into (out of)
the vertex, and $k$ and $\ell$ are gluon momentum flowing into vertex.
In the two-gluon rules, there is \emph{no} summation over~$i$.
As in Ref.~\cite{Mertens:1998wx}, the matrices $t^a$ are
anti-Hermitian, \emph{i.e.}, $U_\mu=\exp\left(g_0t^aA^a_\mu\right)$,
$\sum_{aj} t^a_{ij} t^a_{jk} = - C_F \delta_{ik}$,
and $\tr t^at^b=-\case{1}{2}\delta^{ab}$.

\section{Dirac Algebra}
\label{app:dirac}

To compute the vertex function, there are eight diagrams to consider,
depicted in Fig.~\ref{fig:Feyn}:
\begin{figure}[!b]
	\centering\small
	(a) \includegraphics[width=0.20\textwidth]{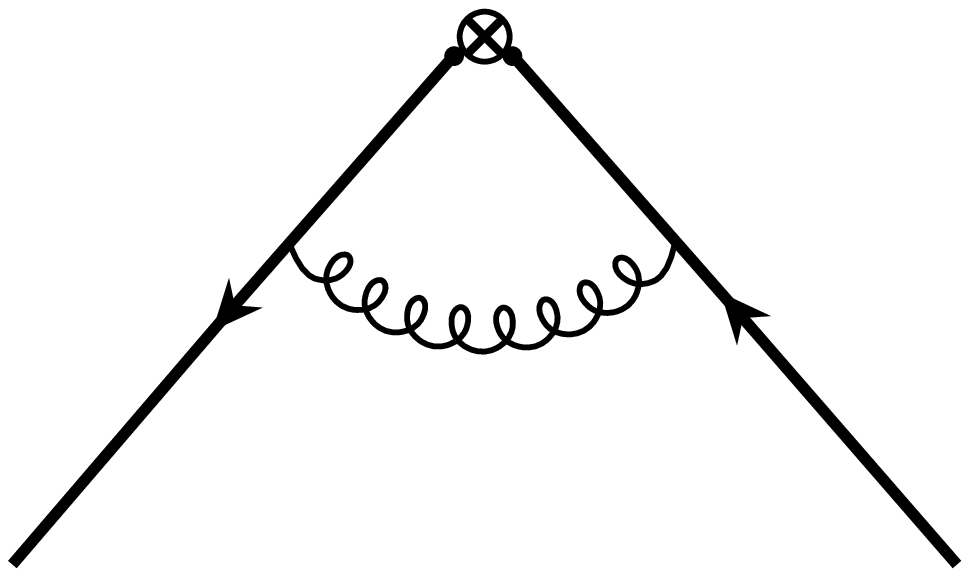} \hfill
	(b) \includegraphics[width=0.20\textwidth]{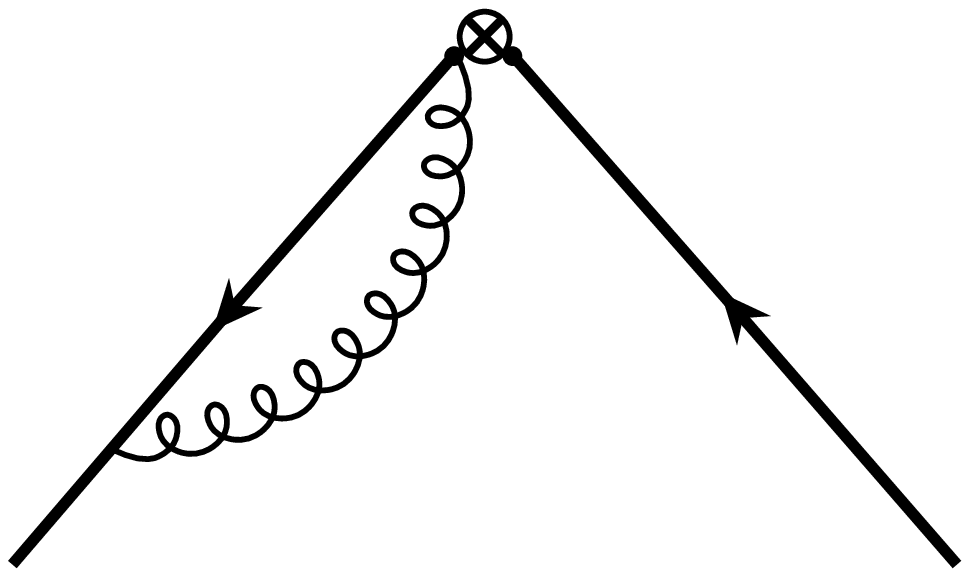} \hfill
	(c) \includegraphics[width=0.20\textwidth]{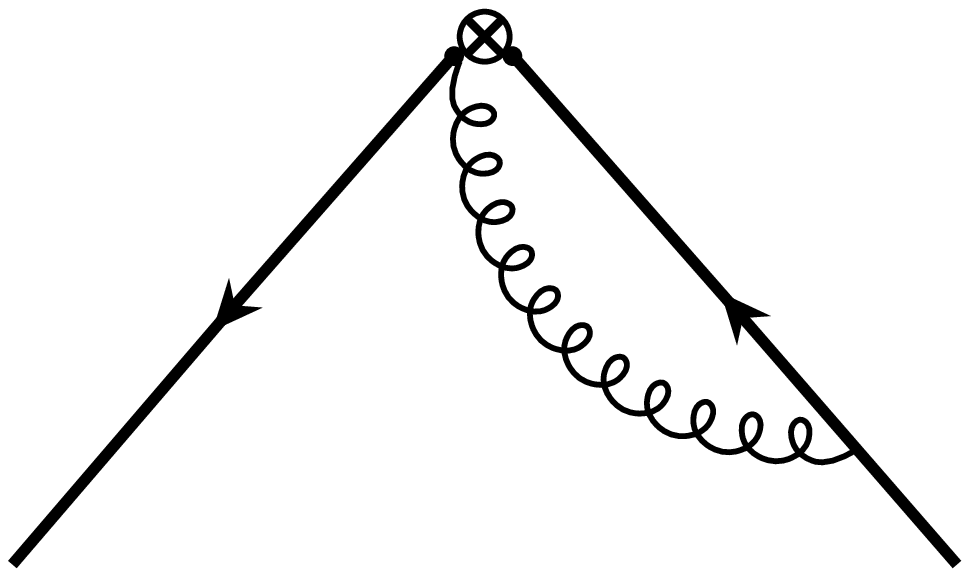} \hfill
	(d) \includegraphics[width=0.20\textwidth]{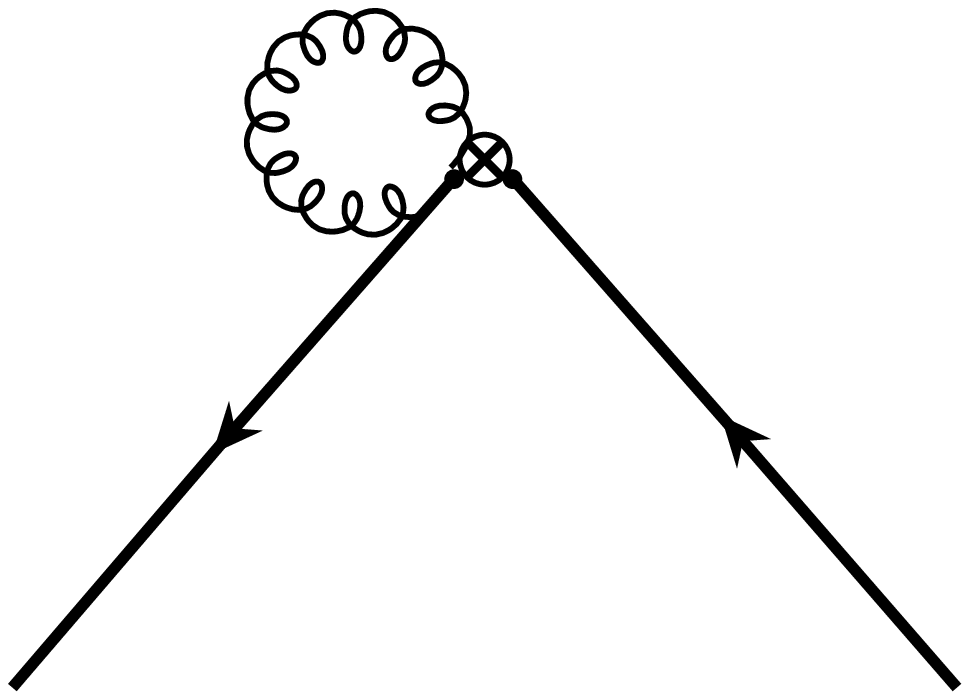} \\
	(e) \includegraphics[width=0.20\textwidth]{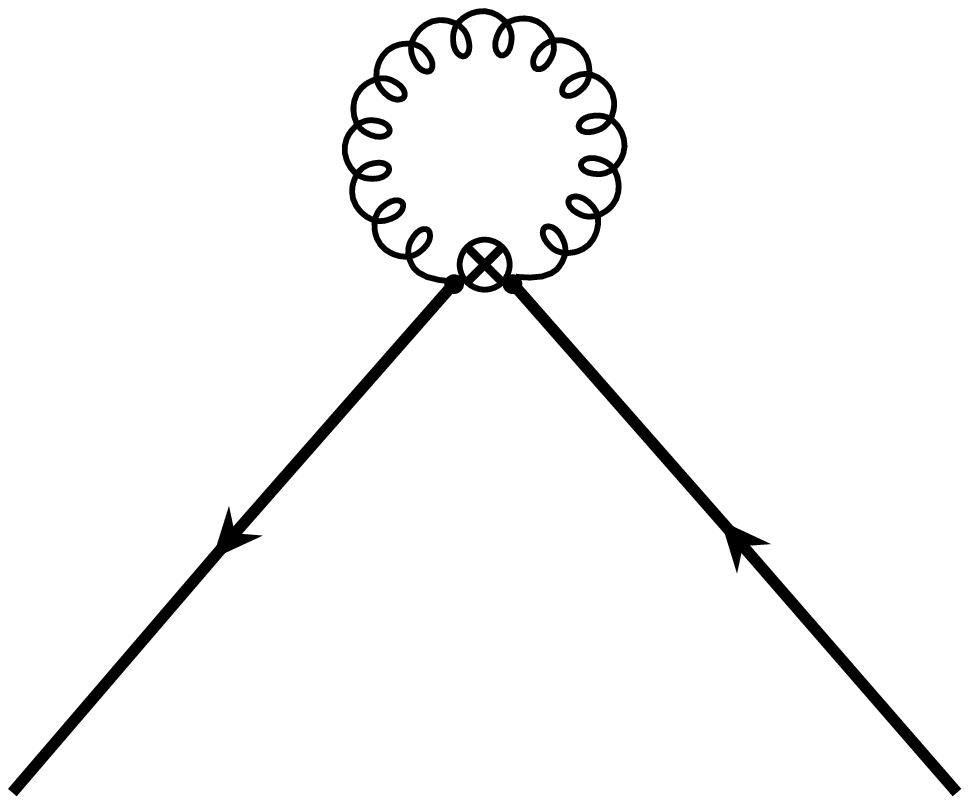} \hfill
	(f) \includegraphics[width=0.20\textwidth]{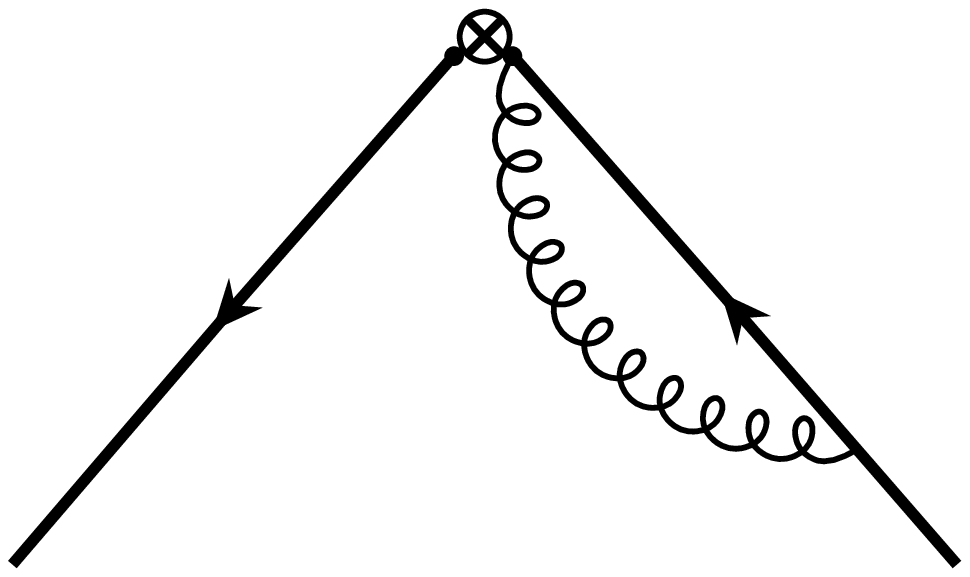} \hfill
	(g) \includegraphics[width=0.20\textwidth]{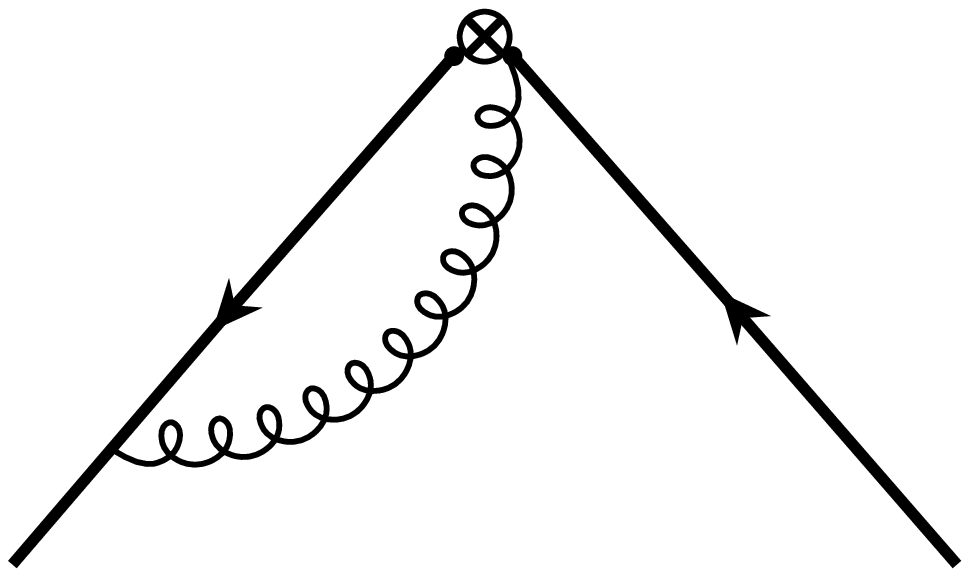} \hfill
	(h) \includegraphics[width=0.20\textwidth]{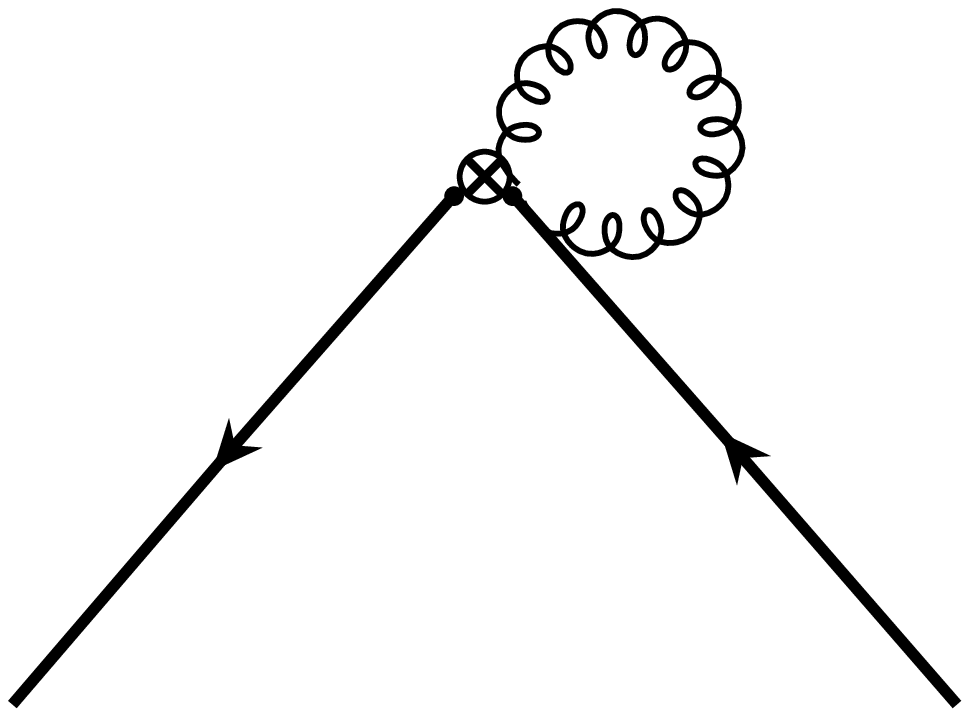}
	\caption[fig:Feyn]{Feynman diagrams for calculating the vertex
	function.
	The $\bullet$ on each side of the $\otimes$ indicates the rotation.}
	\label{fig:Feyn}
\end{figure}
the usual vertex diagram (with the rotation inside),
Fig.~\ref{fig:Feyn}(a);
two diagrams with the gluon connected to the incoming rotation,
Fig.~\ref{fig:Feyn}(b) and~(c);
a tadpole diagram connected to the incoming rotation
[using rule~(\ref{eq:2})], Fig.~\ref{fig:Feyn}(d);
a vertex diagram with a gluon connecting both rotations
[using rule~(\ref{eq:2'})], Fig.~\ref{fig:Feyn}(e);
two diagrams with the gluon connected to the outgoing rotation,
Fig.~\ref{fig:Feyn}(f) and~(g);
and a tadpole diagram connected to the outgoing rotation,
[using rule~(\ref{eq:2''})], Fig.~\ref{fig:Feyn}(h).
The two tadpole diagrams, Figs.~\ref{fig:Feyn}(d) and~(h), vanish for
zero external three-momentum, because $\ell=-k$ and $p_i=0$.

For each non-vanishing diagram, Figs.~\ref{fig:Feyn}(a--c, e--g),
define the integral
\begin{equation}
	I_\Gamma^{\text{(a--c, e--g)}} = - g_0^2 C_F \int \frac{d^4k}{(2\pi)^4}
		\frac{1}{\hat{k}^2}{\cal I}_\Gamma^{\text{(a--c, e--g)}},
\end{equation}
where $k$ is the momentum of the gluon in the loop, and
$\hat{k}_\mu=2\sin(\case{1}{2} k_\mu)$.
Let the incoming massive quark have couplings $m_0$, $r_s$, $\zeta$,
$c_B$, and $c_E$, and external momentum $p$.
Similarly, let the outgoing massless quark have couplings $m'_0=0$,
$r'_s$, $\zeta'$, $c'_B$, and $c'_E$, and external momentum~$p'$.
The internal quark lines carry momentum $p+k$ in and $p'+k$ out.
The integrals~$I$ are obtained directly from the loop diagrams.
Then
\begin{equation}
	Z_J^{[1]} =
		\frac{1}{2} \left(
		{Z_{2b}^{[1]}}_{\rm cont} - {Z_{2b}^{[1]}}_{\rm lat} +
		{Z_{2c}^{[1]}}_{\rm cont} - {Z_{2c}^{[1]}}_{\rm lat}
		\right) +
		\sum_{\rm d} \left(
		{I_\Gamma^{\rm d}}_{\rm cont} - {I_\Gamma^{\rm d}}_{\rm lat}
		\right),
\end{equation}
from Eq.~(\ref{eq:ZJ}).
The relation between the current $J$ and its Dirac matrix~$\Gamma$
is contained in Table~\ref{tab:s}.
\begin{table}
\centering
	\begin{tabular}{ccccc}
	&	  $J$            &      $\Gamma$      & $s_\Gamma$     & \\
		\hline
	&	  $V_\parallel$  &     $\gamma_4$     & $-1$           & \\
	&	  $A_\perp$      & $\gamma_j\gamma_5$ & $+\case{1}{3}$ & \\
	\end{tabular}
	\caption[tab:s]{The factor $s_\Gamma$, defined by
		$\case{1}{3}\sum_r \gamma_r\Gamma\gamma_r=s_\Gamma \Gamma$.}
	\label{tab:s}
\end{table}
The expression relating ${Z_2^{[1]}}_{\rm lat}$ to lattice
self-energy functions is in Ref.~\cite{Mertens:1998wx}.

The most onerous task in evaluating the diagrams is the manipulation of
the Dirac matrices.
A convenient method is to treat each quark line separately, starting
from the initial- or final-state spinor.
Then the spinor, the propagator, and the vertices can be written out in
$2\times 2$ block diagonal form, with Pauli matrices appearing in the
blocks.
Once the Feynman rules are as complicated as in the present calculation,
it is easier to manipulate $2\times 2$ matrices of Pauli matrices than
to manipulate Dirac matrices.
A special advantage of this organization is that the rotation bracket in
Eq.~(\ref{eq:0}) merely ``rotates'' the rest of the leg.
We also obtain ${Z_2^{[1]}}_{\rm lat}$ in this way, with much less
effort than in Ref.~\cite{Mertens:1998wx}.

A further advantage is that the vertex corrections can be expressed
compactly.
The diagram with a gluon connecting the two rotations,
Fig.~\ref{fig:Feyn}(h), is
\begin{equation}
	{\cal I}_\Gamma^{\text{(e)}} = s_\Gamma d'_1 d_1 
		\sum_r \cos^2\case{1}{2}k_r.
\end{equation}
The diagrams with a gluon going from a leg to its own rotation,
Figs.~\ref{fig:Feyn}(b) and~\ref{fig:Feyn}(f), are
\begin{eqnarray}
	{\cal I}_\Gamma^{\text{(b)}} & = & d_1  \frac{\zeta}{D}
		\left[(3-\case{1}{4} \hat{\bbox{k}}^2) L
			+ \case{1}{2} \zeta  \sum_r K_r  S_r^2 \right], \\
	{\cal I}_\Gamma^{\text{(f)}} & = & d'_1 \frac{\zeta'}{D'}
		\left[(3-\case{1}{4} \hat{\bbox{k}}^2) L'
			+ \case{1}{2} \zeta' \sum_r K'_r S_r^2 \right],
\end{eqnarray}
where $S_r=\sin k_r$, and the functions $D$, $L$, and $K_r$ (and primed
analogs) are given in Appendix~\ref{app:useful}.
The diagrams with a gluon going from a leg to the other rotation,
Figs.~\ref{fig:Feyn}(c) and~\ref{fig:Feyn}(g), are
\begin{eqnarray}
	{\cal I}_\Gamma^{\text{(c)}} & = & 
		s_\Gamma d_1 \frac{\zeta'}{D'}
		\left[(3-\case{1}{4} \hat{\bbox{k}}^2) \R{L'}
			+ \case{1}{2} \R{\zeta'} \sum_r K'_r S_r^2 \right], \\
	{\cal I}_\Gamma^{\text{(g)}} & = & 
		s_\Gamma d'_1 \frac{\zeta}{D}
		\left[(3-\case{1}{4} \hat{\bbox{k}}^2) \R{L}
			+ \case{1}{2} \R{\zeta}  \sum_r K_r  S_r^2 \right],
\end{eqnarray}
where~$s_\Gamma$ is given in Table~\ref{tab:s}.

The vertex diagram, Fig.~\ref{fig:Feyn}(a), is complicated.
We find ${\cal I}_\Gamma^{\text{(a)}}=N_\Gamma^{\text{(a)}}/DD'$,
with numerator
\begin{equation}
	N_\Gamma^{\text{(a)}} = 
		\R{U'_0} \R{U_0} - s_\Gamma \R{L'_0} \R{L_0} \bbox{S}^2 
		- \zeta\zeta' X_\Gamma.
\end{equation}
The part $X_\Gamma$ comes from spatial gluon exchange:
\begin{eqnarray}
	X_\Gamma & = &
		- s_\Gamma (3-\case{1}{4} \hat{\bbox{k}}^2) \R{L'} \R{L}
			\\ & &
		+ s_\Gamma^2 (3-\case{1}{4} \hat{\bbox{k}}^2) \R{V'} \R{V}
			\bbox{S}^2 \\ & &
		+ \case{1}{2}   \left( \R{V'}\R{U} - s_\Gamma \R{L'}
		\R{\zeta}\right)
			\sum_r K_rS_r^2
				\\ & &
		+ \case{1}{2}   \left( \R{U'}\R{V} - s_\Gamma \R{\zeta'} \R{L}\right)
			\sum_r K'_rS_r^2
				\\ & &
		+ \case{1}{4}\left( \R{U'}\R{U} - s_\Gamma \bbox{S}^2 \R{\zeta'}
		\R{\zeta}\right)
			\sum_r K'_rK_r\hat{k}_r^2
				\\ & &
		+ \case{1}{8} (1-s_\Gamma^2)
		\left(\hat{\bbox{k}}^2\bbox{S}^2-3\sum_r\hat{k}_r^2S_r^2\right)
		\R{V'} \R{V},
\end{eqnarray}
where the last term is absent for $V_4$ (\emph{i.e.}, when $s_\Gamma^2=1$).
The rotation enters in the ``rotated'' functions
\begin{eqnarray}
	\R{U_0} & = & U_0 + d_1 \bbox{S}^2 L_0, \\
	\R{L_0} & = & L_0 - d_1 U_0, \\
	\R{U} & = & U + d_1 \bbox{S}^2 \zeta, \\
	\R{\zeta} & = & \zeta - d_1 U, \\
	\R{V} & = & V + d_1 L, \\
	\R{L} & = & L - d_1 \bbox{S}^2 V,
\end{eqnarray}
and similarly for primed functions.
Although the vertex diagram is not easy to write down, the rotation
modifies it in a fairly simple way, when using the $2\times 2$ method
described above.

We have verified that these expressions are correct by completely
independent calculation with more common methods for the Dirac algebra.

\section{Useful Functions}
\label{app:useful}

In this appendix we list the functions appearing in
Appendix~\ref{app:dirac} for the action and currents given in
Sec.~\ref{sec:lattice}.
First, let
\begin{eqnarray}
	\mu  & = & 1 + m_0  + \case{1}{2} r_s  \zeta  \hat{\bbox{k}}^2 ,\\
	\mu' & = & 1 + m'_0 + \case{1}{2} r'_s \zeta' \hat{\bbox{k}}^2 .
\end{eqnarray}
from now on a prime means to replace incoming coupling and momentum
with corresponding outgoing coupling and momentum.

When the quark propagator is rationalized it has the denominator
\begin{equation}
	D = 1 - 2\mu\cos(k_4+im^{[0]}_1) + \mu^2 + \zeta^2 \bbox{S}^2,
\end{equation}
where $m_1^{[0]}=\log(1+m_0)$.

In this calculation, the incoming and outgoing heavy quarks both have
zero three-momentum, so both spinors consist only of upper components.
This feature is different from the heavy-light
case~\cite{Harada:2001hl} and explains, at a low level, why the only
dimension-three currents for heavy-heavy currents have
$\Gamma=\gamma_4$ and~$\gamma_j\gamma_5$ only.

To express the useful functions compactly, it is convenient to
introduce first
\begin{eqnarray}
		U	& = & \mu - e^{-m_1^{[0]}+ik_4} , \\
	\bar{U}	& = & \mu - e^{+m_1^{[0]}-ik_4} ,
\end{eqnarray}
because these combinations appear in the other functions.
Then
\begin{eqnarray}
	U_0 & = & U e^{+m_1^{[0]} - ik_4/2} 
		- \case{1}{2} \zeta^2 c_E \cos(\case{1}{2} k_4) \bbox{S}^2 , \\
	L_0 & = & \zeta\left[ e^{+m_1^{[0]} - ik_4/2} 
		+ \case{1}{2}         c_E \cos(\case{1}{2} k_4)
			\bar{U} \right] , \\
	 V  & = & \zeta\left[1 + \case{i}{2} c_E\sin(k_4)\right]
		+ \case{1}{2} c_B U , \\
	 L  & = & - \bar{U} \left[1 + \case{i}{2}
	 c_E\sin(k_4)\right] + \case{1}{2} c_B \zeta \bbox{S}^2 , \\
	K_r & = & r_s - c_B \cos^2(\case{1}{2} k_r) =
	         (r_s - c_B) + \case{1}{4} c_B \hat{k}_r^2 .
\end{eqnarray}
Some related functions appear when the Dirac matrix connects upper and
lower components, as in the case of heavy-light
currents~\cite{Harada:2001hl}.
They will arise for heavy-heavy matching of the dimension-four currents,
for which matrix elements with non-zero three-momentum must be
calculated.


\begin{thebibliography}{99}
%
\bibitem[*]{TO} Present address: Yukawa Institute, Kyoto University.
%
\bibitem{Kronfeld:2000ck}
A.~S.~Kronfeld,
Phys.\ Rev.\  D {\bf 62}, 014505 (2000)
[hep-lat/0002008].
%
\bibitem{Harada:2001hl}
J.~Harada, S.~Hashimoto, K.-I.~Ishi\-kawa, A.~S.~Kron\-feld, T.~Onogi, 
and N.~Yamada,
hep-lat/0112044.
%
\bibitem{Kronfeld:1993jf}
A.~S.~Kronfeld and P.~B.~Mackenzie,
Annu.\ Rev.\ Nucl.\ Part.\ Sci.\ {\bf 43}, 793 (1993)
[hep-ph/9303305].
For recent status, see 
S.~Hashimoto,
Nucl.\ Phys.\ B Proc.\ Suppl.\ {\bf 83}, 3 (2000)
[hep-lat/9909136];
S.~Aoki,
in {\it Proceedings of the XIX International Symposium on Lepton and
Photon Interactions at High Energy},
edited by J.~A. Jaros and M.~E. Peskin,
eConfC {\bf 990809}, 657 (2000)
[hep-ph/9912288];
S.~Ryan,
hep-lat/0111010.
%
\bibitem{El-Khadra:1997mp}
A. X. El-Khadra, A. S. Kronfeld, and P.~B. Mackenzie,
Phys. Rev. D {\bf 55}, 3933 (1997)
[hep-lat/9604004].
%
\bibitem{Wilson:1975hf}
K. G. Wilson,
in {\em New Phenomena in Subnuclear Physics},
edited by A. Zichichi (Plenum, New York, 1977).
%
\bibitem{Isgur:1989vq}
N. Isgur and M.~B. Wise,
Phys. Lett. B {\bf 232}, 113 (1989);
{\bf 237}, 527 (1990).
%
\bibitem{Symanzik:1979ph}
K. Symanzik,
in \emph{Recent Developments in Gauge Theories},
edited by G. 't~Hooft \emph{et al}.\ (Plenum, New York, 1980).
%
\bibitem{Symanzik:1983dc}
K. Symanzik,
in \emph{Mathematical Problems in Theoretical Physics},
edited by R. Schrader \emph{et al}.\ (Springer, New York, 1982);
Nucl.\ Phys.\ B {\bf 226}, 187, 205 (1983).
%
\bibitem{Luscher:1996sc}
M.~L\"uscher, S.~Sint, R.~Sommer, and P.~Weisz,
Nucl.\ Phys.\ B {\bf 478}, 365 (1996)
[hep-lat/9605038].
%
\bibitem{Jansen:1996ck}
K.~Jansen {\it et al.},
Phys.\ Lett.\ B {\bf 372}, 275 (1996)
[hep-lat/9512009];
M.~L\"uscher, S.~Sint, R.~Sommer, P.~Weisz, and U.~Wolff,
Nucl.\ Phys.\ B {\bf 491}, 323 (1997)
[hep-lat/9609035].
%
\bibitem{Sheikholeslami:1985ij}
B.~Sheikholeslami and R.~Wohlert,
Nucl.\ Phys.\  B {\bf 259}, 572 (1985).
%
\bibitem{Kronfeld:1995nu}
A.~S.~Kronfeld,
Nucl.\ Phys.\ B Proc.\ Suppl.\  {\bf 42}, 415 (1995)
[hep-lat/9501002].
%
\bibitem{Hashimoto:2000yp}
S.~Hashimoto \emph{et al.},
Phys.\ Rev.\  D {\bf 61}, 014502 (2000)
[hep-ph/9906376].
%
\bibitem{Hashimoto:2001ds}
S.~Hashimoto \emph{et al.},
hep-ph/0110253;
J.~N.~Simone \emph{et al.},
Nucl.\ Phys.\ B Proc.\ Suppl.\  {\bf 83}, 334 (2000)
[hep-lat/9910026].
%
\bibitem{El-Khadra:2001rv}
A.~X.~El-Khadra, A.~S.~Kronfeld, P.~B.~Mackenzie,
S.~M.~Ryan, and J.~N.~Simone,
Phys.\ Rev.\ D {\bf 64}, 014502 (2001)
[hep-ph/0101023].
%
\bibitem{Kronfeld:1999tk}
A.~S. Kron\-feld and S.~Ha\-shi\-mo\-to,
Nucl.\ Phys.\ B Proc.\ Suppl.\  {\bf 73}, 387 (1999)
[hep-lat/9810042].
%
\bibitem{Brodsky:1983gc}
S.~J.~Brodsky, G.~P.~Lepage, and P.~B.~Mackenzie,
Phys.\ Rev.\  D {\bf 28}, 228 (1983).
%
\bibitem{Lepage:1993xa}
G.~P.~Lepage and P.~B.~Mackenzie,
Phys.\ Rev.\  D {\bf 48}, 2250 (1993)
[hep-lat/9209022].
%
\bibitem{p:epaps}
For the program, see http://theory.fnal.gov/people/kronfeld/LatHQ2QCD/
or the EPAPS Document accompanying this paper and
Ref.~\cite{Kronfeld:1999tk}.
With the latter method, go to the EPAPS
home page http://www.aip.org/pubservs/epaps.html ,
or ftp.aip.org in the directory /epaps/.
See the EPAPS home page for more information.
(Of course, EPAPS is available only after publication.)
%
\bibitem{Eichten:1987xu}
E. Eichten, Nucl. Phys. B Proc. Suppl. {\bf 4}, 170 (1987).
%
\bibitem{Eichten:1990zv}
E. Eichten and B. Hill,
Phys. Lett. B {\bf 234}, 511 (1990);
{\bf 240}, 193 (1990).
%
\bibitem{Grinstein:1990mj}
B.~Grinstein,
Nucl.\ Phys.\ B {\bf 339}, 253 (1990).
%
\bibitem{Georgi:1990um}
H.~Georgi,
Phys.\ Lett.\ B {\bf 240}, 447 (1990).
%
\bibitem{Eichten:1990vp}
E.~Eichten and B.~Hill,
Phys.\ Lett.\ B {\bf 243}, 427 (1990).
%
\bibitem{Caswell:1986ui}
W.~E.~Caswell and G.~P.~Lepage,
Phys.\ Lett.\ B {\bf 167}, 437 (1986);
G.~T.~Bodwin, E.~Braaten and G.~P.~Lepage,
Phys.\ Rev.\ D {\bf 46}, 1914 (1992)
[hep-lat/9205006].
%
\bibitem{Lepage:1987gg}
G.~P. Lepage and B.~A. Thacker,
Nucl. Phys. B Proc. Suppl. {\bf 4}, 199 (1987);
B.~A. Thacker and G.~P. Lepage,
Phys.\ Rev.\ D {\bf 43}, 196 (1991);
G.~P. Lepage, L. Magnea, C.~Nakhleh, U. Magnea, and K. Hornbostel,
\emph{ibid.} {\bf 46}, 4052 (1992)
[hep-lat/9205007].
%
\bibitem{Hashimoto:1996in}
S.~Hashimoto and H.~Matsufuru,
Phys.\ Rev.\ D {\bf 54}, 4578 (1996)
[hep-lat/9511027].
%
\bibitem{Sloan:1998fc}
J.~H.~Sloan,
Nucl.\ Phys.\ B Proc.\ Suppl.\  {\bf 63}, 365 (1998)
[hep-lat/9710061].
%
\bibitem{Luke:1992cs}
M.~Luke and A.~V.~Manohar,
Phys.\ Lett.\ B {\bf 286}, 348 (1992);
M.~Neubert,
Phys.\ Lett.\ B {\bf 306}, 357 (1993);
R.~Sundrum,
Phys.\ Rev.\ D {\bf 57}, 331 (1998).
%
\bibitem{Trueman:1995ca}
T.~L.~Trueman,
Z.\ Phys.\ C {\bf 69}, 525 (1996)
[hep-ph/9504315].
%
\bibitem{Davies:1993ec}
C.~T.~H.~Davies and B.~A.~Thacker,
Phys.\ Rev.\ D {\bf 48}, 1329 (1993).
%
\bibitem{Boyle:2000fi}
P. Boyle and C. Davies,
Phys.\ Rev.\ D {\bf 62}, 074507 (2000)
[hep-lat/0003026].
%
\bibitem{Mertens:1998wx}
B. P. G. Mertens, A. S. Kronfeld, and A.~X. El-Khadra,
Phys. Rev. D {\bf 58}, 034505 (1998)
[hep-lat/9712024].
%
\bibitem{Kronfeld:1998di}
A.~S.~Kronfeld,
Phys.\ Rev.\ D {\bf 58}, 051501 (1998)
[hep-ph/9805215].
%
\bibitem{Sroczynski:2000hg}
Z.~Sroczynski,
hep-lat/0011059;
Nucl.\ Phys.\ B Proc.\ Suppl.\  {\bf 83}, 971 (2000)
[hep-lat/9910004].
%
\bibitem{Neubert:1995qt}
M.~Neubert,
Phys.\ Lett.\ B {\bf 341}, 367 (1995)
[hep-ph/9409453].
%
\bibitem{Gabrielli:1991us}
E.~Gabrielli \emph{et al.},
Nucl.\ Phys.\ B {\bf 362}, 475 (1991).
%
\bibitem{Martinelli:1983mw}
G.~Martinelli and Y.~Zhang,
Phys.\ Lett.\ B {\bf 123}, 433 (1983).
%
\bibitem{Lepage:1977sw}
G.~P.~Lepage,
J.\ Comput.\ Phys.\  {\bf 27}, 192 (1978);
Cornell University report CLNS-80/447 (unpublished, 1980).
%
\bibitem{Kawabata:1995th}
S.~Kawabata,
Comput.\ Phys.\ Commun.\  {\bf 88}, 309 (1995).
%
\bibitem{Hornbostel:2001ey}
K.~Hornbostel, G.~P.~Lepage and C.~Morningstar,
Nucl.\ Phys.\ B Proc.\ Suppl.\  {\bf 94}, 579 (2001)
[hep-lat/0011049].
%
\bibitem{Simone:2001zv}
J.~N. Simone, private communication.
These values of $Z_{V_\parallel}^{\text{NP}}$ were computed as part of
Ref.~\cite{El-Khadra:2001rv}.
%
\bibitem{Luke:1990eg}
M.~E. Luke,
Phys.\ Lett.\ B {\bf 252}, 447 (1990).
%
\bibitem{Kronfeld:1993dd}
A.~S.~Kronfeld and B.~P.~Mertens,
Nucl.\ Phys.\ B Proc.\ Suppl.\ {\bf 34}, 495 (1994)
[hep-lat/9312042].
%
\bibitem{Kuramashi:1998tt}
Y. Kuramashi,
Phys.\ Rev.\ D {\bf 58}, 034507 (1998)
[hep-lat/9705036].
%
\bibitem{Ishikawa:1997xh}
K.-I. Ishikawa, S. Aoki, S. Hashimoto, H. Matsufuru, T. Onogi,
and N. Yamada,
Nucl. Phys. B Proc. Suppl. {\bf 63}, 344 (1998)
[hep-lat/9711005]; 
K.-I. Ishikawa, T. Onogi, and N. Yamada,
Nucl.\ Phys.\ B Proc.\ Suppl.\ {\bf 83}, 301 (2000)
[hep-lat/9909159].
%
\bibitem{Harada:2001ei}
J.~Harada, A.~S.~Kronfeld, H.~Matsufuru, N.~Nakajima and T.~Onogi,
Phys.\ Rev.\ D {\bf 64}, 074501 (2001) [hep-lat/0103026].
%
\bibitem{Kronfeld:1985zv}
A.~S.~Kronfeld and D.~M.~Photiadis,
Phys.\ Rev.\ D {\bf 31}, 2939 (1985).
%
\end{thebibliography}
\end{document}